\documentclass[pra,aps,psfig,epsf,eqsecnum,twocolumn,amsmath,showpacs,amssymb,groupedaddress]{revtex4}
\usepackage{bm}
\usepackage{amsmath}

\DeclareMathOperator\arcsinh{ArcSinh}

\DeclareMathOperator\am{am}

\DeclareMathOperator\const{const}

\usepackage[pdftex]{graphicx}
\begin{document}

\title{Quench dynamics of the spin-imbalanced Fermi-Hubbard model in one dimension} 
\author{Xiao Yin}
\email[]{xiao.yin@colorado.edu}

\author{Leo Radzihovsky}
\email[]{radzihov@colorado.edu}
\affiliation{Department of Physics, Center for Theory of Quantum Matter, University of Colorado, Boulder, CO 80309\\
and Kavli Institute for Theoretical Physics, University of California, Santa Barbara, CA 93106}
\date{\today}
\date{\today}

\begin{abstract}
  We study a nonequilibrium dynamics of a one-dimensional spin-imbalanced
  Fermi-Hubbard model following a quantum quench of on-site
  interaction, realizable, for example, in Feshbach-resonant atomic
  Fermi gases. We focus on the post-quench evolution starting from the
  initial BCS and Fulde-Ferrell-Larkin-Ovchinnikov (FFLO) ground
  states and analyze the corresponding spin-singlet, spin-triplet,
  density-density, and magnetization-magnetization correlation
  functions. We find that beyond a light-cone crossover time, rich
  post-quench dynamics leads to thermalized and pre-thermalized stationary states that display strong dependence on the initial ground state. For initially
  gapped BCS state, the long-time stationary state resembles
  thermalization with the effective temperature set by the initial value
  of the Hubbard interaction. In contrast, while the initial gapless
  FFLO state reaches a stationary pre-thermalized form, it remains far
  from equilibrium.  We suggest that such post-quench dynamics can be
  used as a fingerprint for identification and study of the FFLO
  phase.

\end{abstract}


\maketitle

\section{Introduction}

\subsection{Background and motivation}

Progress in trapping and cooling of Feshbach-resonant (FR) atomic
gases has enabled extensive studies of strongly interacting quantum
matter in a broad range of previously unexplored regimes
\cite{Bloch,Ketterle,GRaop,SRaop,Giorgini,RadzihovskysheehyRPP}.  A
large variety of realized states includes s-wave paired fermionic
superfluids (SF)
\cite{RegalGreiner,Zwierlein,Kinast,Bartenstein,Bourdel} and the
associated Bardeen-Cooper-Schrieffer (BCS) to Bose-Einstein
condensation (BEC) crossover
\cite{Eagles,Leggett,Nozieres,DeMelo,Timmermans2,Holland,Ohashi,
AGR04,GRaop,Stajic,Chinsuperfluidity,Partridge}.

Atomic species number imbalance $m = n_\uparrow-n_\downarrow$
(corresponding to magnetization, $m$, conjugate to a Zeeman field, $h$
(a pseudo-spin chemical-potential difference) in a solid-state
context) \cite{Zwierlein2,Hulet1,Shin,Nascimbene} frustrates
Feshbach-resonant BCS pairing of a two-component
($\uparrow,\downarrow$) Fermi gas, driving quantum phase transitions
from a fully paired superfluid to a variety of other possible ground
states
\cite{Combescot,VincentLiuPRL,Bedaque,Caldas,Reddy,Cohen,SRPRL06,PaoCH,DTSon,SRPRB07,Castorina,Sedrakian,Bulgac,Dukelsky}. In
addition to ubiquitous phase separation \cite{Reddy,SRaop}, a
weakly-imbalanced attractive Fermi gas is predicted to exhibit the
enigmatic Fulde-Ferrell-Larkin-Ovchinnikov state (FFLO) \cite{SRaop},
first proposed in the context of solid-state superconductors over 45
years ago \cite{Fulde,Larkin} and studied extensively
\cite{Casalbuoni} in problems ranging from heavy-fermion
superconductors \cite{Radovan,Capan} to dense nuclear matter
\cite{Alford,Bowers}. Fundamentally, the FFLO state is a Cooper-pair
density wave (PDW) \cite{Berg,SRaop}, characterized by a finite
center-of-mass momentum $Q = k_{F\uparrow} - k_{F\downarrow}$, set by
imposed species imbalance (pseudo-magnetization). Akin to a paired
supersolid \cite{AndreevSS, Thouless, KimChenSS}, the state
spontaneously ``breaks'' gauge and translational symmetry, i.e., it is
a periodically paired gapless superfluid (superconductor),
characterized by a spatially periodic Ginzburg-Landau order parameter,
coupled to gapless quasi-particles.

In three dimensions the simplest form of this state is quite
fragile, predicted to occupy only a narrow sliver of the
interaction-versus-imbalance BCS-BEC phase diagram \cite{SRaop} and
consistent with early imbalanced trapped Fermi gas experiments
\cite{Zwierlein2,Hulet1}. In contrast FFLO ground state is
significantly more stable in lattice systems \cite{Koponen08, LohTrivedi} and in
quasi-one-dimension \cite{Parish, Sun13}, and in fact in one dimension is generic at any nonzero
imbalance
\cite{LutherEmery,JaparidzeNersesyan,Giamarchi,Tsvelik,KunYang}. Though so
far it has eluded a definitive observation, some promising solid-state
\cite{Capan, Radovan, Lortz} and quasi-one-dimensional atomic \cite{Hulet2} candidate systems
have recently been realized.

Distinguished by their coherence and tunability Feshbach-resonant
gases \cite{Bloch, Chin2, Zwierlein, GRaop} also enabled
experimental studies of highly $\it nonequilibrium$ many-body states,
with a particular focus on the dynamics following a quantum
Hamiltonian quench $\hat H_i\rightarrow\hat H_f$ \cite{Cardy, CazalillaReview, Polkovnikov,Langen,Mitra,Yin1, Yin2, Makotyn}.
These have raised numerous fundamental questions on thermalization
under unitary time evolution $|\hat{\psi}(t)\rangle=e^{-i H_f
  t}|\hat{\psi}_i(0)\rangle$ of a closed quantum system vis-\'a-vis
eigenstate thermalization hypothesis \cite{Srednicki,Rigol}, role of
conservation laws and obstruction to full equilibration of integrable
models argued to instead be characterized by a generalized Gibbs
ensemble (GGE), and emergence of statistical mechanics description of
stationary states \cite{Kinoshita,Rigolprl}.
These questions of post-quench dynamics have been extensively explored
theoretically in a large number of systems \cite{AGR04, Barankov04, AltmanVishwanath05,Yuzbashyan06,CazalillaPRA,CazalillaJOP,Cardy2,
  Mitra2, Mitra3, GurarieIsing13,Sondhi13,NatuMueller13,HungChin13,Yin1,
  Bacsi13,Essler14,NessiCazalilla14,Yin2,Foster10, Foster11, Marcuzzi16}.

Motivated by these studies and by the experimental progress 
 toward a realization of the 1D FFLO state in a Feshbach-resonant atomic Fermi gas (showing indirect experimental evidence through species-resolved density profiles) \cite{Hulet2}, here we study the interaction-quench dynamics of a 1D (pseudo-) spin-imbalanced attractive Fermi-Hubbard model \cite{ RieggerMeisnerPRA15}. We utilize
the power of bosonization and re-fermionization available in one dimension to
treat the low-energy dynamics. We study a variety of space-time
correlation functions following the interaction quench from the
fully-gapped BCS and from the gapless FFLO states to the
noninteracting state. We predict stationary pre-thermalized states
emerging beyond a crossover time set by the light-cone
dynamics \cite{Cardy}, that in the case of the initial BCS state can be
associated with an effective thermalization at temperature set by the
interaction energy of the initial state. In contrast, the FFLO state
never thermalizes, as expected \cite{Cardy2} due to its gapless nature.
We suggest that such post-quench dynamics can be used as a
finger-print for identification and study of the FFLO phase.

\subsection{Outline}
The rest of the paper is organized as follows.  We conclude the
Introduction with a summary of our key results. In Sec. II,
starting with a generic one-dimensional spin-imbalanced attractive Hubbard model,
we recall the basics of bosonization and re-express the Hubbard model
and various correlation functions in the bosonization language. In
Sec. III and IV we briefly review the equilibrium properties of
the charge and spin sectors of the model, with the former described by the
gapless Luttinger model and the latter characterized by the
sine-Gordon model, that (for the Luttinger spin parameter
$K_\sigma=1/2$) can be treated exactly via the Luther-Emery (LE)
approach \cite{LutherEmery}. The latter also provides a clear physical
picture for the formation of the spin-gap as well as the
commensurate-incommensurate (CI) phase transition
\cite{PokrovskyTalapov} between the BCS (spin-gapped) and the FFLO
(spin-gapless) phase driven by a Zeeman field
\cite{LutherEmery,JaparidzeNersesyan,KunYang,Giamarchi}. In Sec. V,
we study the spin sector correlators for the generic case away from
the LE point by adding quantum fluctuations around the semiclassical
soliton lattice solution. In Sec. VI, we summarize the equilibrium
properties of the 1D Hubbard model, by combining the charge and spin
sector correlations to compute the spin-singlet, spin-triplet,
density-density, and magnetization-magnetization correlation functions
in coordinate- and momentum spaces for the BCS state and FFLO state,
respectively. Our key new results for the quench dynamics begin with
Secs. VII and VIII, where we compute post-quench dynamics for the
charge and spin sectors for the initial BCS and FFLO states.  In
Sec. IX we combine these results to predict the post-quench
dynamics of the 1D spin-imbalanced attractive Fermi-Hubbard model. We
conclude in Sec. X with the discussion of these results and
relegate the details of the calculations to appendixes.

\subsection{Summary of results}
Before turning to the derivation and analysis, we briefly summarize
the key results of our work, here and throughout the paper utilizing
units such that $\hbar=1$ and $k_B=1$. Using bosonization, we have
calculated an array of correlation functions of the 1D spin-imbalanced
attractive Fermi-Hubbard model in equilibrium and for the
interaction-quench dynamics.
 
We recall that the system in question is well
known \cite{Giamarchi,Tsvelik,Guan13} to exhibit two qualitatively distinct
phases, the fully gapped BCS (``Commensurate,'' C) and the
spin-gapless FFLO (``Incommensurate,'' I) states, for pseudo-Zeeman
field (flavor chemical potential difference) $h < h_c$ and $h > h_c$,
respectively.  These balanced and imbalanced phases are separated by a
CI (Pokrovsky-Talapov) \cite{PokrovskyTalapov} phase transition at $h_c$, set by the attractive interaction strength $U$.

\begin{figure}[!htb]
 \centering
  \includegraphics[width=90mm]{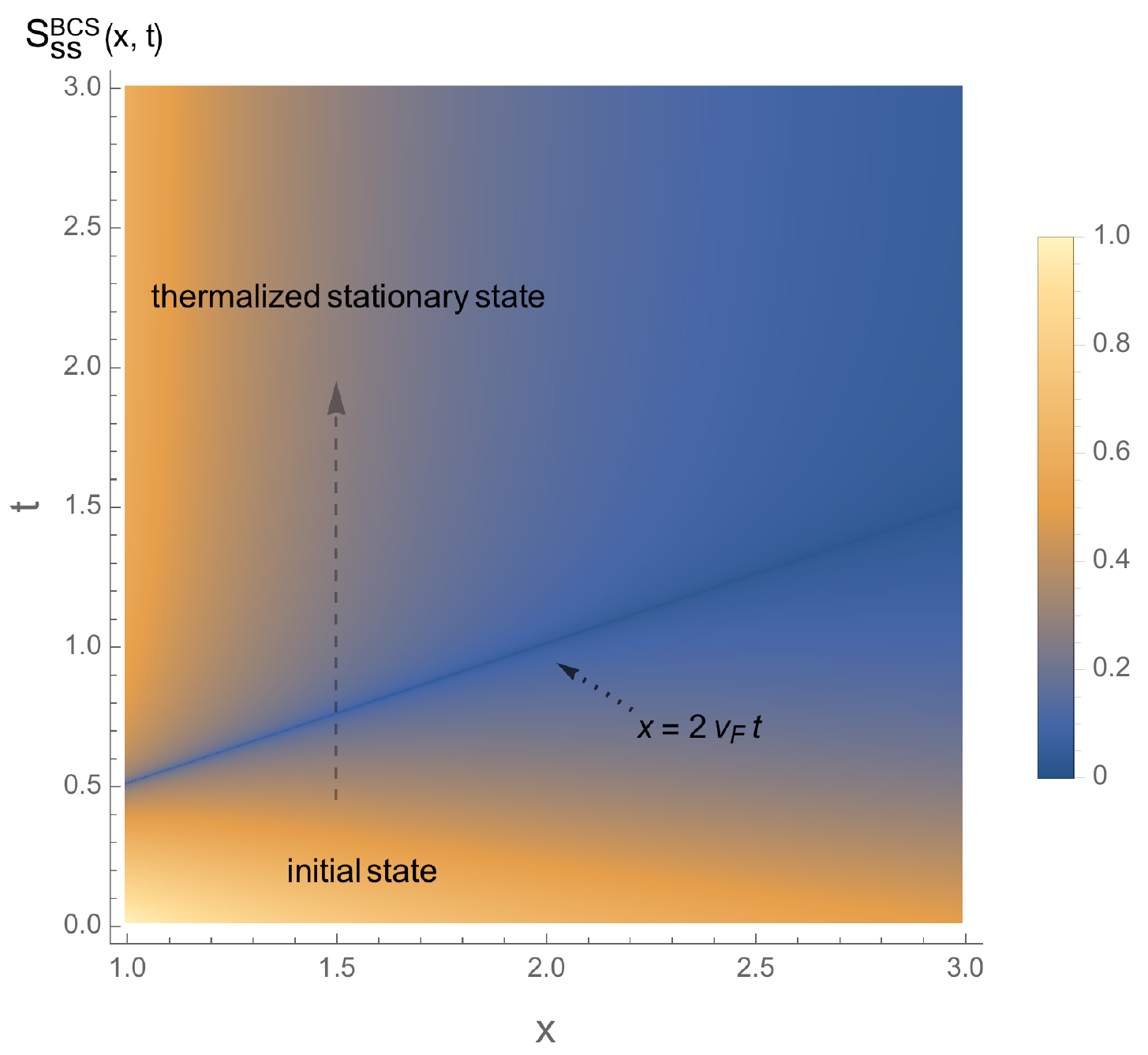}
  \caption{Space-time intensity plot of spin-singlet correlation
    function $S^{\mbox{\tiny\itshape BCS}}_{ss}(x,t)$ following a
    $U\to 0$ quench at $t=0$ from the BCS state. For short times below
    the light-cone boundary ($x=2v_F t$, appearing as a line of nodes
    feature), the correlation $S^{\mbox{\tiny\itshape BCS}}_{ss}(x,t)$
    approaches the initial state one and thus varies similarly with
    $x$ for different $t$. For long times above the boundary, the
    time-dependence drops out and the system develops into thermalized
    stationary state, with correlations in agreement with the one of
    the post-quench free-fermion state at a finite temperature $T$
    fixed by energy conservation.}
\label{IntensityBCSss}
\end{figure}

\begin{figure}[!htb]
 \centering
  \includegraphics[width=80mm]{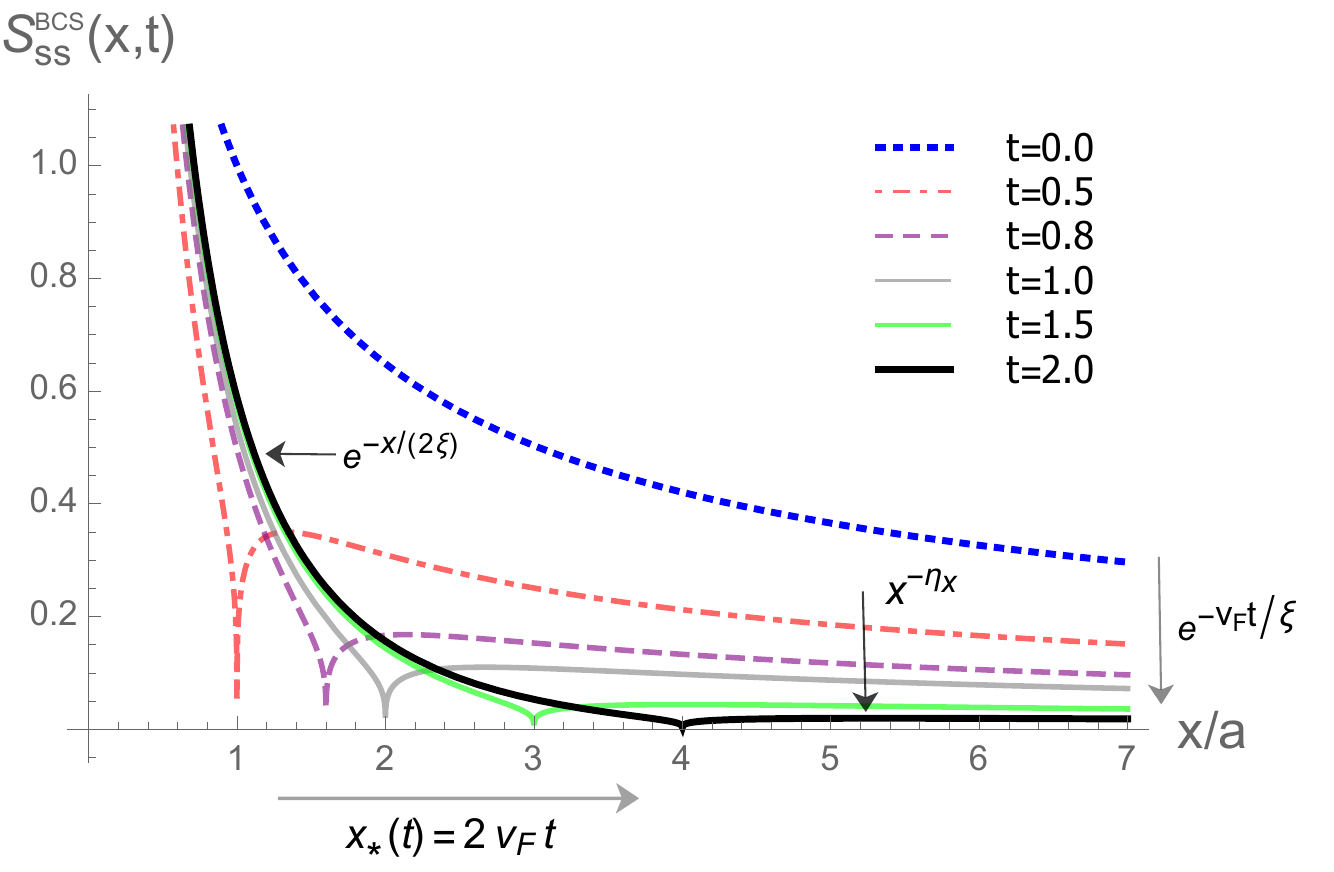}
 \caption{Spin-singlet pairing correlation function
   $S^{\mbox{\tiny\itshape BCS}}_{ss}(x,t)$ following a $U\to 0$
   quench at $t=0$ from the BCS state for a series of times.  The
   light-cone crossover, $x_*(t)=2v_F t$ separates exponential
   correlations $e^{-x/(2\xi)}$ at short scales, $x<x_*(t)$ from the
   power-law correlations $x^{-\eta_x}$ at long scales,
   $x>x_*(t)$. The latter power-law spatial correlations decay
   exponentially in time, as indicated by a vertical arrow on the
   right.}
\label{DynamicsBCSnqpairrealspace}
\end{figure}

Our key new results are the post-quench dynamical correlation
functions, that qualitatively depend on the phase of the initial
state. For interaction quench from the BCS spin gapped to the
noninteracting Fermi gas, we find the spin-singlet pairing correlation
function at time $t$ after the quench (illustrated as an intensity
space-time plot in Fig.~\ref{IntensityBCSss} and as a fixed time cuts
in Fig.~\ref{DynamicsBCSnqpairrealspace}) to be given by
\begin{align}
\begin{split}
S^{\mbox{\tiny\itshape BCS}}_{ss}(x,t)&\sim\begin{cases} \left(\frac{a}{x}\right)^{\eta_{x}}\left(\frac{a}{2v_Ft}\right)^{\eta_t}e^{-\frac{v_F t}{\xi }}, &x\gg 2v_F t,\\ 
\left(\frac{a}{x}\right)^{\eta_x+\eta_t}e^{-\frac{x}{2\xi}} , & x\ll 2v_F t,\end{cases}
\end{split}
\label{summarydynamicsBCSss}
\end{align} 
where $\xi$ and $v_F$ are the correlation length and Fermi velocity
respectively.  Here $a$ is the UV cutoff set by the lattice spacing. The space- and time- power-law exponents, $\eta_{x,t}$ satisfy $1/2<\eta_x<1$ and $0<\eta_t<3/4$. The crossover from correlations in the initial state
takes place at the light-cone crossover time $t_*(x)\equiv x/(2v_F)$,
such that at longer times, $t\gg t_*(x)$, a stationary
state emerges, characterized by exponentially short-ranged spatial
correlations. These are to be distinguished from the power-law $1/x^2$
$T=0$ post-quench correlations of a noninteracting Fermi gas.

However, the exponentially short-ranged spin part indeed resembles the
equilibrium free Fermi gas correlations at a finite temperature
$T\sim U$ [see Eq.~\eqref{thermaltemperature}], indicating thermalization
for a quench from the spin-gapped BCS state.

\begin{figure}[!htb]
 \centering
  \includegraphics[width=90mm]{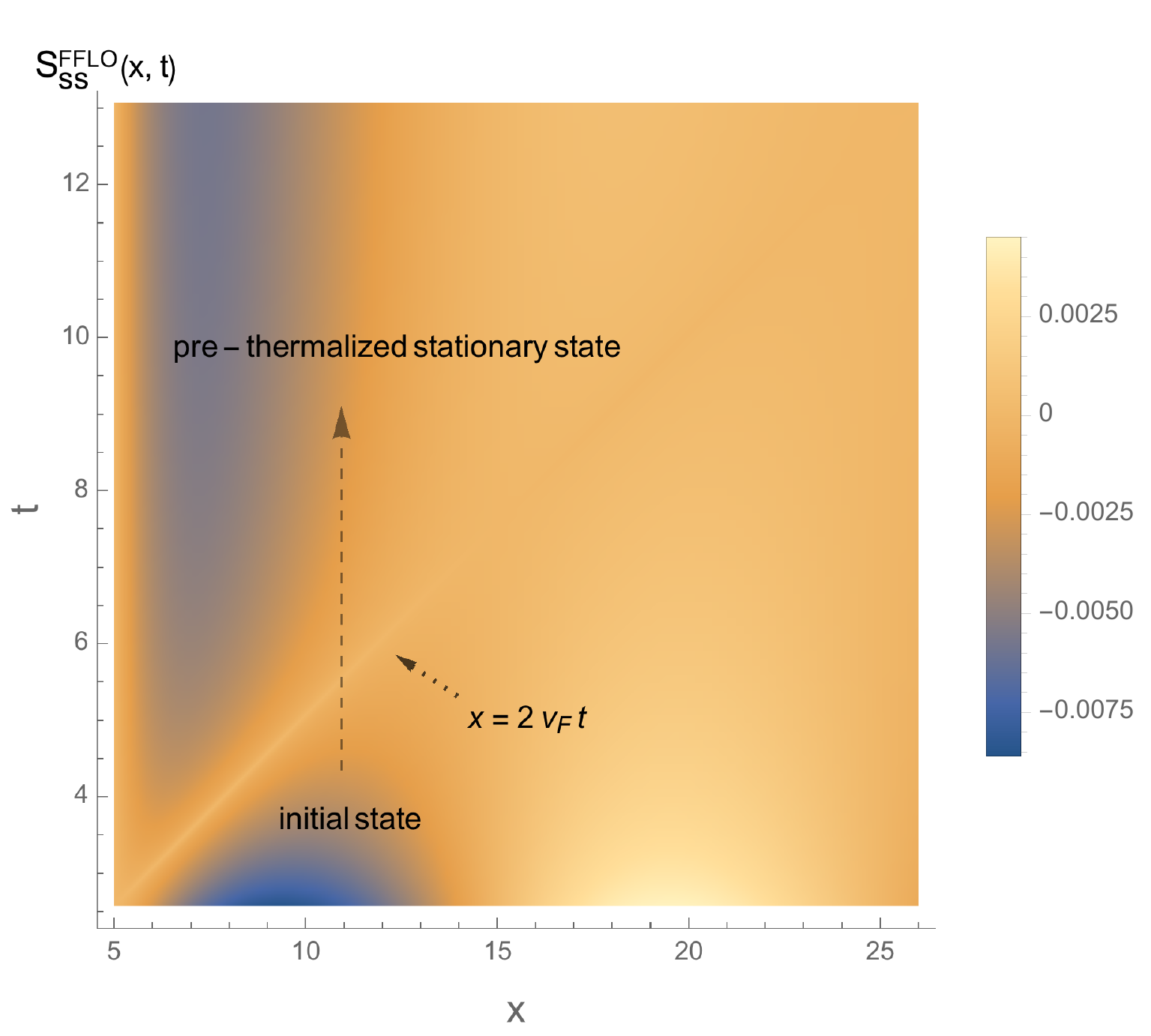}
  \caption{Space-time intensity plot of spin-singlet pairing
    correlation function $S^{\mbox{\tiny\itshape FFLO}}_{ss}(x,t)$
    following a $U\to 0$ quench at $t=0$ from the FFLO state.  In
    contrast to the BCS initial state, the correlations display
    spatial oscillations at the characteristic wavevector
    $k_{FFLO}$. A light-cone boundary $x=2v_F t$ distinguishing short-
    and long- time dynamics is visible.}
\label{IntensityFFLOss}
\end{figure}

\begin{figure}[!htb]
 \centering
  \includegraphics[width=80mm]{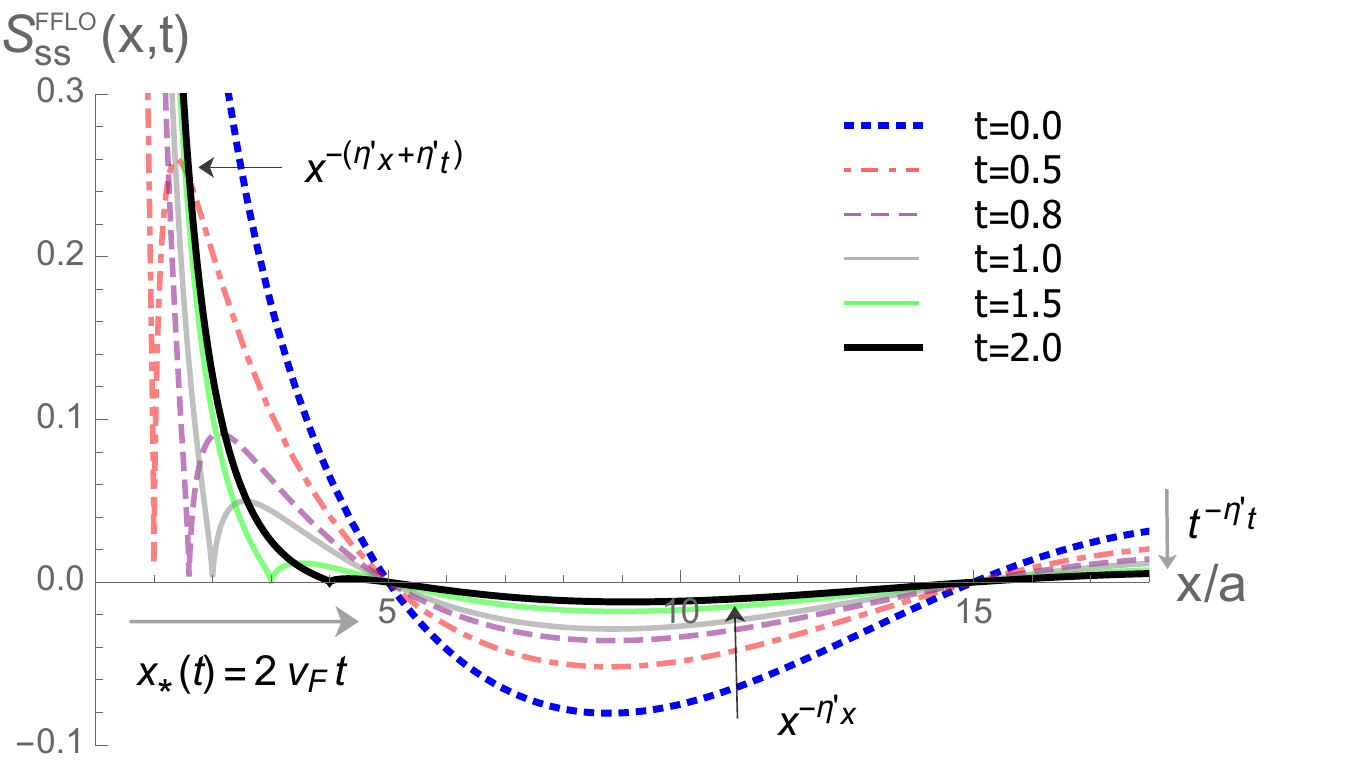}
  \caption{Spin-singlet pairing correlation function
    $S^{\mbox{\tiny\itshape FFLO}}_{ss}(x,t)$ following a $U\to 0$
    quench at $t=0$ from the FFLO state for a series of times. A
    nonzero momentum spatial oscillations at $k=k_{FFLO}$ persist as
    the signature of the FFLO state, contrasting with
    Fig.~\ref{DynamicsBCSnqpairrealspace} for the BCS state. The
    light-cone boundary, $x_*(t)=2v_F t$ separates a spatial power-law
    envelope $x^{-(\eta'_x+\eta'_t)}$ at short scales, $x<x_*(t)$ from
    a shorter-range power-law $x^{-\eta'_x}$ at long scales,
    $x>x_*(t)$. In the latter regime the overall amplitude decays as a
    power-law with time, indicated by a vertical arrow on the right.}
\label{DynamicsFFLOnqpairrealspace}
\end{figure}

On the other hand, following the quench from an initial FFLO
spin-gapless state to a noninteracting Fermi gas, we find that the
dynamical spin-singlet pairing correlation function has the following
asymptotics
\begin{align}
\begin{split}
S^{\mbox{\tiny\itshape FFLO}}_{ss}(x,t) &\sim\cos(k_{FFLO}x)\begin{cases} \left(\frac{a}{x}\right)^{\eta'_x}\left(\frac{a}{2v_{F} t}\right)^{\eta'_{t}}
 , & x\gg 2v_F t, \\ 
\left(\frac{a}{x}\right)^{\eta'_x+\eta'_t}, & x\ll 2v_F t,\end{cases}
\end{split}
\end{align}
with the full expression given in Sec.~\ref{quenchHubbardFFLO}, the
intensity space-time profile illustrated in Fig.~\ref{IntensityFFLOss}, 
and fixed time cuts plotted in
Fig.~\ref{DynamicsFFLOnqpairrealspace}. We observe that spatial
oscillations at $k=k_{FFLO}$, characteristic of the FFLO initial state
persist for all times, though at the light-cone time, $t_*(x)$ the
spatial power-law amplitude of the initial state crosses over to a
shorter-range power-law correlations. Despite that for longer times
$t>t_*(x)$ the dephasing leads to a development of a stationary state,
integrability together with the state's gapless nature forbids full
thermalization to exponential correlations of a free Fermi gas.

\begin{figure}[!htb]
 \centering
  \includegraphics[width=80mm]{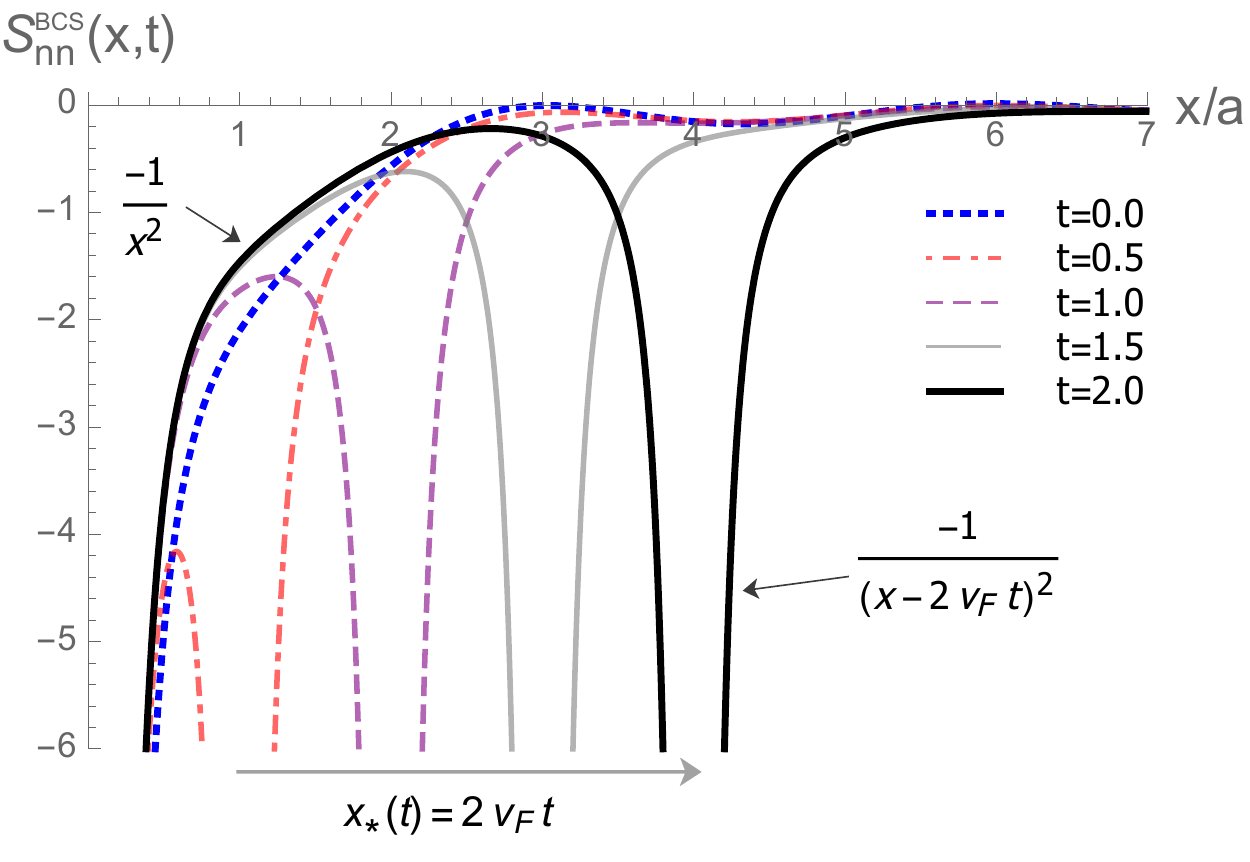}
 \caption{Density-density correlation function $S^{\mbox{\tiny\itshape
       BCS}}_{nn}(x,t)$ following a $U\to 0$ quench at $t=0$ from the
   BCS state for a series of times. The dominant features is the
   moving power-law peak $-1/(x-2v_Ft)^{2}$ at the light-cone
   boundary.}
\label{DynamicsBCSdensitydensityreal}
\end{figure}

We also computed the post-quench evolution of the density-density
correlation function for quench from both spin-gapped BCS state and spin-gapless FFLO states. For the spin-gapped BCS state case, the density-density correlation is given by
\begin{align}
\label{DynamicsBCSdensitydensityrealeqn}
\begin{split}
&S^{\mbox{\tiny\itshape BCS}}_{nn}(x,t)\sim \\
& \begin{cases}- \gamma_x\frac{a^2}{x^{2}}+\left(\frac{ a }{x}\right)^{\gamma_x} \left(\frac{2v_{F} t}{a}\right)^{\gamma_{t}}
  e^{-\frac{v_F t}{\xi}}\cos(2k_F x)
  , & x\gg 2v_F t,\\
   -\frac{\gamma_t}{2}\frac{a^2}{(x-2v_Ft)^{2}} , &x\approx 2v_F t,\\
 - (\gamma_x-\gamma_t)\frac{a^2}{x^{2}} +\left(\frac{ a }{x}\right)^{\gamma_x-\gamma_t}  e^{-\frac{x}{2\xi}}\cos(2k_F x)
 , &x\ll 2v_\sigma t,\end{cases}
\end{split}
\end{align}
and illustrated in Fig.~\ref{DynamicsBCSdensitydensityreal}. Here the
space- and time- power-law exponents, $\gamma_{x,t}$ satisfy
$0<\gamma_t<1<\gamma_x<2$. A striking feature of
$S^{\mbox{\tiny\itshape BCS}}_{nn}(x,t)$ is the divergent peak
associated with the light-cone boundary $x=2v_Ft$. It delineates the
short-time ($t<t_*(x)$) correlation of the initial state (that decay
in time) and the long time ($t>t_*(x)$) regime where the stationary state
emerges. In contrast to long-time pairing correlations, here the
exponentially suppressed spin-gapped correlations are dominated by the
gapless charge $1/x^2$ correlations. Similar results for the quench from the FFLO state are presented in Sec.~\ref{quenchHubbardFFLO}. 

Another interesting quantity is the magnetization and its
correlations. In contrast to the mean-field oscillatory magnetization
profile (qualitatively valid in higher dimensions),
enhanced quantum fluctuations of the 1D geometry completely wash out
this feature. However, they are manifested in the
magnetization-magnetization correlation function which we illustrate in Sec.~\ref{quenchHubbard}.

We next turn to the analysis of the 1D spin-imbalanced attractive
Fermi-Hubbard model that leads to the above and a number of other
results.

\section{Model and its bosonization}
\label{sectionmodel}
One-dimensional spin-imbalanced Fermi gas can be well-described by the
Fermi-Hubbard model
\begin{align}
\label{spinimbalanceHubbard}
\begin{split}
\hat H=-t\sum_{\langle ij\rangle,\sigma}\hat c^\dagger_{i,\sigma}\hat c_{j,\sigma}-U\sum_j \hat n_{j\uparrow}\hat n_{j\downarrow}+h\sum_j(\hat n_{j\uparrow}-\hat n_{j\downarrow}),
\end{split}
\end{align}
where $t$ is the nearest-neighbor hopping matrix element, $U$ the
on-site attractive interaction and $h=(\mu_\downarrow-\mu_\uparrow)/2$
the pseudo-Zeeman field. Throughout the paper, we work in the theoretically more convenient grand-canonical ensemble with the pseudospin- (species) imbalance $m=n_{j\uparrow}-n_{j\downarrow}$ tuned by h. For experiments conducted with fixed spin-imbalance, our theory can be easily applied by mapping h to m through the m(h) relation that we derived in Sec. \ref{spinsemiclassical}. Despite the accessibility of numerical approaches such as DMRG and exact diagonalization, in one dimension we
can utilize the powerful analytical machinery of
bosonization \cite{Coleman,Giamarchi,Tsvelik} to
obtain asymptotically exact behavior and to gain more physical
insight.

Bosonization allows a representation of fermionic operators in terms
of chiral bosonic phase fields, the phonon ${\hat{\phi}}$ and superfluid
phase ${{\hat{\theta}}} $,
\begin{align}
\begin{split}
 \hat{\psi}_\sigma(x_j)=\frac{1}{\sqrt{2\pi a}}\hat c_{j,\sigma} =  \hat{\psi}_{R\sigma}(x_j)+\hat{\psi}_{L\sigma}(x_j),
\end{split}
\end{align}
where $x_j = a j$ and
\begin{align}
\label{bosonizationpsithetaphi}
\begin{split}
  \hat{\psi}_{r,\sigma}(x)&=\hat U_{r,\sigma}\lim_{a\to0}\frac{1}{\sqrt{2\pi a }}\\
 &\quad\times e^{irk_Fx}e^{-\frac{i}{\sqrt{2}}[r{\hat{\phi}}_\rho(x)-{{\hat{\theta}}} _\rho(x)+\sigma(r{\hat{\phi}}_\sigma(x)-{\hat{\theta}} _\sigma(x))]},
 \end{split}
 \end{align}
 with $r=\pm 1$ for right (R) and left (L) movers, and $\sigma=\pm 1$
 for spin-up and spin-down, respectively. Here $a$ is an
 ultra-violet (UV) cutoff set by the lattice constant and
 $U_{r,\sigma}$ the standard Klein factor that ensures
 anti-commutation of fermionic operators. The bosonic phases satisfy the following
commutation relations
 \begin{align}
\begin{split}
[{\hat{\phi}}_\rho(x),\partial_{x'} {\hat{\theta}} _\rho(x')]&=i\pi\delta(x-x'),\\
[{\hat{\theta}}_\rho(x),\partial_{x'} {\hat{\phi}}_\rho(x')]&=i\pi\delta(x-x'),\\
[{\hat{\phi}}_\sigma(x),\partial_{x'} {\hat{\theta}} _\sigma(x')]&=i\pi\delta(x-x'),\\
[{\hat{\theta}}_\sigma(x),\partial_{x'} {\hat{\phi}}_\sigma(x')]&=i\pi\delta(x-x').
\end{split}
\end{align}

The charge $(\rho)$ and spin ($\sigma$) phase fields are given by
\begin{align}
\begin{split}
{\hat{\phi}}_\rho&=\frac{{\hat{\phi}}_\uparrow+{\hat{\phi}}_\downarrow}{\sqrt{2}},\;\;\;{\hat{\theta}} _\rho=\frac{{\hat{\theta}} _\uparrow+{\hat{\theta}} _\downarrow}{\sqrt{2}},\\
{\hat{\phi}}_\sigma&=\frac{{\hat{\phi}}_\uparrow-{\hat{\phi}}_\downarrow}{\sqrt{2}},\;\;\;{\hat{\theta}} _\sigma=\frac{{\hat{\theta}} _\uparrow-{\hat{\theta}} _\downarrow}{\sqrt{2}},
\end{split}
\end{align}
respectively. Using above relations and taking the continuum limit,
the Hubbard Hamiltonian can be re-expressed in terms of the bosonic
fields,
\begin{align}
\begin{split}
\hat H&=\hat H_\rho+\hat H_\sigma,
\end{split}
\label{chargespinhamiltonian}
\end{align}
to lowest order separating into the charge sector,
\begin{align}
\label{LLH}
\begin{split}
\hat H_\rho&=\frac{v_\rho}{2\pi}\int dx\left[\frac{1}{K_\rho}(\partial_x{\hat{\phi}}_\rho)^2+K_\rho(\partial_x{\hat{\theta}} _\rho)^2\right],
\end{split}
\end{align}
and the spin sector,
\begin{align}
\label{SineGordonH}
\begin{split}
\hat   H_\sigma&=\frac{v_\sigma}{2\pi}\int
  dx\left[\frac{1}{K_\sigma}(\partial_x{\hat{\phi}}_\sigma)^2+K_\sigma(\partial_x{\hat{\theta}} _\sigma)^2
\right]\\
  &\quad-\frac{U}{2\pi^2 a}\int dx
  \cos(\sqrt{8}{\hat{\phi}}_\sigma)-\frac{\sqrt{2}h}{\pi}\int dx\partial_x
  {\hat{\phi}}_\sigma.
\end{split}
\end{align}
Above we have neglected the spin-imbalance-induced spin-charge
coupling that is weak for $m/k_F\sim(k_{F\uparrow} -
k_{F\downarrow})/k_F\ll1$ \cite{VincentLiuPRA,Rizzi}.  The parameters
$K_{\rho,\sigma}$ and $v_{\rho,\sigma}$ are, respectively, the Luttinger
parameters and the velocities for the charge and spin part,
respectively, and are perturbatively [in $U a/(\pi v_F)$] related to
the original Hubbard model parameters through
\begin{align}
\label{bosonizationrelations}
\begin{split}
v_F&=2 t a\sin(k_Fa),\\
v_\rho&=v_F\sqrt{1-\frac{U a}{\pi v_F}},\\
v_\sigma&=v_F\sqrt{1+\frac{U a}{\pi v_F})},\\
\frac{1}{K_\rho}&=\sqrt{1-\frac{U a}{\pi v_F}},\\
\frac{1}{K_\sigma}&=\sqrt{1+\frac{U a}{\pi v_F}}.
\end{split}
\end{align}
For the strong interaction above relations break down, but are well known
to satisfy $K_\rho\to 2$ and $K_\sigma\to 1/2$ in the $U\to\infty$
limit \cite{Giamarchi}.

To probe the system, we focus on a variety of correlation functions of
the spin-singlet and triplet pairing operators:
\begin{subequations}
    \begin{align}
    \label{Ossoperatora}
\hat{O}_{ss}(x)&=\hat{\psi}^\dagger_{R\uparrow} \hat{\psi}^\dagger_{L\downarrow}+\hat{\psi}^\dagger_{L\uparrow}\hat{\psi}^\dagger_{R\downarrow} =\frac{1}{\pi a }e^{-i\sqrt{2}{\hat{\theta}} _\rho}\cos(\sqrt{2}{\hat{\phi}}_\sigma),\\
\hat{O}_{st}(x)&=\hat{\psi}^\dagger_{R\uparrow} \hat{\psi}^\dagger_{L\uparrow}+\hat{\psi}^\dagger_{L\downarrow}\hat{\psi}^\dagger_{R\downarrow}=\frac{1}{\pi a }e^{-i\sqrt{2}{\hat{\theta}} _\rho}\cos(\sqrt{2}{\hat{\theta}} _\sigma),
    \label{Ossoperatorb}
  \end{align}
 \end{subequations}
as well as of the number and magnetization density operators,
 \begin{subequations}
    \begin{align}
 \hat{n}(x)&=\hat{\psi}^\dagger_\uparrow\hat{\psi}_\uparrow+\hat{\psi}^\dagger_\downarrow\hat{\psi}_\downarrow
=-\frac{\sqrt{2}}{\pi}\partial_x {\hat{\phi}}_\rho(x)+\hat{O}_{CDW}+h.c.,\\
\hat{m}(x)&=\hat{\psi}^\dagger_\uparrow\hat{\psi}_\uparrow-\hat{\psi}^\dagger_\downarrow\hat{\psi}_\downarrow=-\frac{\sqrt{2}}{\pi}\partial_x {\hat{\phi}}_\sigma(x)+\hat{O}^z_{SDW}+h.c.,
    \end{align}
  \end{subequations}
  where, respectively, the charge- and spin-density wave operators,
\begin{subequations}
\begin{align}
  \hat{O}_{CDW}(x)&=\hat{\psi}^\dagger_{R\uparrow}
  \hat{\psi}_{L\uparrow}+\hat{\psi}^\dagger_{R\downarrow}
  \hat{\psi}_{L\downarrow}=\frac{e^{-2ik_F x}}{\pi a
  }e^{i\sqrt{2}{\hat{\phi}}_\rho}\cos(\sqrt{2}{\hat{\phi}}_\sigma),\\
  \hat{O}^z_{SDW}(x)&=\hat{\psi}^\dagger_{R\uparrow}
  \hat{\psi}_{L\uparrow}-\hat{\psi}^\dagger_{R\downarrow}\hat{\psi}_{L\downarrow}
  =\frac{e^{-2ik_F x}}{\pi a }e^{i\sqrt{2}{\hat{\phi}}_\rho}i\sin(\sqrt{2}{\hat{\phi}}_\sigma).
\end{align}
\end{subequations}
are defined as the oscillatory part of the density component.  

Thus in the bosonized form, the spin-singlet pairing correlation
is given by
\begin{align}
\label{Sssdefinition}
\begin{split}
S_{ss}(x)&\equiv (\pi a )^2\langle\hat{O}_{ss}(x)\hat{O}^\dagger_{ss}(0)\rangle,\\
&= \langle e^{-i\sqrt{2}[{\hat{\theta}} _\rho(x)-{\hat{\theta}} _\rho(0)]}\rangle
\langle \cos(\sqrt{2}{\hat{\phi}}_\sigma(x))\cos(\sqrt{2}{\hat{\phi}}_\sigma(0))\rangle,
\end{split}
\end{align}
and its triplet counterpart is given by
 \begin{align}
 \label{Sstdefinition}
\begin{split}
S_{st}(x)&\equiv (\pi a )^2 \langle\hat{O}_{ts}(x)\hat{O}^\dagger_{ts}(x)\rangle,\\
&=\langle e^{-i\sqrt{2}[{\hat{\theta}} _\rho(x)-{\hat{\theta}} _\rho(0)]}\rangle
 \langle \cos(\sqrt{2}{\hat{\theta}} _\sigma(x))\cos(\sqrt{2}{\hat{\theta}} _\sigma(0))\rangle.\end{split}
\end{align}

Similarly, the density-density and magnetization-magnetization 
correlation functions are given by 
\begin{subequations}
\begin{align}
S_{nn}(x)&\equiv  (\pi a )^2\langle \hat{n}(x) \hat{n}(0) \rangle
\nonumber\\
&=2a^2\langle\partial_x {\hat{\phi}}_\rho(x)\partial_{x'} {\hat{\phi}}_\rho(0)\rangle+2\cos(2k_F x)\langle e^{i\sqrt{2}[{\hat{\phi}}_\rho(x)-{\hat{\phi}}_\rho(0)}\rangle\nonumber\\
&\quad\times\langle
\cos(\sqrt{2}({\hat{\phi}}_\sigma(x))\cos(\sqrt{2}{\hat{\phi}}_\sigma(0))\rangle,\\
S_{mm}(x)&\equiv (\pi a )^2\langle \hat{m}(x) \hat{m}(0) \rangle
\nonumber\\
&=2a^2\langle\partial_x {\hat{\phi}}_\sigma(x)\partial_{x'} {\hat{\phi}}_\sigma(0)\rangle-2\cos(2k_F x)\langle e^{i\sqrt{2}[{\hat{\phi}}_\rho(x)-{\hat{\phi}}_\rho(0)}\rangle\nonumber\\
&\quad\times\langle \sin(\sqrt{2}{\hat{\phi}}_\sigma(x))\sin(\sqrt{2}{\hat{\phi}}_\sigma(0))\rangle.
\label{densitydensitycorrelationdefinition}
\end{align}
\end{subequations}

Having recalled the basic formalism, in the next few sections we will
utilize it to compute the ground-state properties of the
spin-imbalanced attractive Fermi-Hubbard model, and will then utilize
these to calculate the post-quench dynamics of the corresponding
correlators.

\section{charge sector}
\label{chargesector}

As mentioned in last section, one prominent feature of 1D system is
the spin-charge separation. Utilizing this property, we
briefly review the equilibrium charge sector results in this section
to further define our notations and set stage for dynamical analysis
of future sections.

The charge sector is described by the quadratic Hamiltonian
\eqref{LLH} that can therefore be easily diagonalized.
Expressing ${\hat{\phi}}_\rho$ and ${\hat{\theta}} _\rho$ in terms of bosonic creation
and annihilation operators $\hat b^\dagger_p$ and $\hat b_p$ 
\begin{subequations}
\label{phitheta}
 \begin{align}
 {\hat{\phi}}(x)=-\frac{i\pi}{L}\sum_{p\neq0}\left(\frac{L|p|}{2\pi}\right)^{1/2}\frac{1}{p}e^{- a |p|/2-ipx}(\hat b^\dagger_p+\hat b_{-p}),\\
 {\hat{\theta}} (x)=\frac{i\pi}{L}\sum_{p\neq0}\left(\frac{L|p|}{2\pi}\right)^{1/2}\frac{1}{|p|}e^{- a |p|/2-ipx}(\hat b^\dagger_p-\hat b_{-p}),
\end{align}
\end{subequations}
the resulting Hamiltonian
\begin{align}
\label{LLHKbform}
\begin{split}
\hat H_\rho&=\frac{1}{4}\sum_{p\neq0}v_\rho |p|\Big[(1/K_\rho+K_\rho)(\hat b^\dagger_p \hat b_p+\hat b^\dagger_{-p} \hat b_{-p})\\
&\quad+(1/K_\rho-K_\rho)(\hat b^\dagger_p \hat b^\dagger_{-p}+\hat b_p \hat b_{-p})\Big]
\end{split}
\end{align}
is straightforwardly diagonalized by a bosonic Bogoliubov transformation
\begin{align}
\label{bogoliubovinia}
\begin{split}
\begin{pmatrix}\hat b_p \\ \hat b^\dagger_{-p}\end{pmatrix}=\begin{pmatrix}\cosh\beta&-\sinh\beta\\ -\sinh\beta&\cosh\beta\end{pmatrix}\begin{pmatrix}\hat \chi_p \\ \hat \chi^\dagger_{-p}\end{pmatrix},
\end{split}
\end{align}
giving
 \begin{align}
 \label{Hfreeb}
\hat H_0=\sum_{p\neq0}v_\rho|p|\hat \chi^\dagger_p \hat \chi_p
 \end{align}
with $e^{-2\beta}=K_\rho$.

Using the zero-temperature ground-state distribution $\langle \hat
\chi_{p}\hat \chi^\dagger_{p'}\rangle=\delta_{p,{p'}}$, one then obtains
\begin{subequations}
\begin{align}
\label{quadraticaveragephi}
\langle e^{\sqrt{2}i ({\hat{\phi}}_{\rho}(x)-{\hat{\phi}}_{\rho}(0))}\rangle&=e^{-\langle[{\hat{\phi}}_{\rho}(x)-{\hat{\phi}}_{\rho}(0)]^2\rangle},\\
&\sim e^{K_\rho\int^\infty_0{dp}/{p}e^{- a  p}[1-\cos(px)]}\nonumber,\\
&\sim\left(\frac{ a }{x}\right)^{K_\rho},
\label{quadraticaveragephib}
\end{align}
\end{subequations}
and
\begin{subequations}
\begin{align}
\label{quadraticaveragetheta}
\langle e^{\sqrt{2}i ({\hat{\theta}} _{\rho}(x)- {\hat{\theta}} _{\rho}(0))}\rangle&= e^{- \langle[{\hat{\theta}} _{\rho}(x)-{\hat{\theta}} _{\rho}(0)]^2\rangle},\\
&\sim e^{{1}/{K_\rho}\int^\infty_0{dp}/{p}e^{- a  p}[1-\cos(px)]},\nonumber\\
&\sim \left(\frac{ a }{x}\right)^{1/K_\rho},
\label{quadraticaveragethetab}
\end{align}
\end{subequations}
where in Eqs.~\eqref{quadraticaveragephi} and 
\eqref{quadraticaveragetheta}, we have used Wick's theorem,
\begin{align}
\label{wardrelation}
\langle e^{i\hat{A}}\rangle=e^{-\frac{1}{2}\langle \hat{A}^2\rangle},
\end{align}
valid for any free (i.e., Guassian) field operator $\hat A$.

The other component that enters the density-density correlation
function can now also be straightforwardly calculated to be
\begin{align}
\begin{split}
\partial_x\partial_{x'}\langle{\hat{\phi}}_\rho(x) {\hat{\phi}}_\rho(x')\rangle
&=-\frac{1}{2}\partial_x\partial_{x'}\langle({\hat{\phi}}_\rho(x)-{\hat{\phi}}_\rho(x'))^2\rangle,\\
&\sim -\frac{K_\rho}{2(x-x')^2}.
\end{split}
\end{align}
We next turn to the analysis of the spin sector of the model.


\section{spin sector: Luther-Emery exact analysis}
\label{LEapproach}

The analysis of the spin sector Hamiltonian \eqref{SineGordonH} is a
bit more challenging due to the cosine nonlinearity (associated with
the attractive pairing interaction), the so-called sine-Gordon
model. While in principle the model is integrable, its correlation
functions are still difficult to compute and generically approximate
methods (perturbation theory, semiclassics, and renormalization group)
need to be employed. However, for a spin Luttinger parameter $K_\sigma
= 1/2$ (the LE point) the model is exactly solvable through a mapping
(re-fermionization) onto free spinless fermions (the solitons), with
the cosine nonlinearity reducing to a mass that backscatters between
the left and right movers
\cite{LutherEmery,JaparidzeNersesyan,CazalillaJOP}. The
commensurate-incommensurate (PT) BCS-FFLO transition
\cite{PokrovskyTalapov}, driven by the Zeeman field $h$ then maps onto
a simple conduction band filling with chemical potential $h$ and gap
set by the Hubbard $U$ (see Fig.~\ref{bandmagneticown}). The LE
approach thus provides a clear physical picture of the spin-gapped BCS
and gapless FFLO paired states and serves as a benchmark for other
approximate solutions needed away from the $K_\sigma = 1/2$ point.

The essential component of LE is that the nonlinearity
$\cos(2\sqrt{2}{\hat{\phi}}_\sigma)$ in the spin sector, Eq.~\eqref{SineGordonH}
(generated from the interaction of the original fermions) for
$K_\sigma=1/2$ can be ``re-fermionized'' into a quadratic
backscattering mass term,
\begin{align}
 \label{LEtransfer}
\begin{split}
\hat c^\dagger_R(x)\hat c_L(x)=\frac{1}{2\pi a }e^{i2{\hat{\phi}}(x)},
\end{split}
\end{align}
between new left and right moving, spinless Dirac fermions ($r =
R,L$),
\begin{align}
\label{spinlesscphitheta}
\begin{split}
\hat c_{r}(x)&=U_{r}\lim_{ a \to0}\frac{1}{\sqrt{2\pi a }} e^{-i(r{\hat{\phi}}(x)-{\hat{\theta}} (x))}.
\end{split}
\end{align}
This can be seen by rescaling $\sqrt{2}{\hat{\phi}}_\sigma\equiv{\hat{\phi}}$ and
${\hat{\theta}} _\sigma/\sqrt{2}\equiv{\hat{\theta}} $ (to retain the canonical
commutation relation) into new bosonic fields, controlled by an
effective Luttinger parameter $K\equiv 2K_\sigma$. 

For $K=1$ ($K_\sigma = 1/2$), the bosonic Hamiltonian can be
re-fermionized into a noninteracting massive Thirring model \cite{Coleman},
\begin{subequations}
 \label{HLutherequilibrium}
\begin{align}
\hat H&=\sum_p (v_\sigma p+h)\hat c^\dagger_{Rp}\hat c_{Rp}+(-v_\sigma p+h)\hat c^\dagger_{Lp}\hat c_{Lp}\nonumber\\
&\quad-\frac{U}{\pi}\sum_p \hat c^\dagger_{Rp}\hat c_{Lp}+h.c.,\\
&\equiv\sum_p  \hat{\psi}^\dagger(p)  H(p) \hat \hat{\psi}(p),
\end{align}
\end{subequations}
where
 \begin{align}
\begin{split}
 \hat{\psi}(p)=\begin{bmatrix} \hat c_{R}(p)\\\hat c_{L}(p)\\\end{bmatrix},\;\;\;\;
H(p)=\begin{bmatrix} v_\sigma p+h&-U/\pi\\-U/\pi&-v_\sigma p+h\\\end{bmatrix},
\end{split}
\end{align}


The latter is diagonalized by a Bogoliubov transformation
\begin{align}
\label{LEbogoliubov}
\begin{split}
{\hat{\psi}}(p)=\begin{bmatrix} \hat c_u(p)\\\hat c_{l}(p)\\\end{bmatrix}=\begin{bmatrix} \cos\beta_p&\sin\beta_p\\-\sin\beta_p&\cos\beta_p\\\end{bmatrix}\begin{bmatrix} \hat c_R(p)\\\hat c_{L}(p)\\\end{bmatrix},
\end{split}
\end{align}
where $\tan 2\beta_p=-U/(\pi v_\sigma p)$ and lower-index $u,l$ denotes
the upper and lower bands, respectively. This gives
\begin{align}
 \label{HLEdiagonal}
\begin{split}
\hat H=\sum_p \omega(p)[\hat c^\dagger_{u}(p)\hat c_{u}(p)-\hat c^\dagger_{l}(p)\hat c_{l}(p)],
\end{split}
\end{align}
with a spin-gapped excitation spectrum
\begin{align}
\omega(p)=\pm\sqrt{v^2_\sigma p^2+ U^2/\pi^2},\ \ \ \Delta\equiv U/\pi.
\end{align}

\begin{figure}[!htb]
 \centering
\label{fig:fig3}
\end{figure}

\begin{figure}[!htb]
 \centering
  \includegraphics[width=80mm]{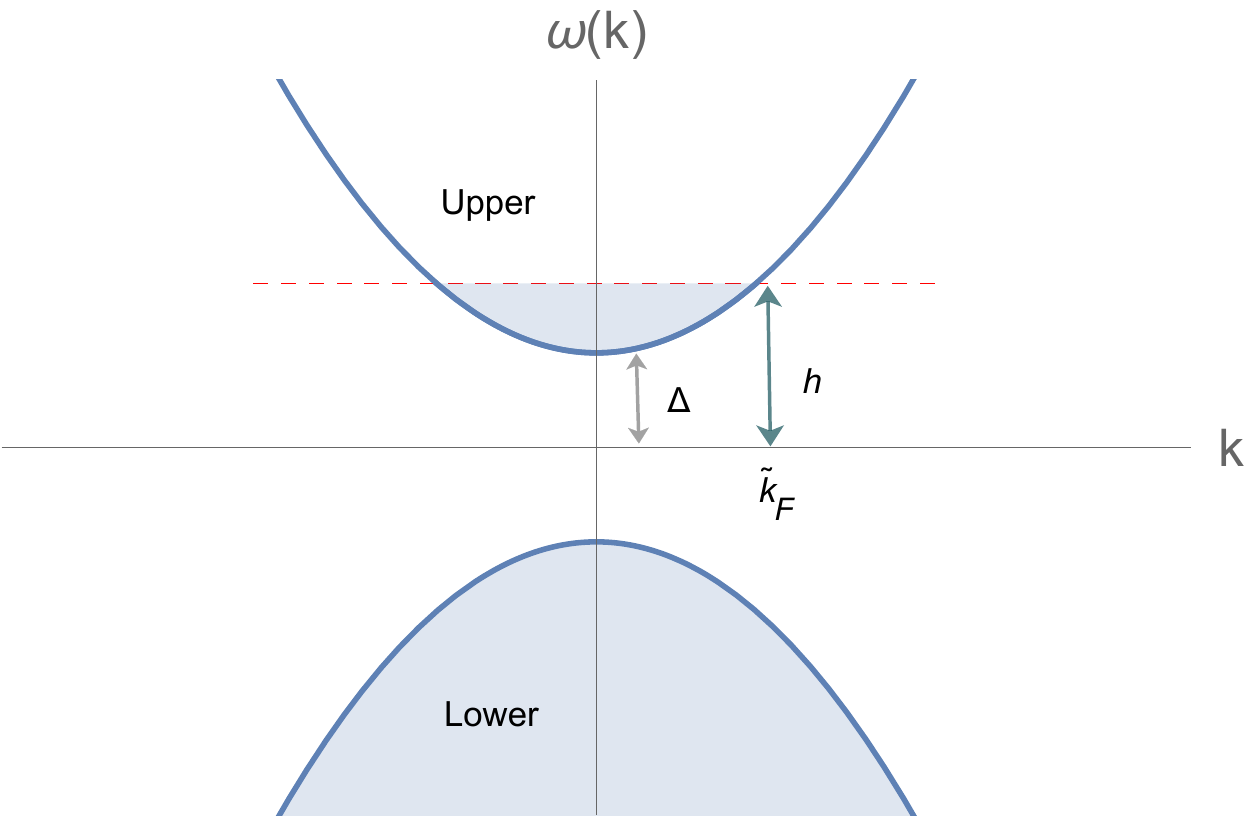}
  \caption{Band structure for spinless fermions for a spin gap
    $\Delta$ with the chemical potential $h$. For $h > h_c = \Delta$,
    the lower-band is completely filled, and the upper band is
    partially filled upto the Fermi momentum $\tilde{k}_F(h)$
    (determined by the effective chemical potential $h$), that gives
    the magnetization (species imbalance) density.}
\label{bandmagneticown}
\end{figure}
This band structure is illustrated in Fig.~\ref{bandmagneticown}.

\subsection{BCS state}
Thus, as illustrated in Fig.~\ref{bandmagneticown}, for the effective
chemical potential $h$ less than the critical spin-gap value $h_c =
\Delta$, i.e., for $h<\Delta\equiv U/\pi$, only the lower band is
filled, 
\begin{subequations}
\label{nklustatesa}
\begin{align}
\langle \hat c^\dagger_{l}(p)\hat c_{l}(p)\rangle&=1,\\
\langle\hat c^\dagger_{u}(p)\hat c_{u}(p)\rangle&=0,
\end{align}
\end{subequations}
the system is fully gapped and this ground state corresponds to the
spin-gapped BCS (soliton vacuum) phase, with a vanishing magnetization
(number density of fermions in the upper band).  The corresponding
occupation of the spinless fermions $\hat c_{R,L}$ is then given by
the Bogoliubov transformation Eq.~\eqref{LEbogoliubov}
\begin{subequations}
\label{LEequBCS}
\begin{align}
\langle \hat c^\dagger_{R}(p)\hat c_{R}(p)\rangle&=\sin^2 \beta_p,\\
\langle \hat c^\dagger_{L}(p)\hat c_{L}(p)\rangle&=\cos^2 \beta_p,\\
\langle \hat c^\dagger_{R}(p)\hat c_{L}(p)\rangle&=-\frac{1}{2}\sin 2\beta_p.
\end{align}
\end{subequations}

Thus, the correlators of the nontrivial bosonic sine-Gordon theory,
that can be simply expressed in terms of these noninteracting
fermions, can be straightforwardly calculated. For example (leaving
details to Appendix~\ref{LEgreensfunctions})
\cite{LutherEmery,JaparidzeNersesyan}
\begin{align}
\label{LEthetacorrelationBCS}
\begin{split}
\langle e^{i\sqrt{2}[{\hat{\theta}} _\sigma(x)-{\hat{\theta}} _\sigma(0)]}\rangle
&\sim \frac{a^2}{ x^2}e^{-2x/ \xi},
\end{split}
\end{align}
where 
\begin{align}
\xi=\pi v_\sigma/U
\label{xi}
\end{align}
is the correlation length. Also, the long-wavelength part of the
magnetization correlator is given by
\begin{align}
    \begin{split}
      \langle \partial_x{\hat{\phi}}_\sigma(x)\partial_{x'}{\hat{\phi}}_\sigma(0)\rangle&\sim
      -\frac{1}{2\pi\xi x}e^{-2x/\xi}.
    \end{split}
  \end{align}

\subsection{FFLO state}
On the other hand, for the chemical potential $h$ larger than the spin
gap $\Delta$, the upper band fills partially up to a Fermi momentum,
 $\tilde k_F = \sqrt{h^2-U^2/\pi^2}/v_\sigma\sim \sqrt{h - h_c}$,
\begin{align}
\label{nklustatesmagnetic}
\begin{split}
\langle \hat c^\dagger_{l}(p)\hat c_{l}(p)\rangle&=1,\\
\langle \hat c^\dagger_{u}(p)\hat c_{u}(p)\rangle&=\Theta(\tilde{k}_F-|k|),\\
\end{split}
\end{align}
corresponding to a proliferation of solitons into the spin-gapless
FFLO ground state, characterized by a nonzero magnetization,
 \begin{align}
    \begin{split}
\bar{m}&
=\frac{\sqrt{2}}{\pi}\tilde k_F
=\frac{\sqrt{2}}{\pi v_\sigma}\sqrt{h^2-U^2/\pi^2}\sim \sqrt{h - h_c}.
    \end{split}
  \end{align}
The ${\hat{\theta}} _\sigma$ correlation functions are then given by
\begin{align}
\label{LEthetacorrelationFFLO}
\begin{split}
\langle e^{i\sqrt{2}[{\hat{\theta}} _\sigma(x)-{\hat{\theta}} _\sigma(0)]}\rangle
&\sim (\tilde k_F\xi)^2\frac{a^2}{x^2},
\end{split}
\end{align}
and 
\begin{align}
    \begin{split}
      \langle \partial_x{\hat{\phi}}_\sigma(x)\partial_{x'}{\hat{\phi}}_\sigma(0)\rangle&
\sim -\frac{\sin^2(\tilde k_F x)}{\pi^2 x^2},
    \end{split}
  \end{align}
  which, as expected resembles the density-density correlator of free
  spinless fermions, but differs drastically from the exponentially
  decaying spin-gapped BCS result.

\section{Spin sector: semiclassical approach}
\label{spinsemiclassical}

Away from the Luther-Emery point, the interaction between LE fermions
precludes an exact solution for arbitrary $K_\sigma\neq 1/2$, but the
structure of the phase diagram and nature of the phases are expected to
remain qualitatively the same.  We instead proceed with a
semiclassical approach by studying quantum fluctuations about the
classical (time-independent) saddle-point solution ${\phi}_\sigma^0$ for the model
\eqref{SineGordonH}, satisfying the sine-Gordon equation, with an
intrinsic length scale $\lambda = \sqrt{\frac{\pi a
    v_\sigma}{8K_\sigma U }}$.

\subsection{BCS state}
\label{modelspingapped}
\begin{figure}[!htb]
 \centering
  \includegraphics[width=80mm]{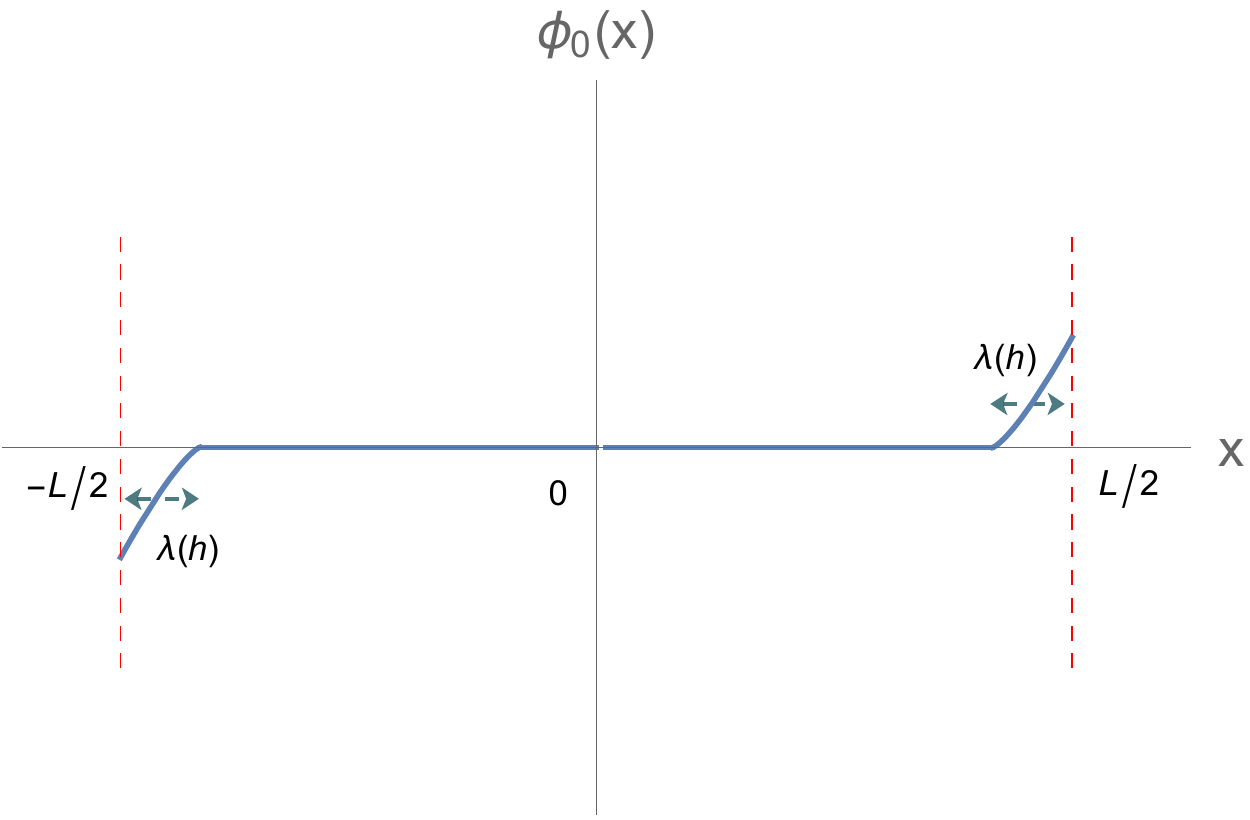}
  \caption{Schematic illustration of the classical spin-gapped
    solution $\phi^0_\sigma(x)$ for a finite system of size $L$, (nearly)
    vanishing in the bulk, but with a finite tilt (magnetization)
    penetrating a length $\lambda(h) < \lambda$ at the edges.}
\label{bcsphiedgeeffect}
\end{figure}

For $h < h_c$, the stable state is a soliton vacuum (spin-gapped BCS)
ground state, characterized by ${\phi}_\sigma^0 = 0$ in the bulk. With
free boundary conditions in a finite length $L$ system, for a finite
$h$, the tilt (magnetization) $\partial_x\phi^0_\sigma(\pm L/2) =
\sqrt{2}K_\sigma h/ v_\sigma$ ``penetrates'' into the sample within an
$h$-dependent length $\lambda(h)$ (a fraction of a soliton), as
schematically illustrated in Fig.~\ref{bcsphiedgeeffect}. At the
classical level (ignoring quantum fluctuations) $h_c^{(cl)}$ is
determined by the condition that the tilt at the boundary is
$~1/\lambda$, i.e., $h_c^{(cl)}\sim v_\sigma/\lambda\sim \sqrt{U}$, a
point beyond which a soliton of width $\lambda$ can enter the bulk.


We observe that the critical field $h_c^{(cl)}$ and characteristic
length $\lambda$ differ qualitatively from their LE (exact)
counterparts of $h_c = U/\pi$ and correlation length
$\xi=v_\sigma/h_c$, \eqref{xi}. This is a consequence of quantum
fluctuations.

Within the spin-gapped BCS state ${\hat{\phi}}_\sigma$ is localized in a minimum
of the cosine potential (at its classical solution ${\phi}_\sigma^0 = 0$), and
can be safely expanded in small quadratic fluctuations governed by \cite{CazalillaJOP}
 \begin{align}
 \label{phiexpansionhamiltonianphitheta}
\begin{split}
\delta \hat H_\sigma&=\frac{v_\sigma}{2\pi}\int dx[\frac{1}{K_\sigma}(\partial_x{\hat{\phi}}_\sigma)^2+K_\sigma(\partial_x{\hat{\theta}} _\sigma)^2]\\
&\quad+\frac{2 U}{\pi^2 a}\int dx({\hat{\phi}}_\sigma)^2-\frac{\sqrt{2}h}{\pi}\int dx\partial_x{\hat{\phi}}_\sigma.
\end{split}
\end{align}
The corresponding Hamiltonian in terms of bosonic operators $\hat b_p,
\hat b^\dagger_p$ can be straightforwardly diagonalized via
\begin{align}
\label{Bogoliubovmatrixphiexpansionequilibrium}
\begin{split}
\begin{bmatrix}\hat b_{p}\\ \hat b^\dagger_{-p}\\\end{bmatrix}=\begin{bmatrix} u_{p}&-v_{p}\\-v_{p}&u_{p}\\\end{bmatrix}\begin{bmatrix}\hat  \alpha_{p}\\ \hat  \alpha ^\dagger_{-p}\\\end{bmatrix},
\end{split}
\end{align}
with $\delta\hat H_\sigma=\sum_p \omega(p)\hat \alpha^\dagger_p\hat
\alpha _p$, where
\begin{align}
\label{bogoliubovuvomega}
\begin{split}
u^2_p&
=\frac{1}{2}\left(\frac{|p|+{1}/(2\lambda^2 p)}{\sqrt{p^2+1/\lambda^2}}+1\right),\\
v^2_p&
=\frac{1}{2}\left(\frac{|p|+{1}/(2\lambda^2 p)}{\sqrt{p^2+1/\lambda^2}}-1\right),\end{split}
\end{align}
and the spectrum is given by
 \begin{align}
\begin{split}
\omega_p=\pm v_\sigma \sqrt{p^2+1/\lambda^2},
\end{split}
\end{align}
with the gap $\Delta_0=v_\sigma/\lambda$.

At $T=0$, in the ground state, 
\begin{subequations}
\label{bdistributionphiexpansionequilibrium}
\begin{align}
\langle \hat b^\dagger_{p}\hat b_{p}\rangle&=\langle \hat b^\dagger_{-p}\hat b_{-p}\rangle=v^2_p,\\
\langle \hat b^\dagger_p \hat b^\dagger_{-p}\rangle&=\langle \hat b_p \hat b_{-p}\rangle=-u_pv_p,
\end{align}
\end{subequations}
which (relegating the details to Appendix \ref{spingappedexpansion})
gives
\begin{subequations}
\label{BCSthetaphicorrelator}
\begin{align}
\langle e^{i\sqrt{2}[{\hat{\phi}}_\sigma(x)-{\hat{\phi}}_\sigma(0)]}\rangle&\sim \const,\\
\langle e^{i\sqrt{2}[{\hat{\theta}} _\sigma(x)-{\hat{\theta}} _\sigma(0)]}\rangle& \sim e^{-x/\xi_0},
\end{align}
\end{subequations}
with $\xi_0=2\lambda K_\sigma/\pi$, and demonstrating the stability of the
spin-gapped state to quantum fluctuations.

The leading exponential decay in the correlator in
\eqref{BCSthetaphicorrelator} agrees qualitatively with one predicted by
the exact LE analysis \eqref{LEthetacorrelationBCS} $K_\sigma=1/2$,
though at this level of calculation misses the subdominant power-law
prefactor.

However, one obvious discrepancy is that the semiclassically computed
correlation length $\xi_0=\sqrt{a v_\sigma K/2\pi U}\sim\sqrt{1/U}$ in
Eq.~\eqref{BCSthetaphicorrelator} differs from $\xi = \pi
v_\sigma/U\sim 1/U$ in Eq.~\eqref{LEthetacorrelationBCS}, obtained
using the exact LE approach. This can be understood by noting that for
small $U$ (such that $\xi_0 \gg a$), for high momentum modes $\xi_0^2
k^2 > (k a)^{2K_\sigma}$ the cosine pinning potential is weaker than
the elastic (density interaction) energy and thus contribute
divergently and nonperturbatively to renormalize the correlation
length. Indeed a standard RG analysis gives $\xi_0\rightarrow
a(\xi_0/a)^{2/(2-2K_\sigma)}\sim 1/U^{1/(2-2K_\sigma)}$, which for
$K_\sigma=1/2$ scales as $1/U$, reassuringly consistent with LE.

\subsection{FFLO state} 
For $h > h_c$, the spin-gapped ground state ${\phi}_\sigma^0=0$ is
unstable to soliton proliferation in the bulk, leading to the FFLO. At
the semiclassical level this takes place when the $h$-imposed tilt at
the boundary reaches $1/\lambda$, allowing solitons to penetrate into
the bulk; more generally the transition is associated with the soliton
(LE fermions in the upper band) gap closing.

The corresponding semiclassical solution to the sine-Gordon equation
is a soliton lattice, extensively studied in the
literature \cite{Hanna}, is illustrated in Fig.~\ref{solitonlatticedilute}
with spacing $d = \bar{m}^{-1}$ and width $\lambda$,
\begin{align}
\label{phisolitonclassical}
\begin{split}
{\phi}_\sigma^0(x)=\frac{1}{\sqrt{2}}\am(x/(\sqrt{2}\lambda k),k).
\end{split}
\end{align}
Above $am(x,k)$ is the Jacobi amplitude function, parameterized by the
$h$-dependent parameter $k$ (ranging from $0$ to $1$, not to be
confused with the momentum $k$) that controls the soliton
density. Minimizing the energy over $k$ gives the equation that
determines $k$ (see also Fig. 8):
\begin{align}
\label{solitondensityequation}
\begin{split}
\frac{E(k)}{k}=\frac{h}{h_c},
\end{split}
\end{align}
where $h_c=2v_\sigma/(\pi\lambda K_\sigma)$ 
and $E(k)$ the Jacobi elliptic function (not to be confused with the
spectrum $E$).
\begin{figure}[!htb]
 \centering
  \includegraphics[width=80mm]{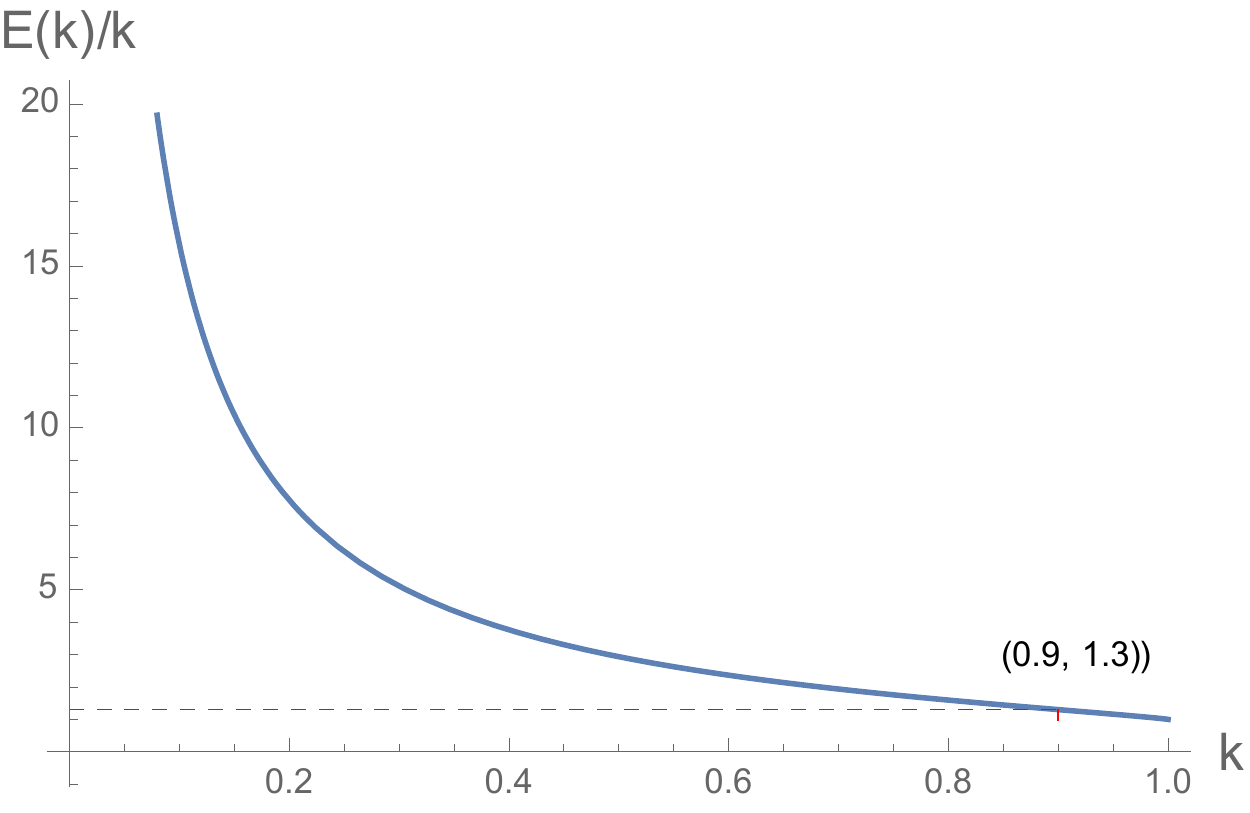}
  \caption{Plot of $E(k)/k$ [Eq.~\eqref{solitondensityequation}] as a
    function of the index $k$, with the soliton solution characterized
    by $0<k<1$ stable for $h > h_c$. For $h/h_c=1.3$, indicated by the
    red dashed line, the corresponding index is $k=0.9$.}
\label{Ekoverk}
\end{figure}
At large $h\gg h_c$ the solution quickly approaches a sloped straight
line, corresponding to a dense soliton lattice, with $\bar m \sim
h/(h_c\lambda)$ deep in the incommensurate phase with
${\phi}_\sigma^0(x) = \bar m x$.
\begin{figure}[!htb]
 \centering
  \includegraphics[width=80mm]{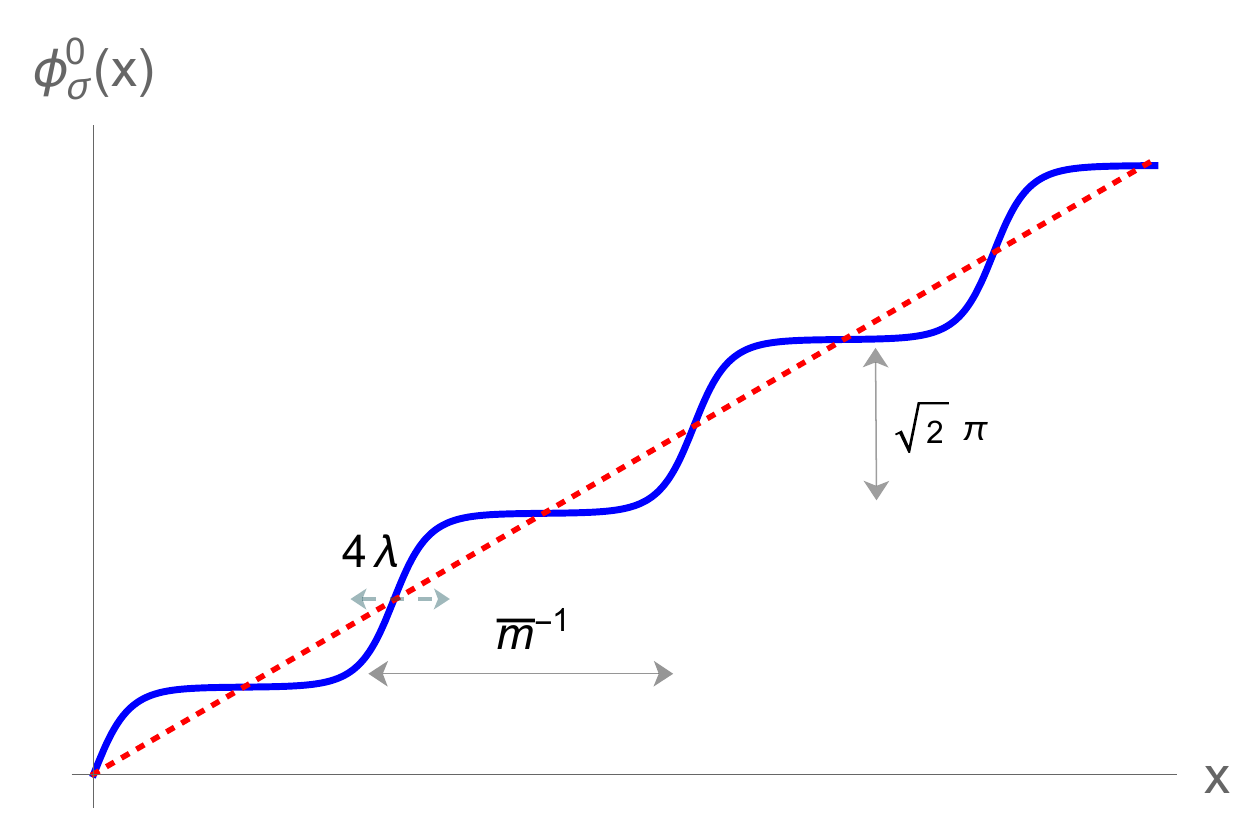}
  \caption{Schematic plot of the soliton lattice solution
    ${\phi}_\sigma^0$ for \eqref{phisolitonclassical} as a function of
    $x$. $\bar{m}^{-1}$ and $\lambda$ are two length scales that
    characterize the average soliton lattice, the soliton spacing
    (density $\bar{m}$) and width, respectively. The straight line is
    a guide to an eye.}
\label{solitonlatticedilute}
\end{figure}

\begin{figure}[!htb]
 \centering
  \includegraphics[width=80mm]{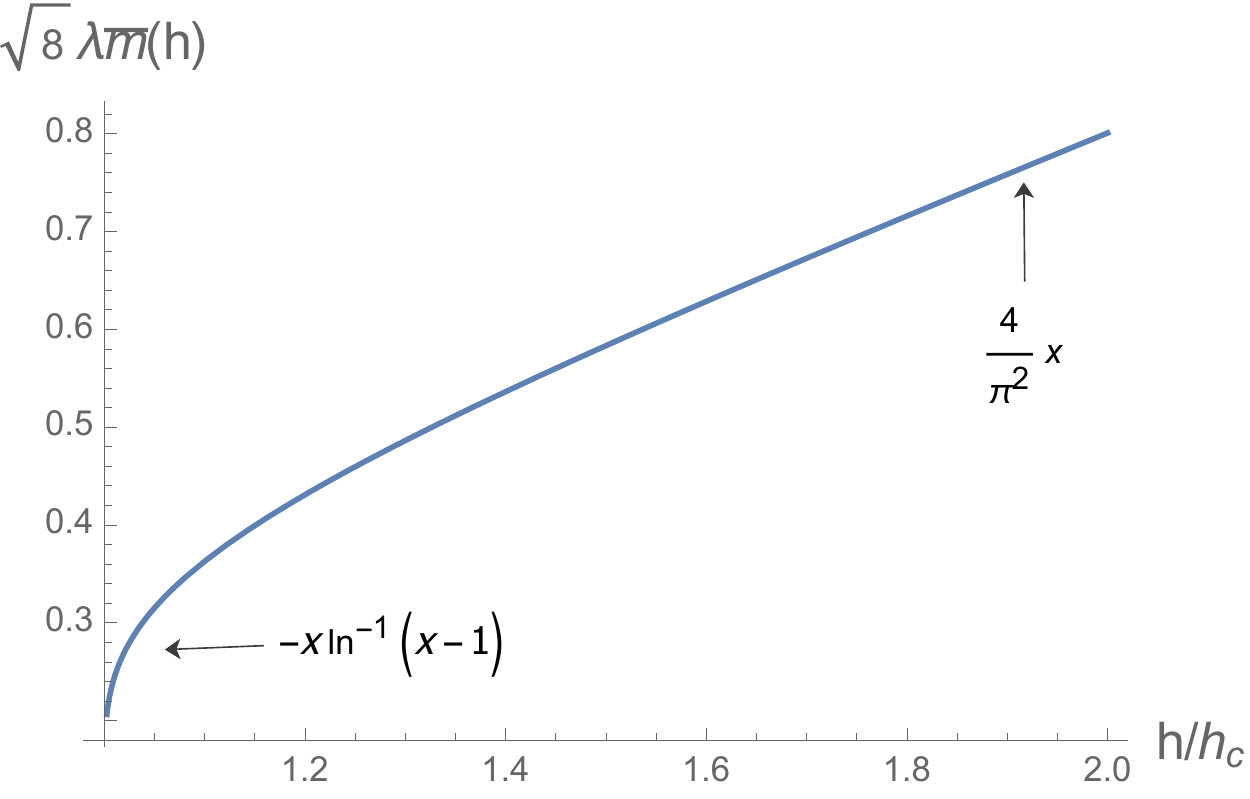}
  \caption{Plot of magnetization $\bar{m}_{cl}$ (within a classical
    approximation) as a function $h/h_c$ [see
    Eq.~\eqref{classicalWKBM}]. Within the critical region,
    $\bar{m}\propto-x\ln^{-1}(x-1)$ and away from critical region it is
    well approximated by a linear function.}
\label{EquMagnetization}
\end{figure}

The magnetization in the FFLO state is set by the soliton density, which is
illustrated in Fig.~\ref{EquMagnetization}, and given by
\begin{align}
\label{classicalWKBM}
\begin{split}
\bar{m}_{cl}&=\frac{1}{2\sqrt{2}\lambda kK(k)}\\
&\sim\frac{1}{2\sqrt{2}\lambda}\begin{cases} -\frac{h}{h_c}1/\ln(\frac{h}{h_c}-1), &\mbox{for }h/h_c\to 1,\\ 
 \frac{4}{\pi^2}\frac{h}{h_c}, &\mbox{for } h/h_c\to\infty,\end{cases}
\end{split}
\end{align}
with $k(h)$ tuned by the Zeeman field $h$ through
Eq.~\eqref{solitondensityequation}.

This fast logarithmic growth is associated with (in the absence of
fluctuations) the exponential weakness of soliton interactions near
$h_c$ where solitons are dilute. Because of this, the soliton density
rises to nearly densely packed value of order $1/\lambda$, at which
point $\bar m(h)$ is a linear function of $h$.  Quantum (and thermal)
fluctuations qualitatively modify these predictions as is clear from
the exact LE analysis, $\bar{m}\sim\sqrt{h^2-h_c^2}$, and more
generally from an array of fluctuating soliton
world lines \cite{PokrovskyTalapov}.

\begin{figure}[!htb]
 \centering
  \includegraphics[width=80mm]{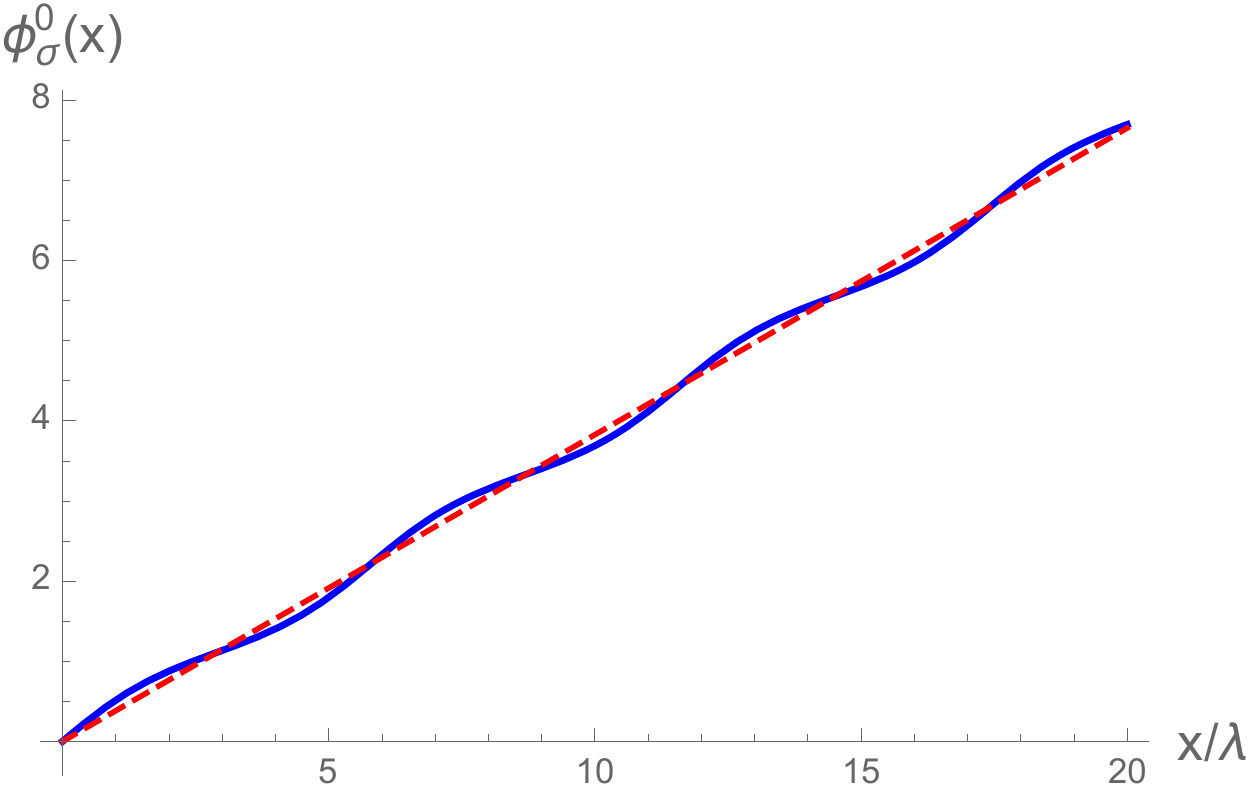}
  \caption{Classical soliton lattice solution ${\phi}_\sigma^0$ \eqref{phisolitonclassical} as a function of $x/\lambda$ for $h/h_c=1.3$, as compared with the approximate dashed straight line solution
    \eqref{phisolitonclassicalapproximate}.}
\label{solitonlattice}
\end{figure}

To include quantum fluctuations about this classical description of
the FFLO state is quite nontrivial. To simplify the analysis, we limit
our study outside of the critical region at $h_c$. As illustrated in
Fig.~\ref{solitonlattice}, in this regime, even for a fairly modest
$h/h_c=1.3$ solitons strongly overlap and the classical solution is given
by 

\begin{align}
\label{phisolitonclassicalapproximate}
\begin{split}
{\phi}_\sigma^0(x)\approx\frac{1}{\sqrt{2}}\am(2K(k)\bar{m}x)
&\approx\frac{h}{\pi h_c}\frac{x}{\lambda}\approx\frac{\pi\bar m}{\sqrt{2}}x.
\end{split}
\end{align}

Quantum fluctuations are naturally included through the phonons
$\tilde{\phi}(x)$ of the soliton lattice, which can thereby be related
to the original bosonic fields,
\begin{align}
\begin{split}
{\hat{\phi}}_\sigma=\frac{1}{\sqrt{2}}(\pi\bar{m}x+\tilde{\phi}(x)),
\label{phi_t}
\end{split}
\end{align}
and correspondingly for the conjugate phase,
\begin{align}
\begin{split}
{\hat{\theta}} _\sigma=\sqrt{2}\tilde{\theta}(x),
\label{theta_t}
\end{split}
\end{align}
crucial for computation of physical observables.  This is consistent
with the LE construction, where at finite $h\gg h_c$ the upper band
spinless fermions around $\pm\tilde k_F$ can be re-bosonized (see Fig.~12), with the
Hamiltonian \cite{Shoucheng},
\begin{align}
\begin{split}
  \tilde H_{FFLO}&=\frac{\tilde{v}}{2\pi}\int
  dx\left[\frac{1}{\kappa}(\partial_x\tilde{\phi})^2
+\kappa(\partial_x\tilde{\theta})^2\right],
\end{split}
\end{align}
where $\tilde{v}$ and $\kappa$ are the new set of Luttinger
parameters. 

\begin{figure}[!htb]
 \centering
  \includegraphics[width=80mm]{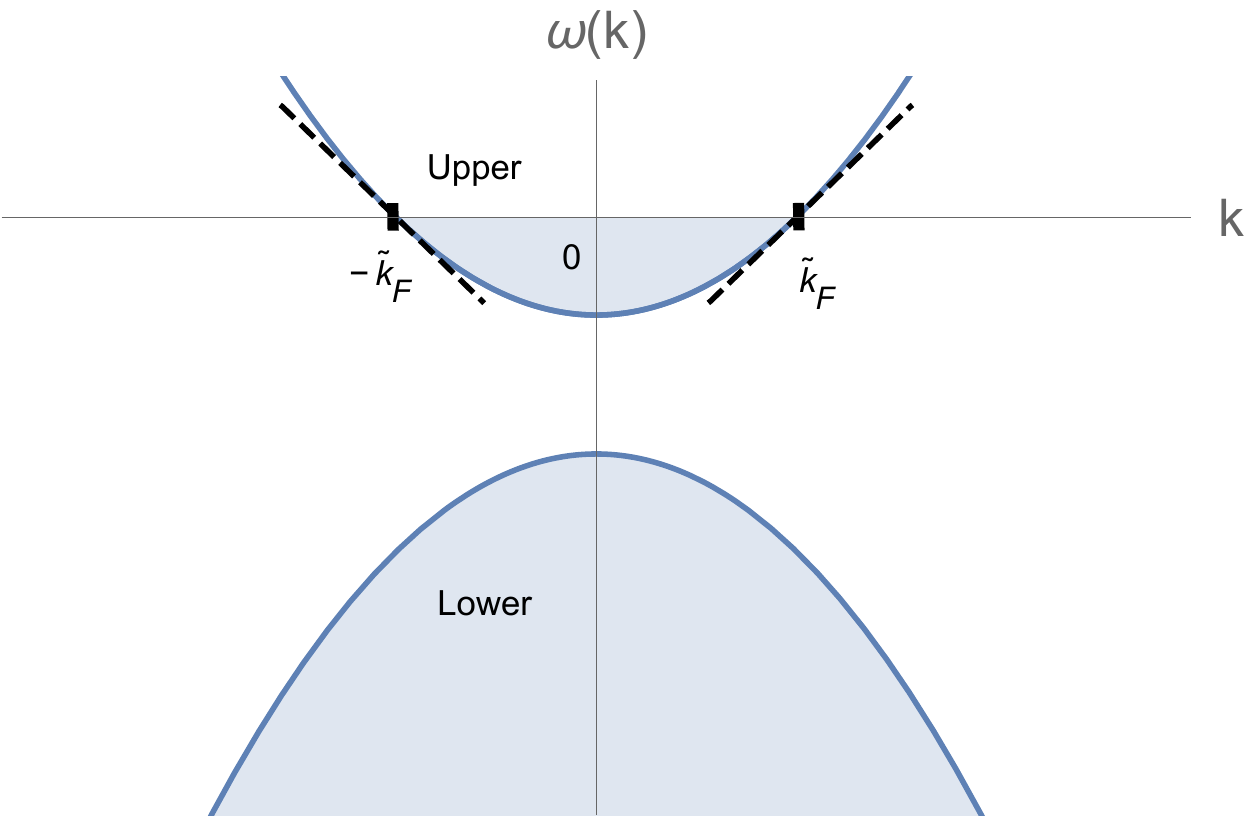}
  \caption{Re-bosonization near the new Fermi points $\pm\tilde k_F$,
    determined by spin-imbalance $h$ for the upper-band spinless
    fermions.}
\label{refermionization}
\end{figure}

At the LE $K_\sigma=1/2$ point the fermions are noninteracting and so
$\kappa = 1$.  Perturbative analysis around the LE point
gives \cite{Schulz}
\begin{align}
\label{schulzformula}
\begin{split}
\kappa=1+\frac{v_\sigma\bar{m}\pi a}{U}\left(2K_\sigma-\frac{1}{2K_\sigma}\right),
\end{split}
\end{align}
Furthermore, near the transition at $h_c$ the LE fermion (soliton) density
$\bar m$ vanishes and we expect $\kappa$ to also approach $1$. 

Utilizing the relation (\ref{phi_t},\ref{theta_t}), standard analysis
of the original bosonic fields then gives ${\hat{\phi}}_\sigma$ and
${\hat{\theta}} _\sigma$ correlators
\begin{subequations}
\label{semiclassicalFFLOcorrelations}
\begin{align}
\langle e^{i\sqrt{2}({\hat{\phi}}_\sigma(x)-{\hat{\phi}}_\sigma(0))}\rangle
&\sim e^{i\pi\bar{m}x}\left|\frac{ a }{x}\right|^{\kappa/2},\\
\langle e^{i\sqrt{2}({\hat{\theta}} _\sigma(x)-{\hat{\theta}} _\sigma(0))}\rangle&\sim {\left|\frac{a}{x}\right|^{2/\kappa}},\\
\frac{{2}}{\pi^2}\langle\partial_x{\hat{\phi}}_\sigma(x)\partial_{x'} {\hat{\phi}}_\sigma(0)\rangle
&\sim-\frac{\kappa}{2\pi^2x^{2}}.
\end{align}
\end{subequations}
Comparing these with the non-interacting fermion correlators also
allows us to infer that $\kappa\rightarrow 2$, as $U\rightarrow 0$.


\section{1D spin-imbalanced Fermi-Hubbard model: Equilibrium}
\label{equproperties}
We can now combine the above results for charge and spin sectors to obtain
the ground-state correlators for the 1D spin-imbalanced attractive
Fermi-Hubbard model \eqref{spinimbalanceHubbard}. 

\subsection{Spin-gapped (commensurate) state: \textit{s}-wave singlet BCS}
For weak Zeeman field $h < h_c = \Delta = U/\pi$, the spin sector is
fully gapped. The spin-singlet pairing correlator (defined in Sec.~\ref{sectionmodel}) is thus given by
\begin{align}
\begin{split}
S^{\mbox{\tiny\itshape BCS}}_{ss}(x)
&\sim\left(\frac{ a }{x}\right)^{1/K_\rho},
\end{split}
\end{align}
with $1<K_\rho<2$ for the Hubbard model. The pairing correlations are longer range and thus (as expected for attractive interactions) are enhanced in the spin-gapped phase relative to the noninteracting fermions with
$S^{\mbox{\tiny\itshape free}}_{ss}(x)\sim\left(\frac{a}{x}\right)^{2}$.

The associated Fourier
transform is the Cooper-pair momentum distribution function
\begin{align}
\begin{split}
n^{pair}_{q}
&\sim \frac{1}{|q|^{1-\eta}},
\end{split}
\end{align}
plotted in the inset of Fig.~\ref{Equnpairfflo}, with $\eta\equiv
1/K_\rho$.

The spin-triplet correlator involves spin correlations and thus
exponentially decays in the spin-gapped BCS state
 \begin{align}
\begin{split}
S^{\mbox{\tiny\itshape BCS}}_{st}(x)
 &\sim\left(\frac{ a }{x}\right)^{\eta}e^{-x/\xi}.
\end{split}
\end{align}

\begin{figure}[!htb]
 \centering
  \includegraphics[width=80mm]{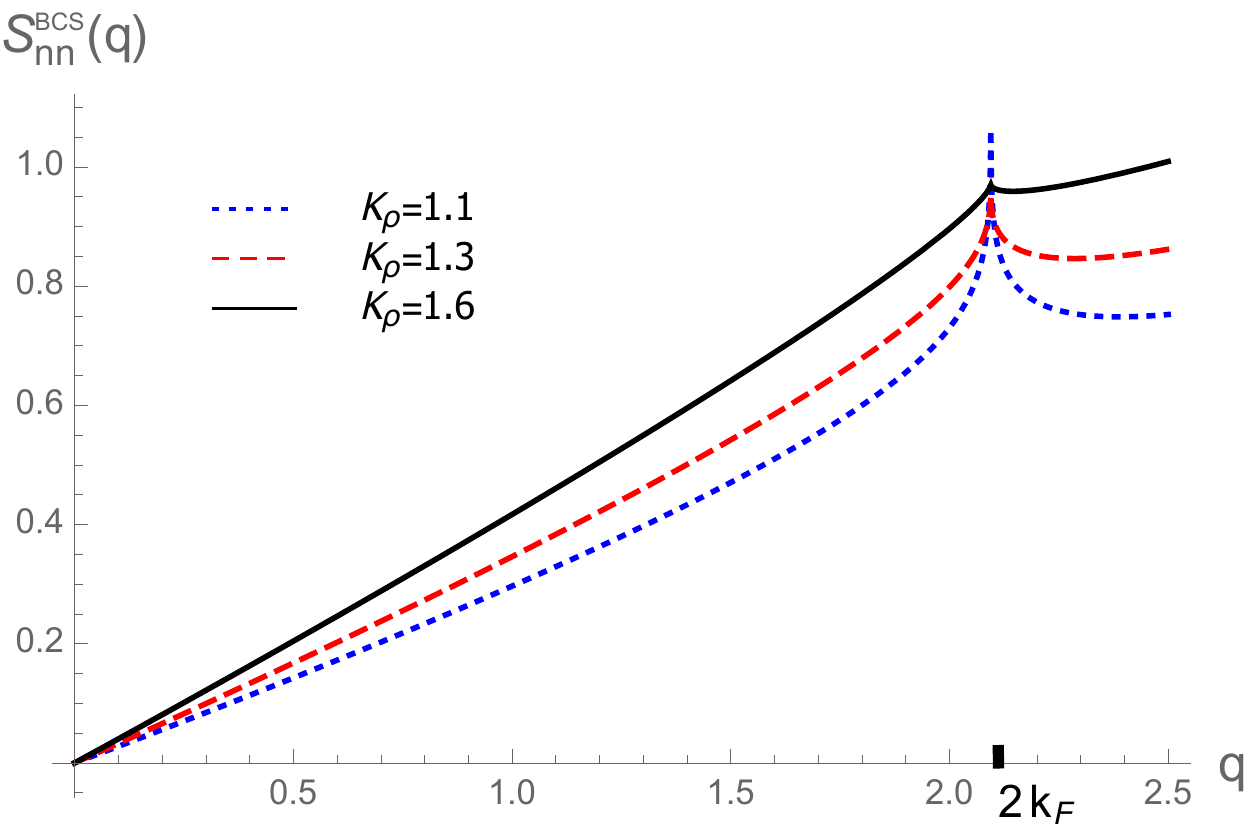}
  \caption{The density structure function $S^{\mbox{\tiny\itshape
        BCS}}_{nn}(q)$ for the BCS state for different interactions:
    $K_\rho=1.1$ (dotted blue), $K_\rho=1.3$ (dashed red) and
    $K_\rho=1.6$ (solid black). A pronounced cusp appears at
    $q=2k_F$.}
\label{EquSnnbcs}
\end{figure}

The density-density correlator
\begin{align}
\begin{split}
S^{\mbox{\tiny\itshape BCS}}_{nn}(x)
&\sim-K_\rho\left(\frac{ a }{x}\right)^{2}
+\cos(2k_F x)\left(\frac{ a }{x}\right)^{K_\rho},
\end{split}
\end{align}
encodes a combination of long-scale power-law charge correlations and
short scale Friedel oscillations. The latter are pronounced in the
structure function
\begin{align}
\begin{split}
S^{\mbox{\tiny\itshape BCS}}_{nn}(q)&\sim K_\rho \sqrt{\frac{\pi}{2}}|q|+\sqrt{\frac{2}{\pi}}\Gamma(1-K_\rho)\sin(\pi K_\rho/2)\\
&\quad\times\left({|q-2k_F|^{K_\rho-1}}+{|q+2k_F|^{K_\rho-1}}\right)
\end{split}
\end{align}
plotted in Fig.~\ref{EquSnnbcs}. The linear behavior at small $q$ and
a cusp at $2k_F$ momentum is in nice agreement with the Monte Carlo
result in \cite{Pour}

\begin{figure}[!htb]
 \centering
  \includegraphics[width=80mm]{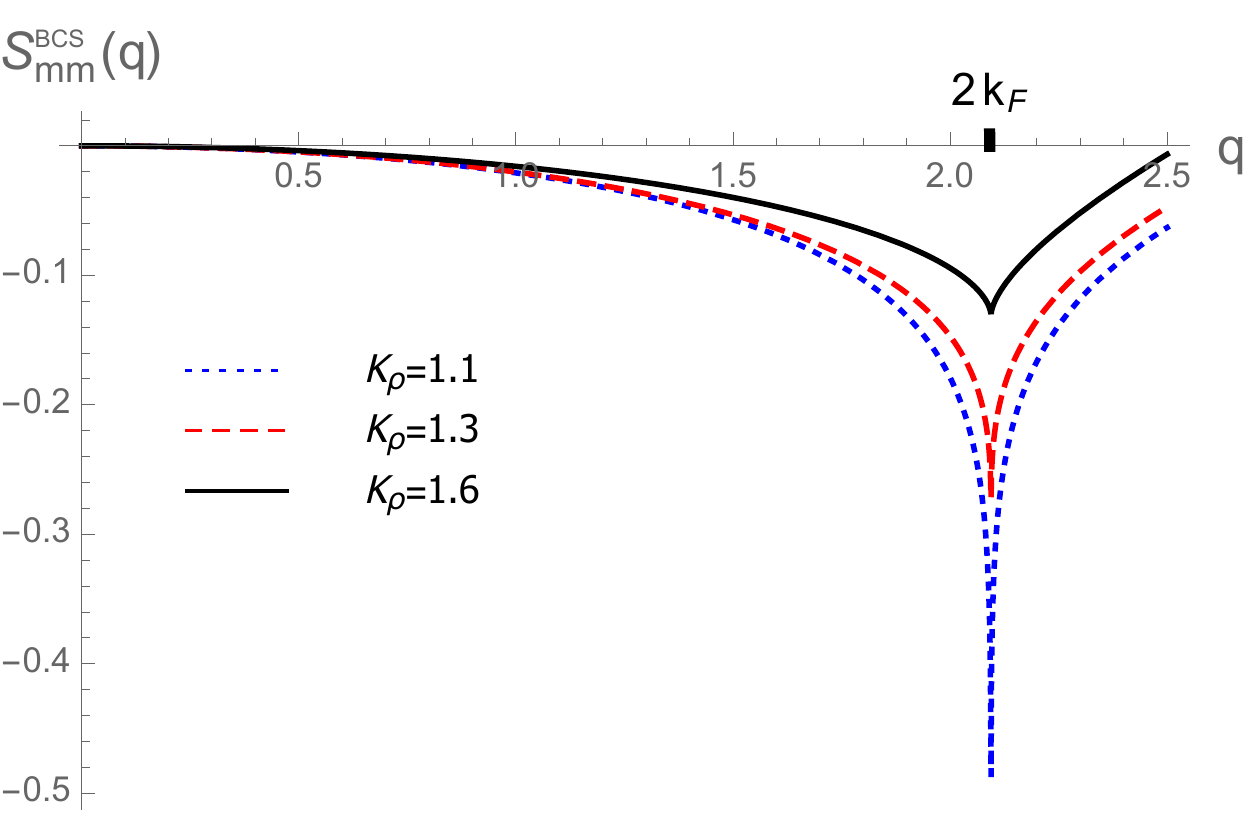}
  \caption{The magnetic structure function $S^{\mbox{\tiny\itshape
        BCS}}_{mm}(q)$ for the BCS state for different interactions:
    $K_\rho=1.1$ (dotted blue), $K_\rho=1.3$ (dashed red), and
    $K_\rho=1.6$ (solid black). An inverted characteristic cusp
    appears at $q=2k_F$.}
\label{EqusmmBCS}
\end{figure}

For BCS state, the average magnetization is zero, but the
magnetization-magnetization correlation does exhibit characteristic
signatures,
\begin{align}
\begin{split}
S^{\mbox{\tiny\itshape BCS}}_{mm}(x)\sim
-\cos(2k_F x)\left(\frac{ a }{x}\right)^{K_\rho},
\end{split}
\end{align}
with the corresponding magnetization structure function
\begin{align}
\begin{split}
S^{\mbox{\tiny\itshape BCS}}_{mm}(q)&\sim-\sqrt{\frac{2}{\pi}}\Gamma(1-K_\rho)\sin(\pi K_\rho/2)\\
&\quad\times\left({|q-2k_F|^{K_\rho-1}}+{|q+2k_F|^{K_\rho-1}}\right),
\end{split}
\end{align}
illustrated in Fig.~\ref{EqusmmBCS}.

\subsection{Spin-incommensurate state: FFLO}
For $h > h_c = \Delta = U/\pi$, the strong Zeeman field exceeds
spin-gap and the ground state becomes unstable to soliton
proliferation in the bulk, leading to the FFLO.
The spin-singlet pairing correlator is thus given by
\begin{align}
 \label{singletequilibriumFFLO}
\begin{split}
S^{\mbox{\tiny\itshape FFLO}}_{ss}(x)
&\sim\left(\frac{ a }{x}\right)^{1/K_\rho+\kappa/2}\cos(k_{FFLO}x),
\end{split}
\end{align}
with $k_{FFLO}=\pi \bar{m}=k_{F\uparrow}-k_{F\downarrow}$. While
correlations still decay as a power-law, they also exhibit an
oscillatory nonzero-momentum signature of the FFLO state. The associated
exponent falls into a range $1 < \eta'\equiv K^{-1}_\rho+\kappa/2 < 2$ with the upper bound and lower bounds reached for $U\to0$ and $U\to \infty$,
respectively. The pairing correlations are thus in a range intermediate between fully paired BCS and unpaired fermions. This is consistent with the idea that FFLO is an intermediate ``mixed'' Zeeman-field ground state (given that FFLO is a compromise between the BCS and the unpaired state), analogous to the Abrikosov lattice of type-II superconductors.

\begin{figure}[!htb]
 \centering
  \includegraphics[width=80mm]{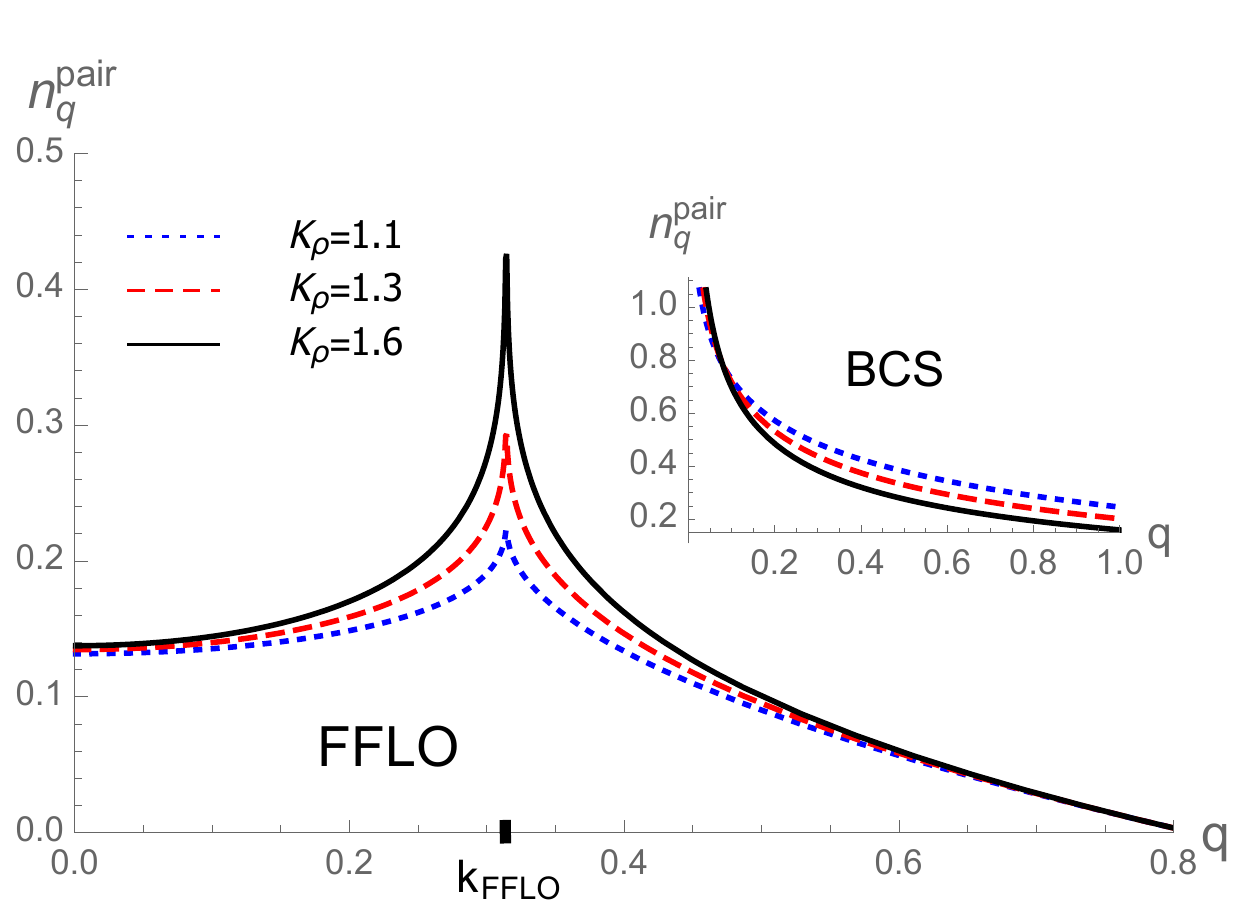}
  \caption{The Cooper-pair momentum distribution function in the FFLO
    ground state for $K_\rho=1.1$ (dotted blue), $K_\rho=1.3$ (dashed
    red), and $K_\rho=1.6$ (solid black) displaying a nonzero momentum
    $q=k_{FFLO}$ peak, narrowing with increasing interaction.  Inset:
    The Cooper-pair momentum distribution for the BCS state, with a
    peak at zero-momentum.}
\label{Equnpairfflo}
\end{figure}

The corresponding pair-momentum distribution function is then given by
\begin{align}
\begin{split}
n^{pair}_{q}&\sim \frac{\Gamma(1-\eta')\sin(\pi \eta'/2)}{\sqrt{2\pi}}
\\&\quad\times\left(|q-k_{FFLO}|^{\eta'-1}+|q+k_{FFLO}|^{\eta'-1}\right)
\end{split}
\end{align}
is plotted in Fig.~\ref{Equnpairfflo}. In contrast with the
spin-gapped BCS state, in the FFLO state the momentum distribution
exhibits a peak at finite momentum
$q=k_{FFLO}=k_{F\uparrow}-k_{F\downarrow}$, that sharpens as the
interaction increases. This is also consistent with the numerical result reported in \cite{Feiguin07}

The triplet-pairing correlator in the FFLO state is given by
 \begin{align}
 \label{tripletequilibriumFFLO}
\begin{split}
S^{\mbox{\tiny\itshape FFLO}}_{st}(x)
&\sim\left(\frac{ a }{x}\right)^{1/K_\rho+2/\kappa},
\end{split}
\end{align}
also decaying as a power-law. This is in sharp contrast with the BCS
case where triplet pairing is exponentially suppressed. Since
$1<\kappa<2$ (based on our discussion in
Sec.~\ref{spinsemiclassical}), $2/\kappa>\kappa/2$, indicating that in
the FFLO ground state, triplet correlations are subdominant to the
singlet ones (see Eq.~\eqref{singletequilibriumFFLO} and
Eq.~\eqref{tripletequilibriumFFLO}).

The density-density correlator in the FFLO state is given by
\begin{align}
\begin{split}
S^{\mbox{\tiny\itshape FFLO}}_{nn}(x)
&\sim-K_\rho\left(\frac{ a }{x}\right)^{2}
\\
&\quad+\frac{1}{2}\left[\cos(2k_{F\uparrow}x)
+\cos(2k_{F\downarrow}x)\right]\left(\frac{ a }{x}\right)^{K_\rho+\kappa/2},
\end{split}
\end{align}
with a two-component Friedel oscillations reflecting two fermionic
populations, $2k_{F\uparrow,\downarrow}$.
 
\begin{figure}[!htb]
 \centering
  \includegraphics[width=80mm]{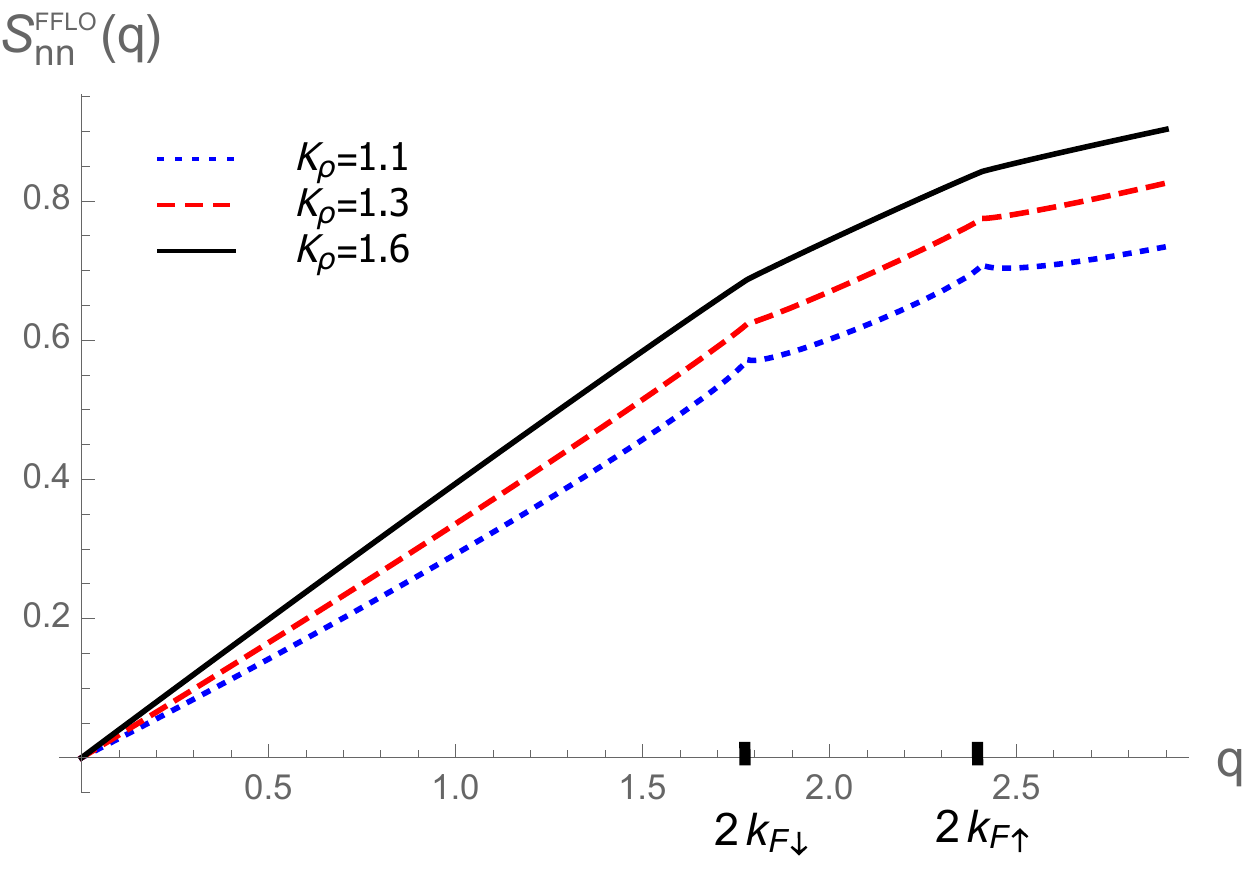}
  \caption{The density structure function $S^{\mbox{\tiny\itshape
        FFLO}}_{nn}(q)$ for the FFLO state for different interactions: $K_\rho=1.1$ (dotted blue), $K_\rho=1.3$  (dashed red) and $K_\rho=1.6$ (solid black), displaying cusps at  $2k_{F\uparrow}$ and $2k_{F\downarrow}$, that reflect
    pseudo-spin-imbalance.}
\label{Equsnnfflo}
\end{figure}

The corresponding structure factor, 
\begin{align}
\begin{split}
S^{\mbox{\tiny\itshape FFLO}}_{nn}(q)&\sim K_\rho \sqrt{\frac{\pi}{2}}|q|+\frac{\Gamma(1-\gamma)\sin(\pi \gamma/2)}{2\sqrt{2\pi}} ({|q-2k_{F\uparrow}|^{\gamma-1}}\\
&\quad+{|q+2k_{F\uparrow}|^{\gamma-1}}+{|q+2k_{F\downarrow}|^{\gamma-1}}+{|q-2k_{F\downarrow}|^{\gamma-1}})
\end{split}
\end{align}
where $\gamma\equiv K_\rho+\kappa/2>1$, is plotted in
Fig.~\ref{Equsnnfflo} for small magnetization $\bar m=\pi/10$, with
$\kappa\sim 1$ according to Eq.~\eqref{schulzformula}. It displays
cusps at $2k_{F\uparrow}$ and $2k_{F\downarrow}$, reflecting the
contribution from pseudo-spin-up and pseudo-spin-down fermions,
respectively and in a qualitative agreement with the DMRG results
\cite{Rizzi}.

\begin{figure}[!htb]
 \centering
  \includegraphics[width=80mm]{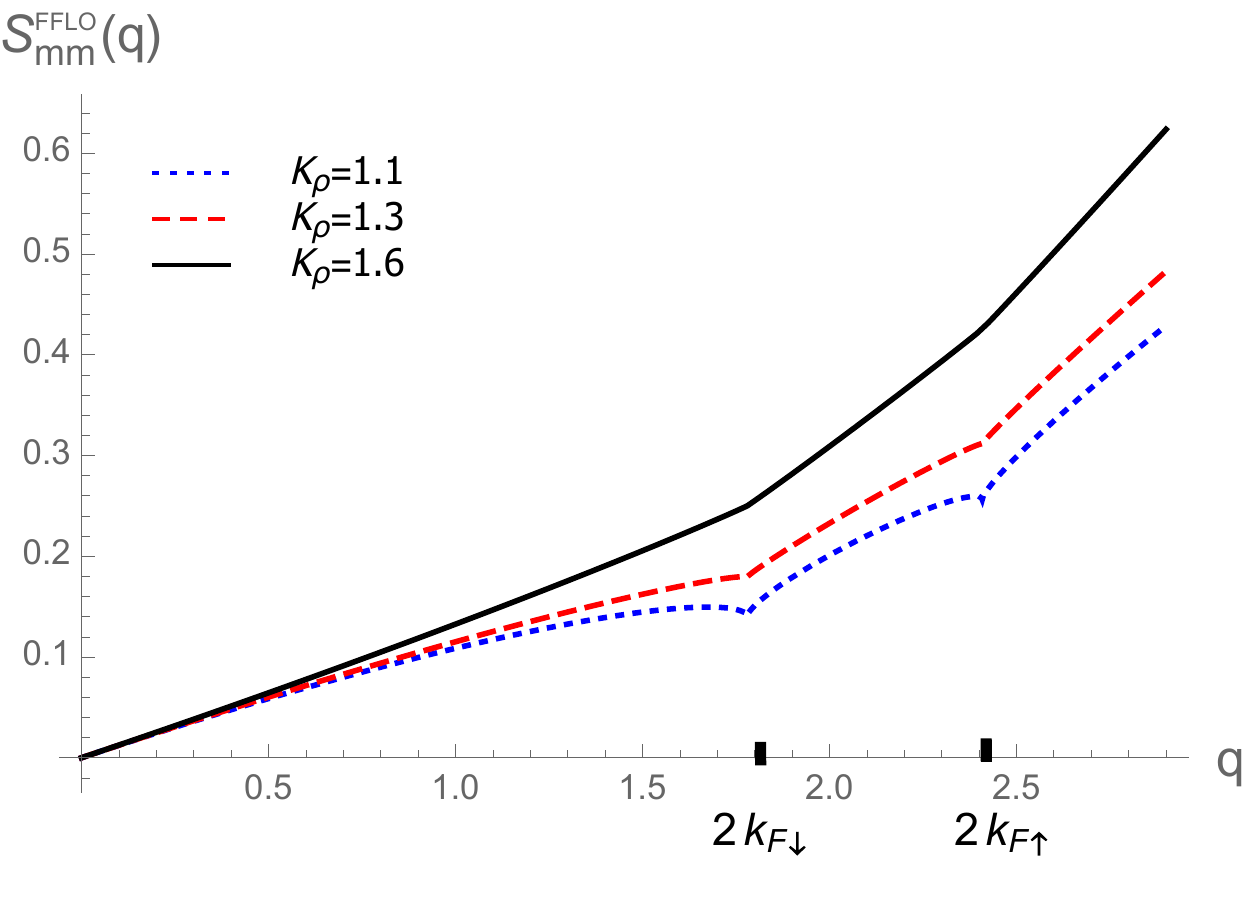}
  \caption{The magnetic structure function $S^{\mbox{\tiny\itshape
        FFLO}}_{mm}(q)$ for the FFLO state for
    different interactions: $K_\rho=1.1$ (dotted blue), $K_\rho=1.3$
    (dashed red) and $K_\rho=1.6$ (solid black).}
\label{Equsmmfflo}
\end{figure}

In the FFLO state the average magnetization density is uniform $\langle \hat
m(x) \rangle=\bar {m}$, as the structure is washed out by strong
quantum fluctuations in one dimension, that precludes spontaneous translational
symmetry breaking. The characteristic short-scale correlations are 
captured by the magnetization-magnetization correlator 
\begin{align}
\begin{split}
  S^{\mbox{\tiny\itshape
      FFLO}}_{mm}(x)&\sim-\frac{\kappa}{2}
\left(\frac{a  }{x}\right)^2\\
&\quad-\frac{1}{2}\left[\cos(2k_{F\uparrow}x)
+\cos(2k_{F\downarrow}x)\right]\left(\frac{a}{x}\right)^{K_\rho+\kappa/2},
\end{split}
\end{align}
that distinguishes FFLO from the BCS state. The corresponding magnetic
structure function is then given by
\begin{align}
\begin{split}
S^{\mbox{\tiny\itshape FFLO}}_{mm}(q)&\sim\frac{\kappa}{2}\sqrt{\frac{\pi}{2}}|q|-\frac{\Gamma(1-\gamma)\sin(\pi \gamma/2)}{2\sqrt{2\pi}} \Big({|q-2k_{F\uparrow}|^{\gamma-1}}\\
&+{|q+2k_{F\uparrow}|^{\gamma-1}}+{|q+2k_{F\downarrow}|^{\gamma-1}}+{|q-2k_{F\downarrow}|^{\gamma-1}}\Big)
\end{split}
\end{align}
and is illustrated in Fig.~\ref{Equsmmfflo}.

The above analysis thus demonstrates that there are clear qualitative
features that distinguish the BCS and the FFLO ground states, even in
the presence of strong quantum fluctuations. 
\section{quench dynamics: charge sector}
\label{quenchdynamicscharge}
We now turn to the nonequilibrium dynamics following a quantum quench
for the 1D Hubbard model. We first study the quench dynamics of the
charge sector \eqref{LLH}. Here we consider a generic quench protocol
of a sudden shift of the Luttinger parameter $K_\rho$ from $K_{\rho0}$
to $K_{\rho1}$, induced by the corresponding quench of the interaction
$U$ possible in Feshbach resonant systems. The quadratic form of the
charge sector Hamiltonian lends itself to an exact analysis, as has
been done in a variety of systems \cite{CazalillaPRA,Mitra2,
  HungChin13, NatuMueller13,Yin1,Cardy2}.

\subsection{Normal modes approach}
As discussed in Sec.~\ref{chargesector}, the pre-quench and post-quench
Hamiltonians can be diagonalized by the following Bogoliubov
transformations,
\begin{align}
\label{bogoliubovini}
\begin{split}
\begin{pmatrix}\hat b_p \\ \hat b^\dagger_{-p}\end{pmatrix}=\begin{pmatrix}\cosh\beta_0&-\sinh\beta_0\\ -\sinh\beta_0&\cosh\beta_0\end{pmatrix}\begin{pmatrix}\hat \chi_p \\ \hat \chi^\dagger_{-p}\end{pmatrix}
\equiv U(0^-)\begin{pmatrix}\hat\chi_p \\ \hat\chi^\dagger_{-p}\end{pmatrix},
\end{split}
\end{align}
and
\begin{align}
\label{bogoliubovfinal}
\begin{split}
\begin{pmatrix}\hat b_p \\\hat  b^\dagger_{-p}\end{pmatrix}=\begin{pmatrix}\cosh\beta_1&-\sinh\beta_1\\ -\sinh\beta_1&\cosh\beta_1\end{pmatrix}\begin{pmatrix}\hat \gamma_p \\ \hat \gamma^\dagger_{-p}\end{pmatrix}\equiv U(0^+)\begin{pmatrix}\hat \gamma_p \\ \hat \gamma^\dagger_{-p}\end{pmatrix},
\end{split}
\end{align}
where ($\hat\chi^\dagger_p,\hat\chi_p$) and ($\hat \gamma^\dagger_p,
\hat \gamma_p$) are the two sets of Bogoliubov quasiparticles,
carrying momentum $p$. The parameters $\beta_0$ and $\beta_1$ are
implicitly given by $e^{-2\beta_0}=K_{\rho0}$ and
$e^{-2\beta_1}=K_{\rho1}$.

The charge sector dynamics after the quench ($t>0$) is thus given by
\begin{align}
\label{timeevolvgamma}
\begin{split}
\begin{pmatrix}\hat \gamma_p(t) \\ \hat \gamma^\dagger_{-p}(t)\end{pmatrix}
=\begin{pmatrix}e^{iv_{\rho1}|p|t}&0\\ 0&e^{-iv_{\rho1}|p|t}\end{pmatrix}\begin{pmatrix}\hat \gamma_p \\\hat  \gamma^\dagger_{-p}\end{pmatrix}\equiv U_T(t)\begin{pmatrix}\hat \gamma_p \\ \hat \gamma^\dagger_{-p}\end{pmatrix},
\end{split}
\end{align}
where $v_{\rho1}$ is the charge velocity for the post-quench
Hamiltonian and $\hat\gamma_p\equiv \hat\gamma_p(t=0)$ and
$\hat\gamma^\dagger_p\equiv\hat\gamma^\dagger_p(t=0)$.

Combining Eqs.~\eqref{bogoliubovini}, \eqref{bogoliubovfinal}, and
\eqref{timeevolvgamma} relates ($\hat b^\dagger_p(t), \hat
b_{-p}(t)$) to the initial pre-quench set of quasi-particles
($\hat\chi_p, \hat \chi^\dagger_{-p}$) via
\begin{align}
\label{chargesectorbogoliubovtime}
\begin{split}
\begin{bmatrix} \hat b_p(t)\\\hat b^\dagger_{-p}(t)\\\end{bmatrix}&=U(0^+)U_T(t)U^{-1}(0^+)U(0^-)\begin{bmatrix}\hat  \chi_{p}\\\hat \chi^\dagger_{-p}\\\end{bmatrix},
\end{split}
\end{align}
where the zero-temperature initial ground-state distribution of ($\hat
\chi^\dagger_p$, $\hat \chi_p$) is given by $\langle \hat \chi_{p}\hat
\chi^\dagger_{p'}\rangle=\delta_{p,{p'}}$.

The post-quench correlators of ${\hat{\phi}}_\rho(t)$ and ${\hat{\theta}} _\rho(t)$ can
now be readily computed. Combining Eqs.~\eqref{phitheta} and
\eqref{chargesectorbogoliubovtime}, and relegating the details to
Appendix \ref{chargemicrodynamics}, we find
\begin{align}
\begin{split}
\label{quenchluttingerphitheta}
&\quad\langle e^{i\sqrt{2}[{\hat{\phi}}_\rho(x,t)-{\hat{\phi}}_\rho(0,t)}\rangle\\
&\sim\left(\frac{ a }{x}\right)^{\frac{1}{2}(\frac{K^2_{\rho1}}{K_{\rho0}}+K_{\rho0})}\left|\frac{x^2-(2v_{\rho1} t)^2}{ (2v_{\rho1} t)^2+a^2}\right|^{\frac{1}{4}(\frac{K^2_{\rho1}}{K_{\rho0}}-K_{\rho0})},\\
&\sim\begin{cases} \left(\frac{ a }{x}\right)^{K_{\rho0}}\left(\frac{1}{(2v_{\rho1} t/a)^2+1}\right)^{\frac{1}{4}(\frac{K^2_{\rho1}}{K_{\rho0}}-K_{\rho0})} , &  x\gg 2v_{\rho1} t, \\ 
\left(\frac{ a }{x}\right)^{\frac{1}{2}(K_{\rho0}+\frac{K^2_{\rho1}}{K_{\rho0}})}, &  x\ll 2v_{\rho1} t, \end{cases}
\end{split}
\end{align}
\begin{align}
\begin{split}
\label{quenchluttingerphitheta2}
&\quad\langle e^{i\sqrt{2}[{\hat{\theta}} _\rho(x,t)-{\hat{\theta}} _\rho(0,t)}\rangle\\
&\sim\left(\frac{ a }{x}\right)^{\frac{1}{2}(\frac{K_{\rho0}}{K^2_{\rho1}}+\frac{1}{K_{\rho0}})}\left|\frac{x^2-(2v_{\rho1} t)^2}{(2v_{\rho1} t)^2+a^2}\right|^{\frac{1}{4}(\frac{K_{\rho0}}{K^2_{\rho1}}-\frac{1}{K_{\rho0}})},\\
&\sim\begin{cases} \left(\frac{ a }{x}\right)^{K^{-1}_{\rho0}}\left(\frac{1}{(2v_{\rho1} t/a)^2+1}\right)^{\frac{1}{4}(\frac{K_{\rho0}}{K^2_{\rho1}}-\frac{1}{K_{\rho0}})} , &  x\gg 2v_{\rho1} t, \\ 
\left(\frac{ a }{x}\right)^{\frac{1}{2}(\frac{K_{\rho0}}{K^2_{\rho1}}+\frac{1}{K_{\rho0}})}, &  x\ll 2v_{\rho1} t, \end{cases}
\end{split}
\end{align}
and
\begin{align}
\label{chargequenchphirho}
\begin{split}
&\quad2a^2\langle\partial_x {\hat{\phi}}_\rho(x,t)\partial_{x'} {\hat{\phi}}_\rho(0,t)\rangle\\
&=-\frac{(K_{\rho0}+\frac{K^2_{\rho1}}{K_{\rho0}})a^2}{2x^{2}}
-\frac{(K_{\rho0}-\frac{K^2_{\rho1}}{K_{\rho0}})a^2}{4(x+2v_{\rho1} t)^2}-\frac{(K_{\rho0}-\frac{K^2_{\rho1}}{K_{\rho0}})a^2}{4(x-2v_{\rho1} t)^2}.
\end{split}
\end{align}

We will combine these dynamical charge sector correlators with the
spin sector ones to calculate observables for the Hubbard model, as we
have done earlier for the equilibrium case. However, before moving on
we will reproduce above charge correlators using a simpler approach,
that will be essential for the spin sector analysis, where
the nonlinearity precludes a direct diagonalization utilized above.

\subsection{Heisenberg equations of motion approach}
A complementary approach to the above momentum eigenmodes Hamiltonian
diagonalization is to directly solve the Heisenberg equation of motion
for the field operators ${\hat{\phi}}_\rho(x,t)$ and ${\hat{\theta}}
_\rho(x,t)$ in terms of the corresponding pre-quench operators,
${\hat{\phi}}_\rho(x,t=0)$ and ${\hat{\theta}} _\rho(x,t=0)$, whose
correlators we have already computed in Sec.~\ref{chargesector}.

Using the charge sector Hamiltonian \eqref{LLH} and the commutation
relations,
 \begin{align}
\begin{split}
[{\hat{\phi}}_\rho(x),\partial_{x'} {\hat{\theta}} _\rho(x')]=i\pi\delta(x-x'),\\
[{\hat{\theta}}_\rho(x),\partial_{x'} {\hat{\phi}}_\rho(x')]=i\pi\delta(x-x'),
\end{split}
\end{align}
the coupled Heisenberg equations of motion for ${\hat{\phi}}_\rho$ and
${\hat{\theta}}_\rho$ are readily obtained:
 \begin{align}
\begin{split}
\dot{\hat{\phi}}_\rho(x,t)&=K_{\rho1}v_{\rho1}\partial_x {\hat{\theta}}_\rho(x,t),\\
\dot{\hat{\theta}}_\rho(x,t)&=\frac{v_{\rho1}}{K_{\rho1}}\partial_x {\hat{\phi}}_\rho(x,t).
\end{split}
\end{align}
As usual, they can be decoupled into wave equations:
 \begin{align}
 \label{eomHeisenberg}
\begin{split}
\ddot{\hat{\phi}}_\rho(x,t)=v^2_{\rho1}\partial^2_x{\hat{\phi}}_\rho(x,t),\\
\ddot{\hat{\theta}}_\rho(x,t)=v^2_{\rho1}\partial^2_x{\hat{\theta}}_\rho(x,t),
\end{split}
\end{align}
with the initial conditions
 \begin{align}
 \label{eomHeisenberginitial}
\begin{split}
{\hat{\phi}}_\rho(x,0)&\equiv{\hat{\phi}}_\rho(x),\\
{\hat{\theta}}_\rho(x,0)&\equiv{\hat{\theta}}_\rho(x),\\
\dot{\hat{\phi}}_\rho(x,0)&=K_{\rho1}v_{\rho1}\partial_x {\hat{\theta}}_\rho,\\
\dot{\hat{\theta}}_\rho(x,0)&=\frac{v_{\rho1}}{K_{\rho1}}\partial_x {\hat{\phi}}_\rho.
\end{split}
\end{align}

The dynamics is now straightforwardly obtained as a linear combination
of the left and right traveling solutions satisfying the above initial
conditions,
\begin{subequations}
 \label{eomheisenbergsolution}
\begin{align}
{\hat{\phi}}_\rho(x,t)&=\frac{1}{2}[{\hat{\phi}}_\rho(x+v_{\rho1}t)+{\hat{\phi}}_\rho(x-v_{\rho1}t)]\nonumber\\
&\quad+\frac{K_{\rho1}}{2}[{\hat{\theta}}_\rho(x+v_{\rho1}t)-{\hat{\theta}}_\rho(x-v_{\rho1}t)],\\
{\hat{\theta}}_\rho(x,t)&=\frac{1}{2}[{\hat{\theta}}_\rho(x+v_{\rho1}t)+{\hat{\theta}}_\rho(x-v_{\rho1}t)]\nonumber\\
&\quad+\frac{1}{2K_{\rho1}}[{\hat{\phi}}_\rho(x+v_{\rho1}t)-{\hat{\phi}}_\rho(x-v_{\rho1}t)],
\end{align}
\end{subequations}
with operators at $t=0$ appearing on the right-hand side.  These thus
allow us to connect the post-quench dynamical correlators (with all
averages taken in the pre-quenched initial ground state) to their
initial $t=0^-$ pre-quench counterparts,
\begin{align}
\label{phicorrelatoreom}
\begin{split}
&\quad\langle e^{i\sqrt{2}[{\hat{\phi}}_\rho(x,t)-{\hat{\phi}}_\rho(0,t)]}\rangle\\
&=\langle e^{i\frac{\sqrt{2}}{2}[{\hat{\phi}}_\rho(x+v_{\rho1}t)+{\hat{\phi}}_\rho(x-v_{\rho1}t)-{\hat{\phi}}_\rho(v_{\rho1}t)-{\hat{\phi}}_\rho(-v_{\rho1}t)]}\rangle\\
&\quad\times \langle e^{i\frac{\sqrt{2}K_{\rho1}}{2}[{\hat{\theta}}_\rho(x+v_{\rho1}t)-{\hat{\theta}}_\rho(x-v_{\rho1}t)-{\hat{\theta}}_\rho(v_{\rho1}t)+{\hat{\theta}}_\rho(-v_{\rho1}t)]}\rangle,\\
&=e^{1/4[2F_{\hat{\phi}}(2v_{\rho1}t)-2F_{\hat{\phi}}(x)-F_{\hat{\phi}}(x+2v_{\rho1}t)-F_{\hat{\phi}}(|x-2v_{\rho1}t|)]}\\
&\quad\times e^{-K^2_{\rho1}/4[2F_{\hat{\theta}}(2v_{\rho1}t)+2F_{\hat{\theta}}(x)-F_{\hat{\theta}}(x+2v_{\rho1}t)-F_{\hat{\theta}}(|x-2v_{\rho1}t|)]},
\end{split}
\end{align}
and
 \begin{align}
 \label{thetacorrelatoreom}
\begin{split}
&\quad\langle e^{i\sqrt{2}[{\hat{\theta}}_\rho(x,t)-{\hat{\theta}}_\rho(0,t)}\rangle\\
&=\langle e^{i\frac{\sqrt{2}}{2}[{\hat{\theta}}_\rho(x+v_{\rho1}t)+{\hat{\theta}}_\rho(x-v_{\rho1}t)-{\hat{\theta}}_\rho(v_{\rho1}t)-{\hat{\theta}}_\rho(-v_{\rho1}t)]}\rangle\\
&\quad\times \langle e^{i\frac{\sqrt{2}}{2K_{\rho1}}[{\hat{\phi}}_\rho(x+v_{\rho1}t)-{\hat{\phi}}_\rho(x-v_{\rho1}t)-{\hat{\phi}}_\rho(v_{\rho1}t)+{\hat{\phi}}_\rho(-v_{\rho1}t)]}\rangle,\\
&=e^{1/4[2F_{\hat{\theta}}(2v_{\rho1}t)-2F_{\hat{\theta}}(x)-F_{\hat{\theta}}(x+2v_{\rho1}t)-F_{\hat{\theta}}(|x-2v_{\rho1}t|)]}\\
&\quad\times e^{-1/(4K^2_{\rho1})[2F_{\hat{\phi}}(2v_{\rho1}t)+2F_{\hat{\phi}}(x)-F_{\hat{\phi}}(x+2v_{\rho1}t)-F_{\hat{\phi}}(|x-2v_{\rho1}t|)]},
\end{split}
\end{align}
with the details evaluated in Appendix \ref{Wickmultiple}.  Using the initial
pre-quench correlators of ${\hat{\phi}}_\rho(x)$ and ${\hat{\theta}}_\rho(x)$, studied
in Sec.~\ref{chargesector},
\begin{align}
\begin{split}
F_{\hat{\phi}}(x)&\equiv\langle ({\hat{\phi}}_\rho(x)- {\hat{\phi}}_\rho(0))^2\rangle\sim K_{\rho0}\ln\frac{x}{a},\\
F_{\hat{\theta}}(x)&\equiv\langle ({\hat{\theta}}_\rho(x)- {\hat{\theta}}_\rho(0))^2\rangle\sim \frac{1}{K_{\rho0}}\ln\frac{x}{a}.
\end{split}
\end{align}
inside Eqs.~\eqref{phicorrelatoreom} and \eqref{thetacorrelatoreom}, we much
more simply reproduce the results
\eqref{quenchluttingerphitheta} and \eqref{quenchluttingerphitheta2}
obtained with the Hamiltonian diagonalization approach.

\section{Quench dynamics of spin sector}
\label{dynamicsspinpart}
We now study the dynamics of the spin sector Hamiltonian
\eqref{SineGordonH} following a $U\to 0$ quench to noninteracting
fermions at $t=0$. This quench protocol is simple enough to be
conveniently implemented in experiments, yet still contains the
essential elements of the nonequilibrium dynamics. It also has the
appeal that the post-quench evolution is exactly solvable, reducing
all the difficulties to the analysis of correlations in the initial
interacting state. 


Thus for the $U\to 0$ quench protocol, the Heisenberg equations of
motion satisfied by ${\hat{\phi}}_\sigma(x,t)$ and ${\hat{\theta}}_\sigma(x,t)$ is
still a linear wave equation \eqref{eomHeisenberg} with initial conditions
 \begin{align}
\begin{split}
{\hat{\phi}}_\sigma(x,0)&\equiv{\hat{\phi}}_\sigma(x),\\
{\hat{\theta}}_\sigma(x,0)&\equiv{\hat{\theta}}_\sigma(x),\\
\dot{\hat{\phi}}_\sigma(x,0)&=v_{F}\partial_x {\hat{\theta}}_\sigma,\\
\dot{\hat{\theta}}_\sigma(x,0)&=v_{F}\partial_x {\hat{\phi}}_\sigma+\frac{\sqrt{2}h}{\pi}.
\end{split}
\end{align}
and the post-quench Luttinger parameters $K_{\sigma}=1$ and
$v_{\sigma}=v_F$ for free fermions.

The solution thus is straightforwardly obtained and is quite similar
to the charge sector,
\begin{subequations}
 \label{eomHeisenbergspinphitheta}
\begin{align}
{\hat{\phi}}_\sigma(x,t)&=\frac{1}{2}[{\hat{\phi}}_\sigma(x+v_{F}t)+{\hat{\phi}}_\sigma(x-v_{F}t)]\nonumber\\
&\quad+\frac{1}{2}[{\hat{\theta}}_\sigma(x+v_{F}t)-{\hat{\theta}}_\sigma(x-v_{F}t)],\\
{\hat{\theta}}_\sigma(x,t)&=\frac{1}{2}[{\hat{\theta}}_\sigma(x+v_{F}t)+{\hat{\theta}}_\sigma(x-v_{F}t)]\nonumber\\
&\quad+\frac{1}{2}[{\hat{\phi}}_\sigma(x+v_{F}t)-{\hat{\phi}}_\sigma(x-v_{F}t)]+\frac{\sqrt{2}h}{\pi}t.
\end{align}
\end{subequations}

The dynamical correlators at time $t$ can then again be expressed in
terms of the pre-quench ones at the initial time,
 \begin{align}
 \label{eomHeisenbergphicorrelator}
\begin{split}
&\quad\langle e^{i\sqrt{2}[{\hat{\phi}}_\sigma(x,t)-{\hat{\phi}}_\sigma(0,t)}\rangle\\
&=e^{1/4[2D_{\hat{\phi}}(2v_{F}t)-2D_{\hat{\phi}}(x)-D_{\hat{\phi}}(x+2v_{F}t)-D_{\hat{\phi}}(|x-2v_{F}t|)]}\\
&\quad\times e^{-1/4[2D_{\hat{\theta}}(2v_{F}t)+2D_{\hat{\theta}}(x)-D_{\hat{\theta}}(x+2v_{F}t)-D_{\hat{\theta}}(|x-2v_{F}t|)]},
\end{split}
\end{align}
and
 \begin{align}
  \label{eomHeisenbergthetacorrelator}
\begin{split}
&\quad\langle e^{i\sqrt{2}[{\hat{\theta}}_\sigma(x,t)-{\hat{\theta}}_\sigma(0,t)}\rangle\\
&=e^{1/4[2D_{\hat{\theta}}(2v_{F}t)-2D_{\hat{\theta}}(x)-D_{\hat{\theta}}(x+2v_{F}t)-D_{\hat{\theta}}(|x-2v_{F}t|)]}\\
&\quad\times e^{-1/4[2D_{\hat{\phi}}(2v_{F}t)+2D_{\hat{\phi}}(x)-D_{\hat{\phi}}(x+2v_{F}t)-D_{\hat{\phi}}(|x-2v_{F}t|)]},
\end{split}
\end{align}
where
\begin{align}
\begin{split}
D_{\hat{\phi}}(x)&\equiv\langle ({\hat{\phi}}_\sigma(x)- {\hat{\phi}}_\sigma(0))^2\rangle,\\
D_{\hat{\theta}}(x)&\equiv\langle ({\hat{\theta}}_\sigma(x)- {\hat{\theta}}_\sigma(0))^2\rangle
\end{split}
\end{align}
are the correlators for ${\hat{\phi}}_\sigma(x)$ and ${\hat{\theta}}_\sigma(x)$ prior
to the quench.  These depend qualitatively on the initial state, as we
now discuss for quenches from the BCS and the FFLO ground states.

\subsection{Quench $U\to0$ from BCS state}
Following the quench from the BCS state to the non-interacting Fermi
gas, the classical field part $\langle{\hat{\phi}}_\sigma(x,t)\rangle$ evolves
according to Eq.~\eqref{eomHeisenbergspinphitheta}, given
approximately by
\begin{align}
\begin{split}
&\langle{\hat{\phi}}_\sigma(x,t)\rangle=\frac{1}{2}[\langle{\hat{\phi}}_\sigma(x+v_{F}t)\rangle
+\langle{\hat{\phi}}_\sigma(x-v_{F}t)\rangle]\\
&\sim\begin{cases} 0 , &  |x|<L/2-\xi'-v_Ft, \\ 
h\frac{\sqrt{2K_\sigma}}{4\pi}\left(x - L/2 +\xi'+v_Ft\right), 
&  |x|>L/2-\xi'-v_Ft.
 \end{cases}
\end{split}
\end{align}
This describes the penetration of magnetization into the bulk via a
ballistic motion of a fraction of a soliton from the edge into the
bulk, and as expected eventually leads to a constant magnetization
(species imbalance) of the noninteracting Fermi gas in the presence of
a finite Zeeman field (chemical potential imbalance) $h$.  However,
this takes a macroscopically long time $t_L\approx L/v_F$ to travel
through the system. From here on, we will focus on the thermodynamic
limit ($L\rightarrow\infty$ but $t$ finite), in the bulk of the sample
and thus neglect these ``edge'' effects. We will thus take
$\langle{\hat{\phi}}_\sigma(x,t)\rangle=0$ in the spin-gapped BCS ground
state.

Using the BCS ground-state correlators from Sec.~\ref{spinsemiclassical},
\begin{align}
\begin{split}
D^{\mbox{\tiny\itshape BCS}}_{\hat{\phi}}(x)&\sim \const.,\\
D^{\mbox{\tiny\itshape BCS}}_{\hat{\theta}}(x)&\sim x/\xi+\const.
\end{split}
\end{align}
inside post-quench ones at time $t$,
Eq.~\eqref{eomHeisenbergphicorrelator},\eqref{eomHeisenbergthetacorrelator},
we obtain
 \begin{align}
 \label{DynamicsBCSphi}
\begin{split}
\langle e^{i\sqrt{2}[{\hat{\phi}}_\sigma(x,t)-{\hat{\phi}}_\sigma(0,t)}\rangle&\sim e^{-(x+2v_Ft-|x-2v_Ft|)/(4\xi)}\\
&\sim \begin{cases} e^{-\frac{2v_Ft}{2\xi}}, &  x\gg 2v_F t,
\\ e^{- \frac{x}{2\xi}} , &  x\ll 2v_F t,\end{cases}
\end{split}
\end{align}
and
\begin{align}
 \label{DynamicsBCStheta}
\begin{split}
\langle e^{i\sqrt{2}[{\hat{\theta}}_\sigma(x,t)-{\hat{\theta}}_\sigma(0,t)}\rangle
&\sim e^{-(3x+|x-2v_Ft|-2v_Ft)/(4\xi_0)}\\
&\sim \begin{cases} e^{-\frac{x-v_Ft}{\xi }}, &  x\gg 2v_F t,
\\ e^{-\frac{x}{2\xi }}, &  x\ll 2v_F t, \end{cases}
\end{split}
\end{align}
These exhibit exponential behavior in time and space and at long time
reach a stationary form. As we will see these differ qualitatively
from a quench from the FFLO state to which we turn next.

\subsection{Quench $U\to 0$ from FFLO state}
The nonzero magnetization in the FFLO state is carried by a soliton
lattice. Thus, following a quench we expect nontrivial dynamics
associated with soliton lattice oscillations and breathing, described
by the solution of the sine-Gordon equation. However, strong 1D
quantum fluctuations wash out this classical dynamics, that will,
however, appear in higher dimensions. Furthermore, as we have seen in
Sec.~\ref{spinsemiclassical}, outside of a narrow range above $h_c$
solitons strongly overlap, leading to a vanishing periodic (in space)
component of the density. The magnetization can therefore be well
approximated by a constant, corresponding to
\begin{align}
\begin{split}
\langle{\hat{\phi}}_\sigma(x,t)\rangle
&\sim\frac{\pi\bar{m}}{\sqrt{2}}x.
\end{split}
\end{align}

Recalling from Sec.~\ref{spinsemiclassical}, the FFLO ground state
pre-quench correlations are given by
 \begin{align}
\begin{split}
D^{\mbox{\tiny\itshape FFLO}}_{\hat{\phi}}(x)&=1/2\langle (\tilde{\phi}_\sigma(x)-\tilde {\phi}_\sigma(0))^2\rangle\sim\frac{\kappa}{2}\ln\frac{x}{a},\\
D^{\mbox{\tiny\itshape FFLO}}_{\hat{\theta}}(x)&=2\langle (\tilde{\theta}_\sigma(x)- \tilde{\theta}_\sigma(0))^2\rangle\sim\frac{2}{\kappa}\ln\frac{x}{a}.
\end{split}
\end{align}
Using them inside post-quench correlators in
Eqs.~\eqref{eomHeisenbergphicorrelator} and \eqref{eomHeisenbergthetacorrelator},
we find
\begin{align}
\begin{split}
\langle e^{i\sqrt{2}[{\hat{\phi}}_\rho(x,t)-{\hat{\phi}}_\rho(0,t)}\rangle&\sim e^{i\pi\bar{m}x}\left(\frac{ a }{x}\right)^{\frac{1}{2}(\frac{2}{\kappa}+\frac{\kappa}{2})}\\
&\quad\times\left|\frac{x^2-(2v_{F} t)^2}{ (2v_{F} t)^2+a^2}\right|^{\frac{1}{4}(\frac{2}{\kappa}-\frac{\kappa}{2})},
\end{split}
\end{align}
\begin{align}
\begin{split}
\langle e^{i\sqrt{2}[{\hat{\theta}}_\rho(x,t)-{\hat{\theta}}_\rho(0,t)}\rangle
&\sim \left(\frac{ a }{x}\right)^{\frac{1}{2}(\frac{2}{\kappa}+\frac{\kappa}{2})}\left|\frac{(2v_{F} t)^2+a^2}{x^2-(2v_{F} t)^2}\right|^{\frac{1}{4}(\frac{2}{\kappa}-\frac{\kappa}{2})},
\end{split}
\end{align}
and 
\begin{align}
\begin{split}
&\quad 2a^2\langle\partial_x{\hat{\phi}}_\sigma(x)\partial_{x'} {\hat{\phi}}_\sigma(0)\rangle\\
&\sim -\frac{(\kappa/2+2/\kappa)a^2}{2x^{2}}-\frac{(\kappa/2-2/\kappa)a^2}{4(x+2v_Ft)^{2}}-\frac{(\kappa/2-2/\kappa)a^2}{4(x-2v_Ft)^{2}}.
\end{split}
\end{align}
These power-law dynamic correlations contrast strongly with those for
the quench from the BCS state and exhibit a long-time stationary
pre-thermalized state.

\section{1D spin-imbalanced Fermi-Hubbard model: Quench dynamics}
\label{quenchHubbard}
We now are in the position to assemble the results from earlier
sections to predict the quench dynamics of the 1D spin-imbalanced
Fermi gas, described by the Fermi-Hubbard model.  As discussed in
Sec.~\ref{dynamicsspinpart}, in this paper we limit our study to a
quench protocol to a vanishing on-site interaction, i.e., a quench
$U\to0$. Such quench is not only simpler to analyze, but also more
straightforwardly implementable in a Feshbach resonant gas by tuning
to a point of a vanishing scattering length and by shutting off the
trap.

\subsection{$U\to 0$ from BCS state}
\label{quenchHubbardBCS}
Combining Eqs.~\eqref{quenchluttingerphitheta2} and
\eqref{DynamicsBCSphi}, we obtain the spin-singlet pairing correlator
[defined via Eqs.~\eqref{Ossoperatora} and \eqref{Sssdefinition}] at time $t$, following a $U\rightarrow0$ quench from an initial BCS ground state,
\begin{align}
\label{quenchBCSss}
\begin{split}
&\quad S^{\mbox{\tiny\itshape BCS}}_{ss}(x,t)\\
&= \langle e^{i\sqrt{2}[{\hat{\theta}}_\rho(x,t)-{\hat{\theta}}_\rho(0,t)]}\rangle\langle \cos(\sqrt{2}{\hat{\phi}}_\sigma(x,t))\cos(\sqrt{2}{\hat{\phi}}_\sigma(0,t))\rangle,\\
&\sim \left(\frac{ a }{x}\right)^{(K_{\rho0}+K^{-1}_{\rho0})/2}\left|\frac{x^2-(2v_F t)^2}{(2v_Ft)^2+a^2}\right|^{(K_{\rho0}-K^{-1}_{\rho0})/4} \\
&\quad\times e^{-(x+2v_Ft-|x-2v_Ft|)/(4\xi)},\\
&\sim\begin{cases} \left(\frac{ a }{x}\right)^{K^{-1}_{\rho0}}\left(\frac{a}{2v_{F} t}\right)^{(K_{\rho0}-K^{-1}_{\rho0})/2} e^{-\frac{v_F t}{\xi }} , & x\gg 2v_F t,\\ 
\left(\frac{ a }{x}\right)^{(K_{\rho0}+K^{-1}_{\rho0})/2}e^{-\frac{x}{2\xi }} , &  x\ll 2v_F t, \end{cases}\\
\end{split}
\end{align}
illustrated in Fig.~\ref{IntensityBCSss} and Fig.~\ref{DynamicsBCSnqpairrealspace}. We observe that despite the initial power-law correlations in the BCS ground state (set by gapless charge fluctuations), at the light-cone time $t_*(x) = x/(2v_F)$ \cite{Cardy}, these correlations crossover to short-ranged stationary
ones set by the correlation length $\xi$.

\begin{figure}[!htb]
 \centering
  \includegraphics[width=80mm]{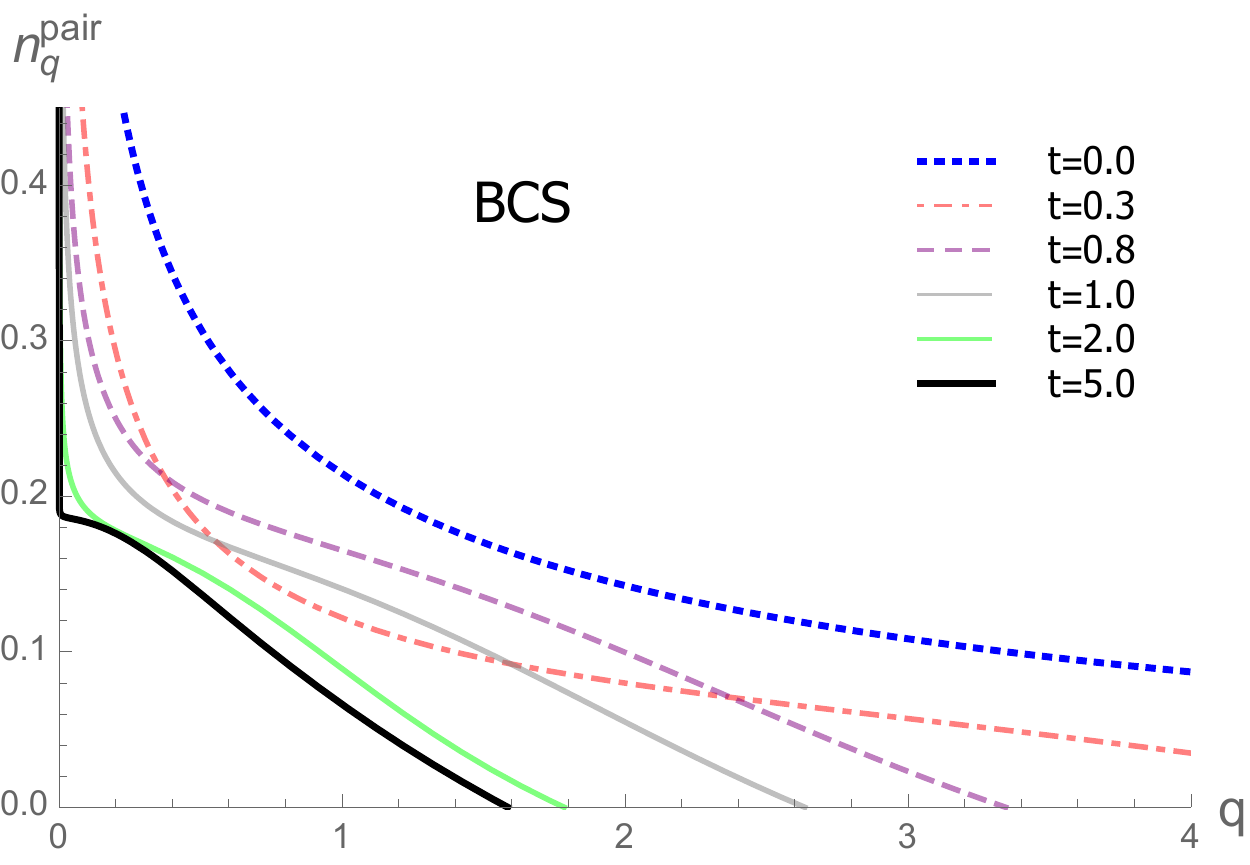}
  \caption{Cooper-pair momentum distribution $n^{pair}_{q}(t)$ for a series of time following the quantum quench $U\to 0$ at $t=0$ from the BCS ground state for $K_\rho=1.6$ at $t=0^-$. It illustrates a decay of the zero-momentum peak following the quench.}
\label{DynamicsBCSnqpair}
\end{figure}

Above dynamics is also reflected in momentum space. We illustrate in
Fig.~\ref{DynamicsBCSnqpair} the spatial Fourier transform of
$S^{\mbox{\tiny\itshape BCS}}_{ss}(x,t)$, namely the Cooper-pair
momentum distribution $n^{pair}_q(t)$. We observe that the strength of
zero-momentum peak decays in time, indicating the collapse of the BCS
pairing following the quench to the noninteracting state. This is
consistent with the exponentially decaying BCS pairing order shown in
Eq.~\eqref{quenchBCSss}.

\begin{figure}[!htb]
 \centering
  \includegraphics[width=90mm]{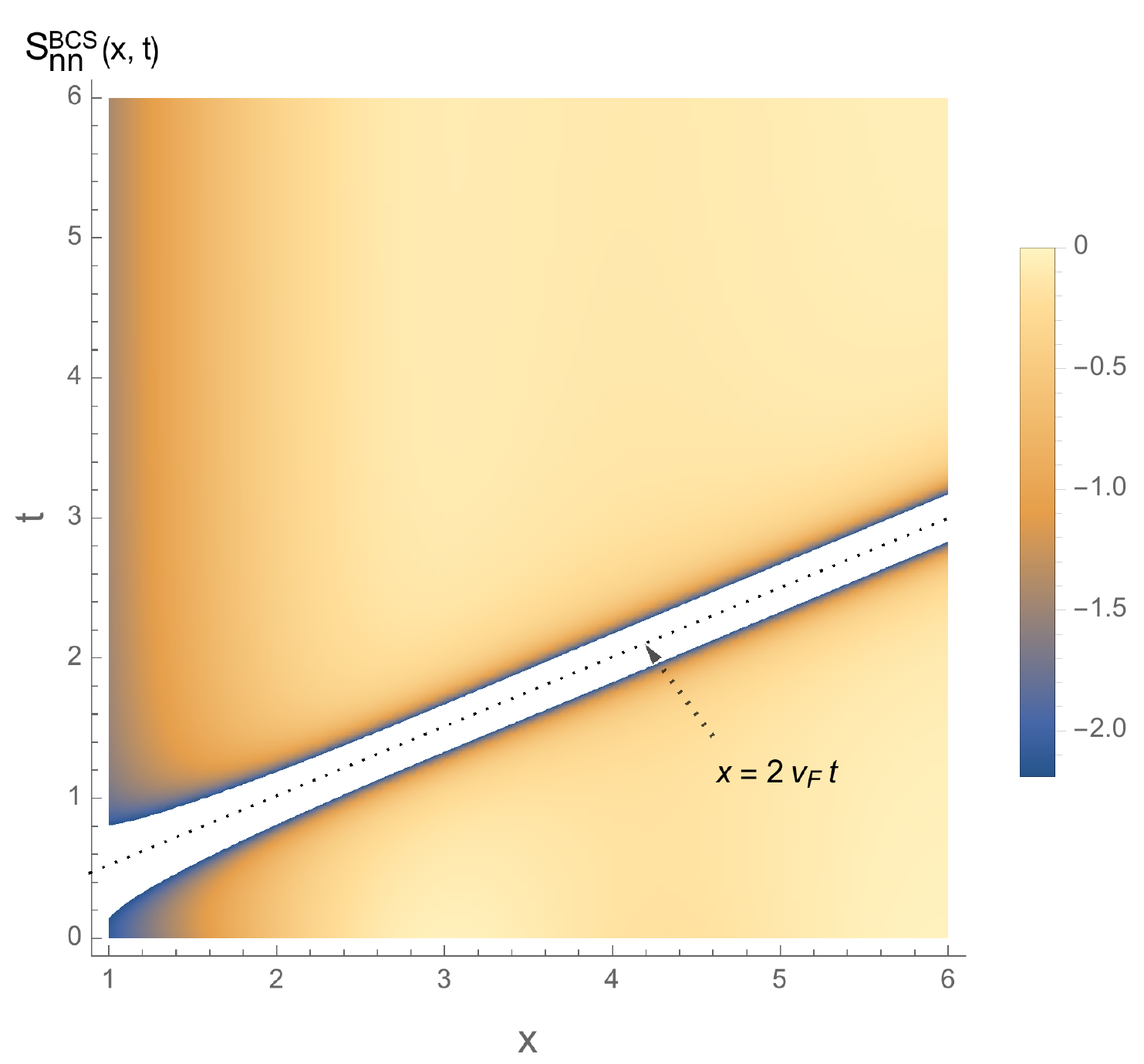}
  \caption{Space-time intensity plot of density-density correlation
    function $S^{\mbox{\tiny\itshape BCS}}_{nn}(q,t)$ following a
    $U\to 0$ quench at $t=0$ from the BCS state for $K_\rho=1.6$ at
    $t=0^-$. The light-cone boundary ($x=2v_F t$, the dotted line is a
    guide to an eye) is a crossover between the early-time
    ground-state correlation to those in the asymptotic long time
    stationary state. }
 \label{IntensityBCSnn}
\end{figure}
\begin{figure}[!htb]
 \centering
  \includegraphics[width=80mm]{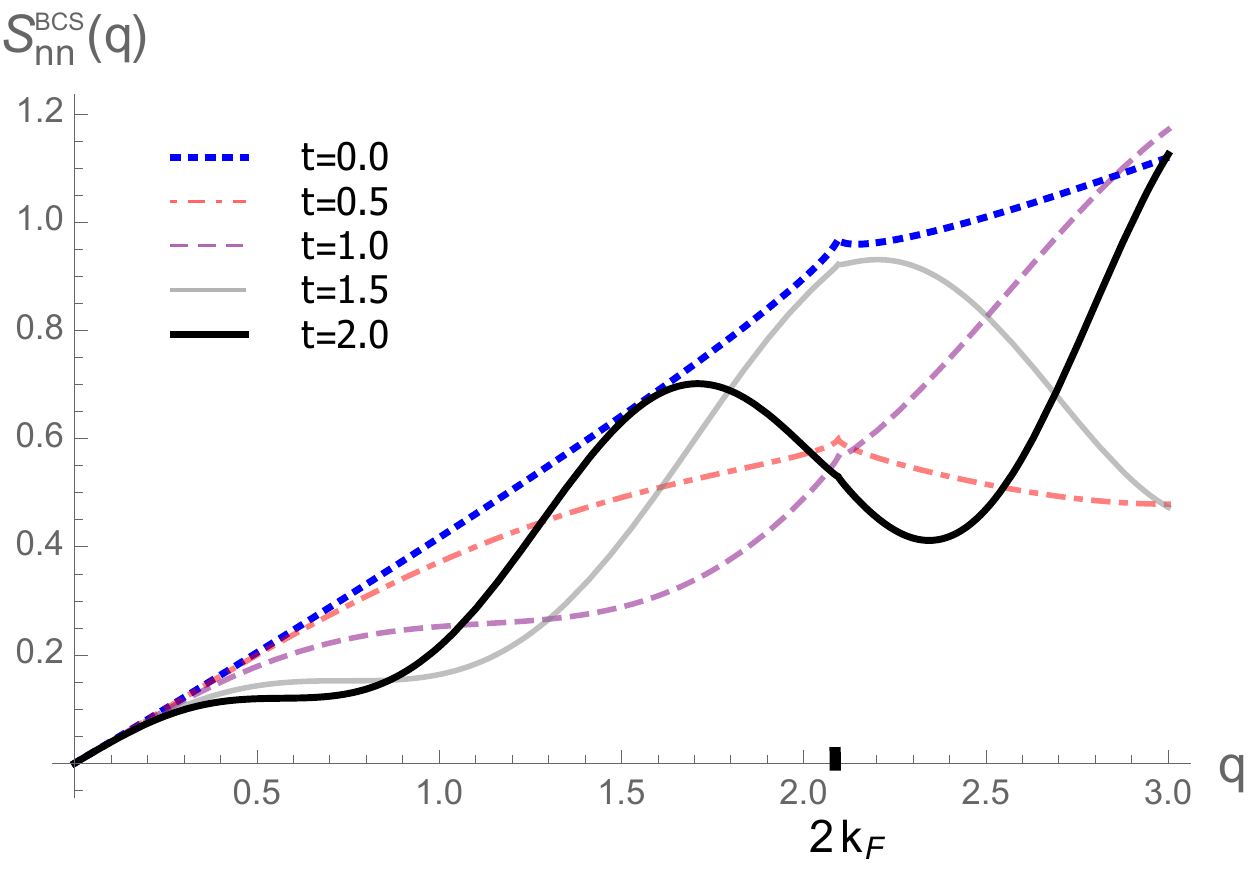}
  \caption{Structure function $S^{\mbox{\tiny\itshape
        BCS}}_{nn}(q,t)$ for a series of time following the quantum
    quench $U\to 0$ at $t=0$ from the BCS spin-singlet ground state for $K_\rho=1.6$ at $t=0^-$, with approximate analytical form given by Eq. \eqref{QuenchBCSSnnq}.}
\label{DynamicsBCSdensitydensitymomentum}
\end{figure}

Combining Eqs.~\eqref{quenchluttingerphitheta2} and \eqref{DynamicsBCStheta}, we obtain the post-quench dynamical spin-triplet correlator (defined via
Eq.~\eqref{Ossoperatorb}\eqref{Sstdefinition})
\begin{align}
\label{quenchBCSst}
\begin{split}
&\quad S^{\mbox{\tiny\itshape BCS}}_{st}(x,t)\\
&=\langle e^{i\sqrt{2}[{\hat{\theta}}_\rho(x,t)-{\hat{\theta}}_\rho(0,t)]}\rangle
  \langle \cos(\sqrt{2}{\hat{\theta}}_\sigma(x,t))\cos(\sqrt{2}{\hat{\theta}}_\sigma(0,t))\rangle \\
&\sim\left(\frac{ a }{x}\right)^{(K_{\rho0}+K^{-1}_{\rho0})/2}\left|\frac{x^2-(2v_F t)^2}{(2v_Ft)^2+a^2}\right|^{(K_{\rho0}-K^{-1}_{\rho0})/4}\\
&\quad\times e^{-(3x+|x-2v_Ft|-2v_Ft)/(4\xi)},\\
&\sim \begin{cases} e^{-\frac{x-v_Ft}{\xi }}\left(\frac{ a }{x}\right)^{K^{-1}_{\rho0}}\left(\frac{a}{2v_{F} t}\right)^{(K_{\rho0}-K^{-1}_{\rho0})/2}  , &  x\gg 2v_F t,\\
 e^{-\frac{x}{2\xi}} \left(\frac{ a }{x}\right)^{(K_{\rho0}+K^{-1}_{\rho0})/2}, &  x\ll 2v_F t. \end{cases}
\end{split}
\end{align}
We observe that it also displays a light-cone crossover from a spatial
exponential decay in the singlet BCS ground state at short times to a stationary
exponential decay with a doubled correlation length at
long time $t > t_*(x)$.  The doubled correlation length seems to indicate
that, compared to the initial singlet BCS ground state, the
triplet-pairing is enhanced as the former is exponentially
suppressed. 

Combining Eqs.~\eqref{quenchluttingerphitheta}, \eqref{chargequenchphirho}, and \eqref{DynamicsBCSphi} we obtain the post-quench density-density
correlation function,
\begin{widetext}
\begin{align}
\label{DynamicsBCSdensitydensityrealeqn}
\begin{split}
S^{\mbox{\tiny\itshape BCS}}_{nn}(x,t)&=2a^2\langle\partial_x {\hat{\phi}}_\rho(x,t)\partial_{x'} {\hat{\phi}}_\rho(0,t)\rangle
+e^{-2ik_F x}\langle e^{i\sqrt{2}[{\hat{\phi}}_\rho(x,t)-{\hat{\phi}}_\rho(0,t)}\rangle\langle \cos(\sqrt{2}{\hat{\phi}}_\sigma(x,t))\cos(\sqrt{2}{\hat{\phi}}_\sigma(0,t)) \rangle+h.c.,\\
 &\sim-\frac{{1}}{2}\left((K_{\rho0}+K^{-1}_{\rho0})\left|\frac{ a }{x}\right|^2+\frac{K_{\rho0}-K^{-1}_{\rho0}}{2}\left[\frac{ a ^2}{(x+2v_F t)^2}+\frac{ a ^2}{(x-2v_F t)^2}\right]\right)\\
 &\quad+\cos(2k_F x) \left(\frac{ a }{x}\right)^{(K_{\rho0}+K^{-1}_{\rho0})/2}\left|\frac{(2v_Ft)^2+a^2}{x^2-(2v_F t)^2}\right|^{(K_{\rho0}-K^{-1}_{\rho0})/4} e^{-(x+2v_Ft-|x-2v_Ft|)/(4\xi)}\\
 &\sim \begin{cases} -K_{\rho0}\left|\frac{ a }{x}\right|^2 +\left(\frac{ a }{x}\right)^{K_{\rho0}}  \left(\frac{2v_{F} t}{a}\right)^{(K_{\rho0}-K^{-1}_{\rho0})/2} e^{-\frac{v_F t}{\xi}}\cos(2k_F x)
  , &  x\gg 2v_F t,\\
- \frac{1}{4}(K_{\rho0}-K^{-1}_{\rho0})\frac{ a ^2}{(x-2v_F t)^2}
 , &  x\approx 2v_\sigma t,\\
- \frac{1}{2}(K_{\rho0}+K^{-1}_{\rho0})\left|\frac{ a }{x}\right|^2+\left(\frac{ a }{x}\right)^{(K_{\rho0}+K^{-1}_{\rho0})/2}  e^{-\frac{x}{2\xi}}\cos(2k_F x)
 , &  x\ll 2v_\sigma t. \end{cases}
\end{split}
\end{align}
\end{widetext}
illustrated in Figs.~\ref{DynamicsBCSdensitydensityreal} and 
\ref{IntensityBCSnn}.  It displays a divergent power-law peak
at the light-cone boundary $x = 2v_F t$, instead of a moving
light-cone node for the spin-singlet \eqref{quenchBCSss} and triplet \eqref{quenchBCSst}
correlators. In the bulk, for early time ($t < t_*(x)$) the correlator
displays spatial correlations of the initial ground state, with a
time-dependent decaying pre-factor. In the long time limit ($t >
t_*(x)$) it crosses over to a time-independent stationary form,
expected for a pre-thermalized state.

The corresponding structure function $S^{\mbox{\tiny\itshape
    BCS}}_{nn}(q)$ is illustrated in
Fig.~\ref{DynamicsBCSdensitydensitymomentum}. It
displays a buildup of oscillations as time evolves. Noting from
Eq.~\eqref{DynamicsBCSdensitydensityrealeqn} that the charge sector
contribution dominates over the spin sector, the structure function
can be well approximated by
\begin{align}
\label{QuenchBCSSnnq}
\begin{split}
S^{\mbox{\tiny\itshape BCS}}_{nn}(q)&\sim\sqrt{\frac{\pi}{2}}|q|\left[(K_{\rho0}+K^{-1}_{\rho0})+(K_{\rho0}-K^{-1}_{\rho0})\cos(2qv_Ft)\right]
\end{split}
\end{align}
qualitatively consistent with the behavior of the full expression
illustrated
Fig.~\ref{DynamicsBCSdensitydensitymomentum}.

\begin{figure}[!htb]
 \centering
  \includegraphics[width=90mm]{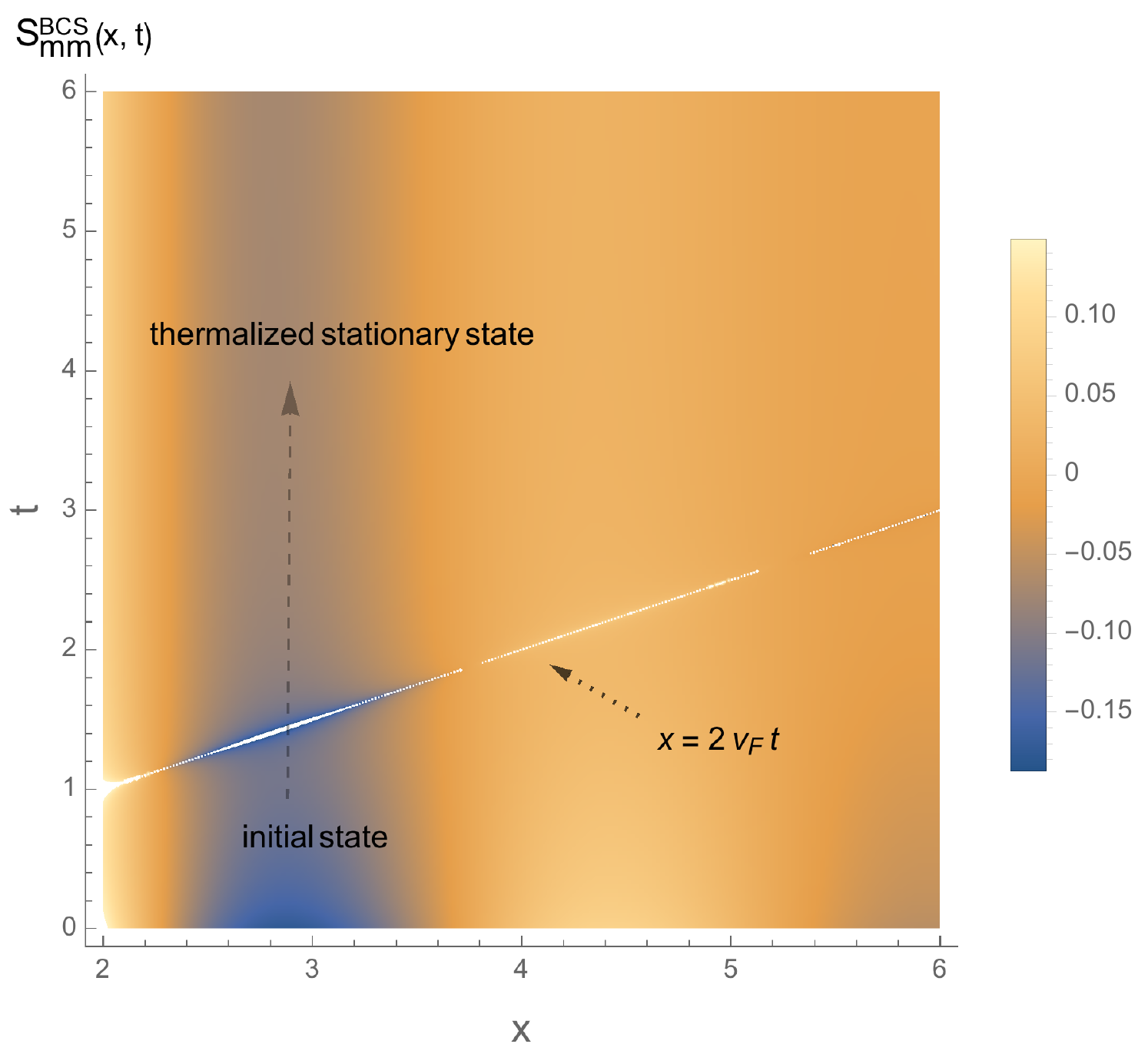}
  \caption{Space-time intensity plot of the
    magnetization-magnetization correlation function
    $S^{\mbox{\tiny\itshape BCS}}_{mm}(x,t)$ following a $U\to 0$
    quench at $t=0$ from the BCS state for $K_\rho=1.6$ at $t=0^-$. At
    the light-cone time $t_*(x)=x/(2v_F)$, the correlation shows a
    crossover from the spatial dependence of the initial ground state
    to the one of the thermalized stationary state.}
 \label{IntensityBCSmm}
\end{figure}

\begin{figure}[!htb]
 \centering
  \includegraphics[width=80mm]{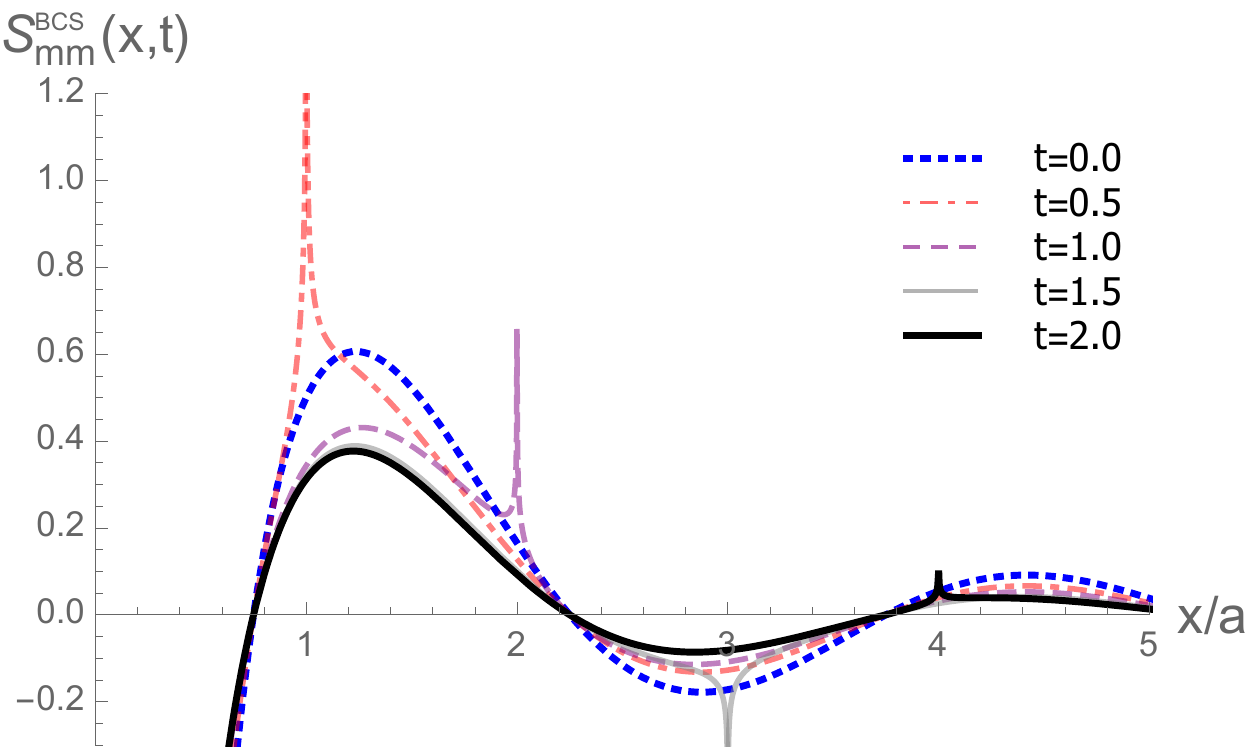}
  \caption{Magnetization-magnetization correlation function
    $S^{\mbox{\tiny\itshape BCS}}_{mm}(x,t)$ following a $U\to 0$
    quench at $t=0$ from the BCS state for a series of times. The peak
    moves as a light-cone wave-front, $x=2v_Ft$.}
\label{DynamicsBCSMMreal}
\end{figure}

The dynamical magnetization-magnetization correlator is given by
\begin{align}
\label{DynamicsBCSMMrealeqn}
\begin{split}
& S^{\mbox{\tiny\itshape BCS}}_{mm}(x)=-\cos(2k_F x) \left(\frac{ a }{x}\right)^{(K_{\rho0}+K^{-1}_{\rho0})/2}\\
&\quad\quad\times\left|\frac{(2v_Ft)^2+a^2}{x^2-(2v_F t)^2}\right|^{(K_{\rho0}-K^{-1}_{\rho0})/4} e^{-(x+2v_Ft-|x-2v_Ft|)/(4\xi)}\\
 &\sim \begin{cases}  -\left(\frac{ a }{x}\right)^{K_{\rho0}}  \left(\frac{2v_{F} t}{a}\right)^{(K_{\rho0}-K^{-1}_{\rho0})/2} e^{-\frac{v_F t}{\xi}}\cos(2k_F x), &x\gg 2v_F t,\\
-\left(\frac{ a }{x}\right)^{(K_{\rho0}+K^{-1}_{\rho0})/2}  e^{-\frac{x}{2\xi}}\cos(2k_F x), &x\ll 2v_\sigma t. \end{cases}
\end{split}
\end{align}
It is illustrated in Figs.~\ref{IntensityBCSmm} and \ref{DynamicsBCSMMreal} and  also shows light-cone dynamics and thermalization. The evolution is more evidently demonstrated in momentum space (see Fig. 23), where following the quench the $2k_F$ quasi-Bragg peak of the BCS ground state is suppressed and rounded into a Lorentzian following the quench.

\begin{figure}[!htb]
 \centering
  \includegraphics[width=80mm]{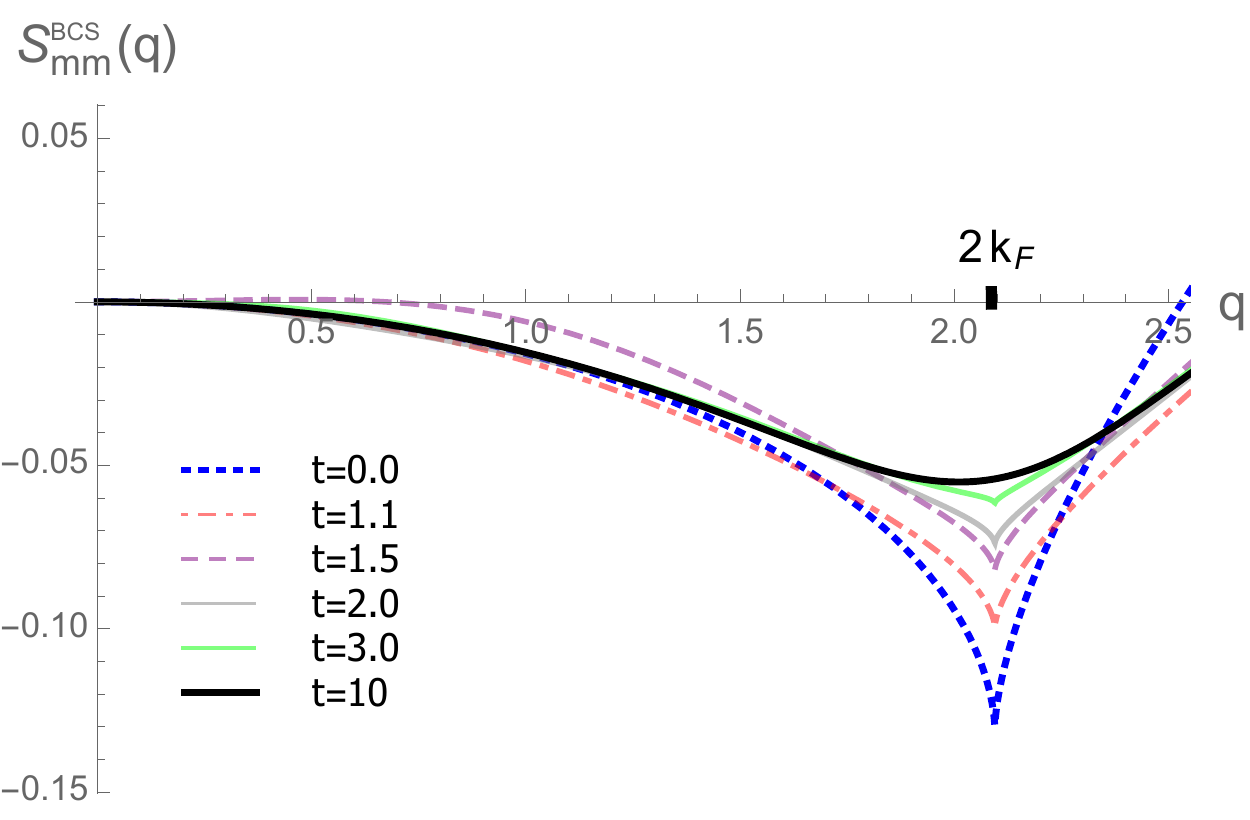}
  \caption{Magnetization structure function $S^{\mbox{\tiny\itshape
        BCS}}_{mm}(q,t)$ for a series of times following a $U\to 0$
    quench at $t=0$ from the BCS state for $K_\rho=1.6$ at
    $t=0^-$. With time, the $2k_F$ peak evolves from the quasi-Bragg
    peak of the ground-state to a Lorentzian at long times.}
\label{DynamicsBCSMMmomentum}
\end{figure}

\subsection{$U\to 0$ from FFLO state}
\label{quenchHubbardFFLO}
Using results of previous sections, Eqs.~\eqref{quenchluttingerphitheta2} and \eqref{DynamicsBCSphi} inside \eqref{Sssdefinition}, we obtain the spin-singlet pairing correlator following a $U\rightarrow0$ quench from an initial FFLO ground state,
\begin{widetext}
\begin{align}
\label{DynamicsFFLOnqpairrealspaceeqn}
\begin{split}
 S^{\mbox{\tiny\itshape FFLO}}_{ss}(x,t)
&\sim\cos(k_{FFLO}x)\left(\frac{a}{x}\right)^{(\frac{\kappa}{2}+\frac{2}{\kappa}+K_{\rho0}+K^{-1}_{\rho0})/2}\left|\frac{x^2-(2v_F t)^2}{(2v_Ft)^2+a^2}\right|^{\frac{1}{4}(K_{\rho0}-K^{-1}_{\rho0}+\frac{2}{\kappa}-\frac{\kappa}{2})},\\
&\sim\cos(k_{FFLO}x)\begin{cases} \left(\frac{a}{x}\right)^{K^{-1}_{\rho0}+\frac{\kappa}{2}}\left(\frac{a}{2v_{F} t}\right)^{\frac{1}{2}(K_{\rho0}-K^{-1}_{\rho0}+\frac{2}{\kappa}-\frac{\kappa}{2})}
 , &  x\gg 2v_F t, \\ 
\left(\frac{a}{x}\right)^{\frac{1}{2}(\frac{\kappa}{2}+\frac{2}{\kappa}+K_{\rho0}+K^{-1}_{\rho0})}, &  x\ll 2v_F t.\end{cases}
\end{split}
\end{align}
\end{widetext}
illustrated in Fig.~\ref{IntensityFFLOss} and~\ref{DynamicsFFLOnqpairrealspace}. It displays a light-cone crossover from the early time ($ t < t_*(x)$) spatial power-law correlations of the FFLO
ground state to a long-time stationary shorter-range power-law
correlations. 

\begin{figure}[!htb]
 \centering
  \includegraphics[width=80mm]{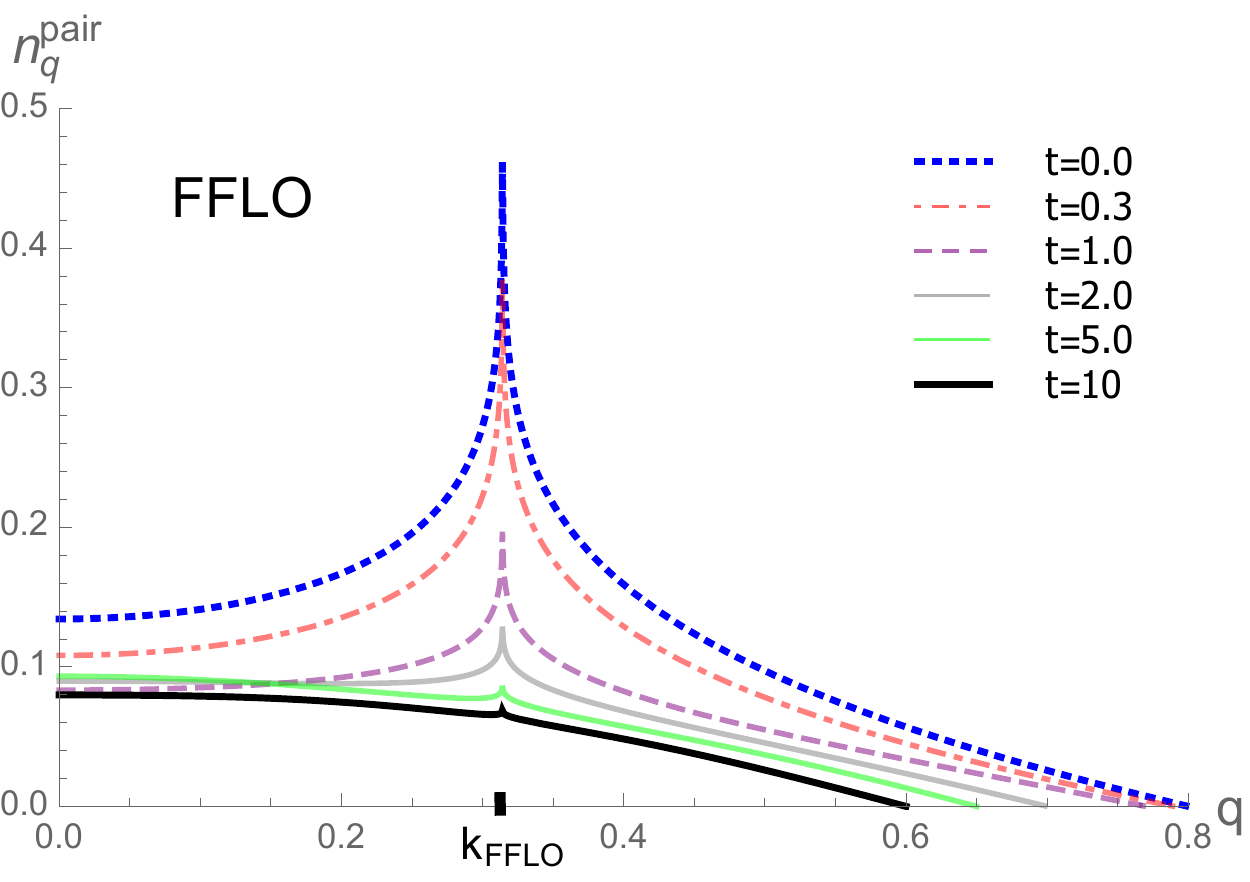}
    \caption{Cooper-pair momentum distribution $n^{pair}_{q}(t)$ for a
      series of time following the quantum quench $U\to 0$ at $t=0$
      from the FFLO ground state for $K_\rho=1.6$ at $t=0^-$. The FFLO
      peak shrinks with increasing time following the quench.}
\label{DynamicsFFLOnqpair}
\end{figure}

\begin{figure}[!htb]
 \centering
  \includegraphics[width=90mm]{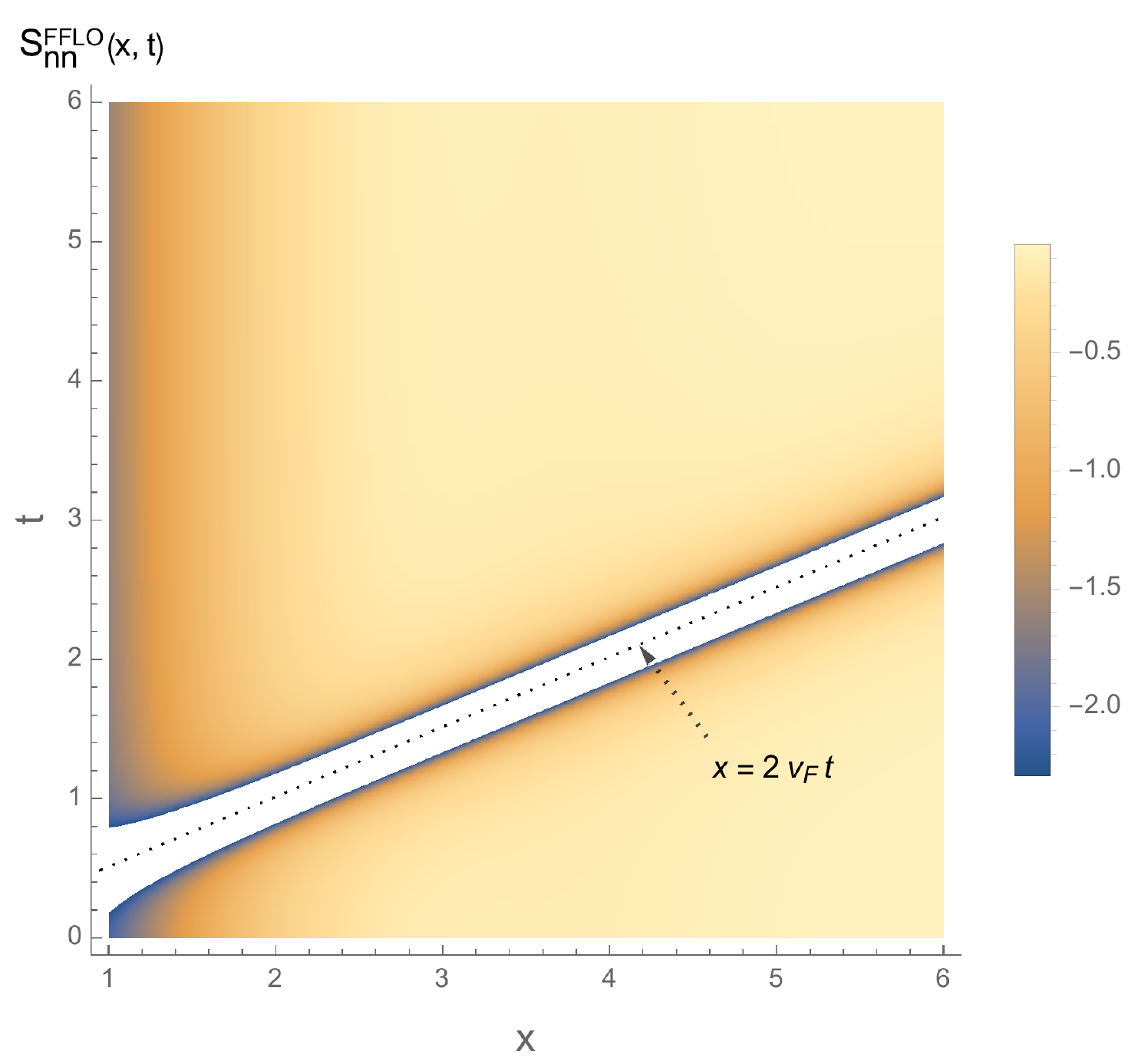}
    \caption{Space-time intensity plot of density-density correlation function
 $S^{\mbox{\tiny\itshape FFLO}}_{nn}(x,t)$ following a $U\to 0$
   quench at $t=0$ from the FFLO state for $K_\rho=1.6$ at $t=0^-$. The light-cone boundary ($x=2v_F t$, the dotted line is a guidance to the eye) separates early-time ground-state correlation from those in the asymptotic large time stationary state.}
 \label{IntensityFFLOnn}
\end{figure}

\begin{figure}[!htb]
 \centering
  \includegraphics[width=80mm]{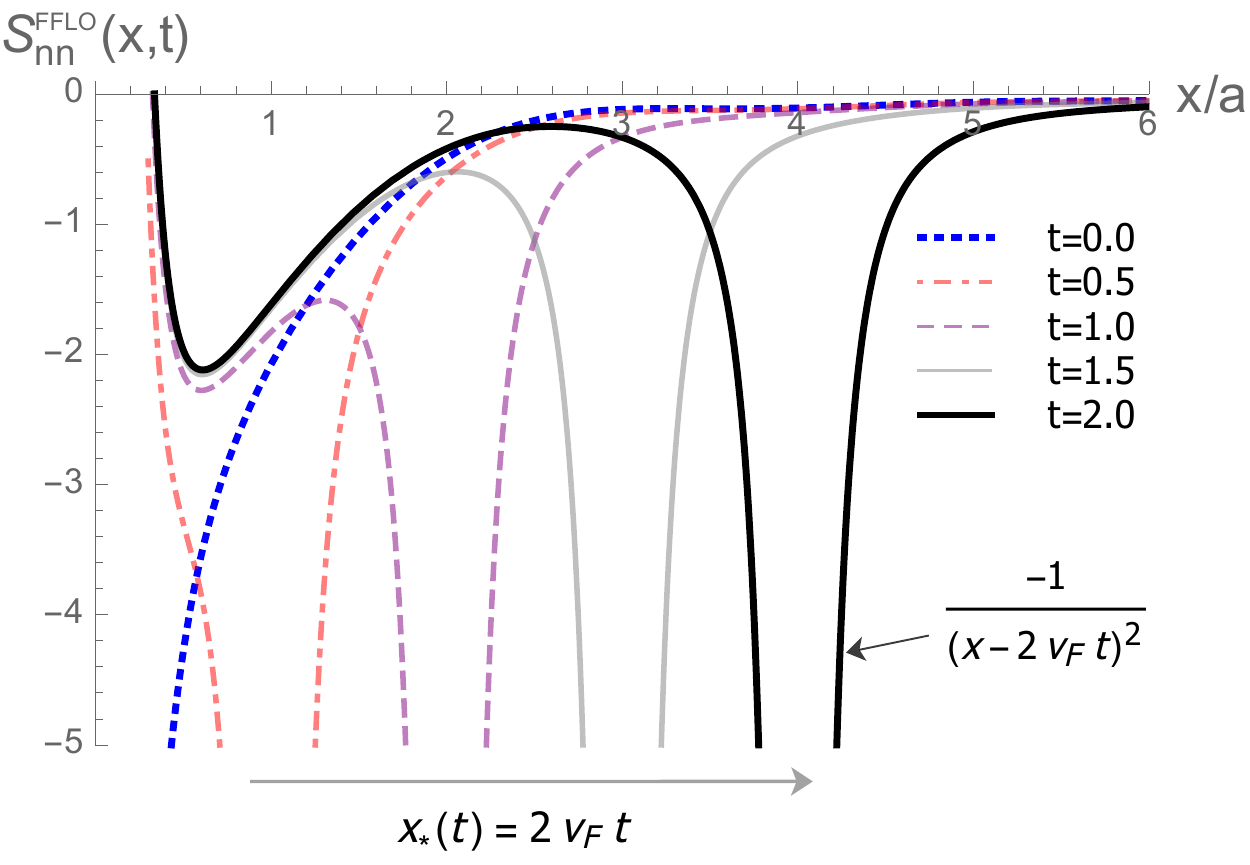}
 \caption{ Density-density correlation function
   $S^{\mbox{\tiny\itshape FFLO}}_{nn}(x,t)$ following a $U\to 0$
   quench at $t=0$ from the FFLO state for a series of times. The
   dominant features is the moving power-law peak $-1/(x-2v_Ft)^{2}$
   at the light-cone boundary, similar to the BCS quench.}
\label{DynamicsFFLOdensitydensityreal}
\end{figure}

We compute the associated Cooper-pair momentum distribution
$n^{pair}_{q}(t)$, and illustrate it in
Fig.~\ref{DynamicsFFLOnqpair}. Following the quench, the finite
momentum peak at $k_{FFLO}$ gradually diminishes, indicating the
weakening of the FFLO pairing correlations. We note this is also in
qualitative agreement with the DMRG result of quenching $U\to-U$ for
small $U$ \cite{RieggerMeisnerPRA15}, as for repulsive $U$ the cosine
nonlinearity becomes irrelevant away from commensurate fillings and
the quench is expected to share similar features with the $U\to0$
result illustrated above.

Similarly, evaluating the spin-triplet correlator we find,
 \begin{align}
\begin{split}
 &\quad S^{\mbox{\tiny\itshape FFLO}}_{st}(x,t)\\
  &\sim\left(\frac{ a }{x}\right)^{\frac{1}{2}(\frac{2}{\kappa}+\frac{\kappa}{2}+K_\rho+K^{-1}_\rho)}\left|\frac{(2v_{F} t)^2+a^2}{x^2-(2v_{F} t)^2}\right|^{\frac{1}{4}(\frac{2}{\kappa}-K_\rho+K^{-1}_\rho-\frac{\kappa}{2})},\\
   &\sim \begin{cases}\left(\frac{ a }{x}\right)^{(\frac{2}{\kappa}+K^{-1}_\rho)}\left(\frac{2v_{F} t}{a}\right)^{\frac{1}{2}(\frac{2}{\kappa}-K_\rho+K^{-1}_\rho-\frac{\kappa}{2})}, &  x\gg 2v_F t,\\ 
\left(\frac{ a }{x}\right)^{\frac{1}{2}(\frac{2}{\kappa}+\frac{\kappa}{2}+K_\rho+K^{-1}_\rho)}, &  x\ll 2v_F t.\end{cases}
\end{split}
\end{align}
which crosses-over from an initial spatial power-law decay of the FFLO
ground state to an asymptotic stationary power-law decay with a
distinct power-law exponent. The latter can be shown to be smaller
than the exponent of the FFLO initial state, indicating triplet pairing is
effectively enhanced in the long-time stationary state, in agreement
with quench from the BCS state.

\begin{figure}[!htb]
 \centering
  \includegraphics[width=80mm]{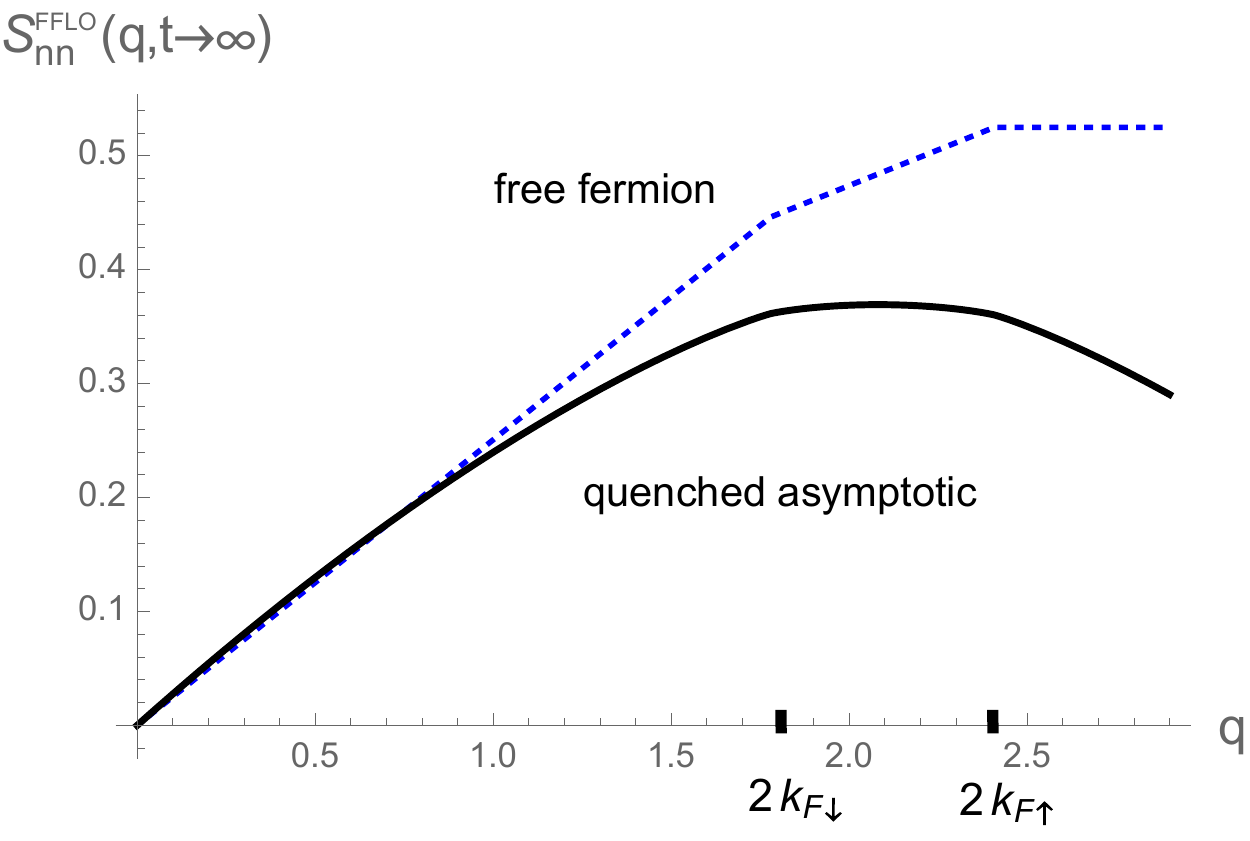}
  \caption{Asymptotic long-time limit of the number density structure
    function $S^{\mbox{\tiny\itshape FFLO}}_{nn}(q,t\to\infty)$
    following the quantum quench $U\to 0$ at $t=0$ from the FFLO
    ground state for $K_\rho=1.6$ at $t=0^-$, as compared to the free
    fermion result.}
 \label{DynamicsFFLOdensitydensitymomentum}
\end{figure}

Combining components found in the previous sections we also computed
the density-density correlation function after a quench from the FFLO
state, illustrated in Figs.~\ref{IntensityFFLOnn} and~\ref{DynamicsFFLOdensitydensityreal}. Because of the identical
  gapless behavior of the charge sector, we find that the correlator
  shares some features with the quench from the BCS state. However,
  the $2k_F$ (and harmonic) Friedel oscillations of the BCS state, for
  the FFLO quench are replaced by $2k_{F\uparrow}$ and
  $2k_{F\downarrow}$ counterparts, corresponding to the imbalanced
  densities of the two fermionic species. Also, for the FFLO quench
  power-law space-time correlations replace the exponential ones.
 The asymptotic long-time limit of the structure function $S^{\mbox{\tiny\itshape FFLO}}_{nn}(q,t\to\infty)$, is also illustrated in Fig.~\ref{DynamicsFFLOdensitydensitymomentum}, clearly differing from the free-fermion result.
 
\begin{widetext}
\begin{align}
\label{DynamicsFFLOdensitydensityrealeqn}
\begin{split}
  S^{\mbox{\tiny\itshape FFLO}}_{nn}(x,t)
      &=2a^2\langle\partial_x {\hat{\phi}}_\rho(x,t)\partial_{x'} {\hat{\phi}}_\rho(0,t)\rangle+e^{-2ik_F x}\langle e^{i\sqrt{2}[{\hat{\phi}}_\rho(x,t)-{\hat{\phi}}_\rho(0,t)}\rangle\langle \cos(\sqrt{2}({\hat{\phi}}_\sigma(x,t)-{\hat{\phi}}_\sigma(0,t))\rangle+h.c.,\\
      &\sim-\frac{{1}}{2}\left((K_{\rho0}+K^{-1}_{\rho0})\left(\frac{ a }{x}\right)^2+\frac{K_{\rho0}-K^{-1}_{\rho0}}{2}\left[\frac{ a ^2}{(x+2v_F t)^2}+\frac{ a ^2}{(x-2v_F t)^2}\right]\right)\\
      &\quad+\cos(k_{FFLO}x)\cos(2k_Fx)\left(\frac{a}{x}\right)^{(K_{\rho0}+K^{-1}_{\rho0}+\frac{\kappa}{2}+\frac{2}{\kappa})/2}\left|\frac{x^2-(2v_F t)^2}{(2v_Ft)^2+a^2}\right|^{(\frac{2}{\kappa}-K_{\rho0}+K^{-1}_{\rho0}-\frac{\kappa}{2})/4},
\\&\sim \begin{cases}-K_{\rho0}\left(\frac{ a }{x}\right)^2+\left(\frac{a}{x}\right)^{K_{\rho0}+\kappa/2}\left(\frac{a}{2v_{F} t}\right)^{(\frac{2}{\kappa}-K_{\rho0}+K^{-1}_{\rho0}-\frac{\kappa}{2})/2}\frac{\cos(2k_{F\uparrow}x)+\cos(2k_{F\downarrow}x)}{2}, &  x\gg 2v_F t ,\\ 
 - \frac{1}{4}(K_{\rho0}-K^{-1}_{\rho0})\frac{ a ^2}{(x-2v_F t)^2}
 , &  x\approx 2v_\sigma t,\\
-\frac{1}{2} (K_{\rho0}+K^{-1}_{\rho0})\left(\frac{ a }{x}\right)^2+\left(\frac{a}{x}\right)^{(K_{\rho0}+K^{-1}_{\rho0}+\frac{\kappa}{2}+\frac{2}{\kappa})/2}\frac{\cos(2k_{F\uparrow}x)+\cos(2k_{F\downarrow}x)}{2} , &  x\ll 2v_F t,\end{cases}
\end{split}
\end{align}
\end{widetext}

\begin{figure}[!htb]
 \centering
  \includegraphics[width=90mm]{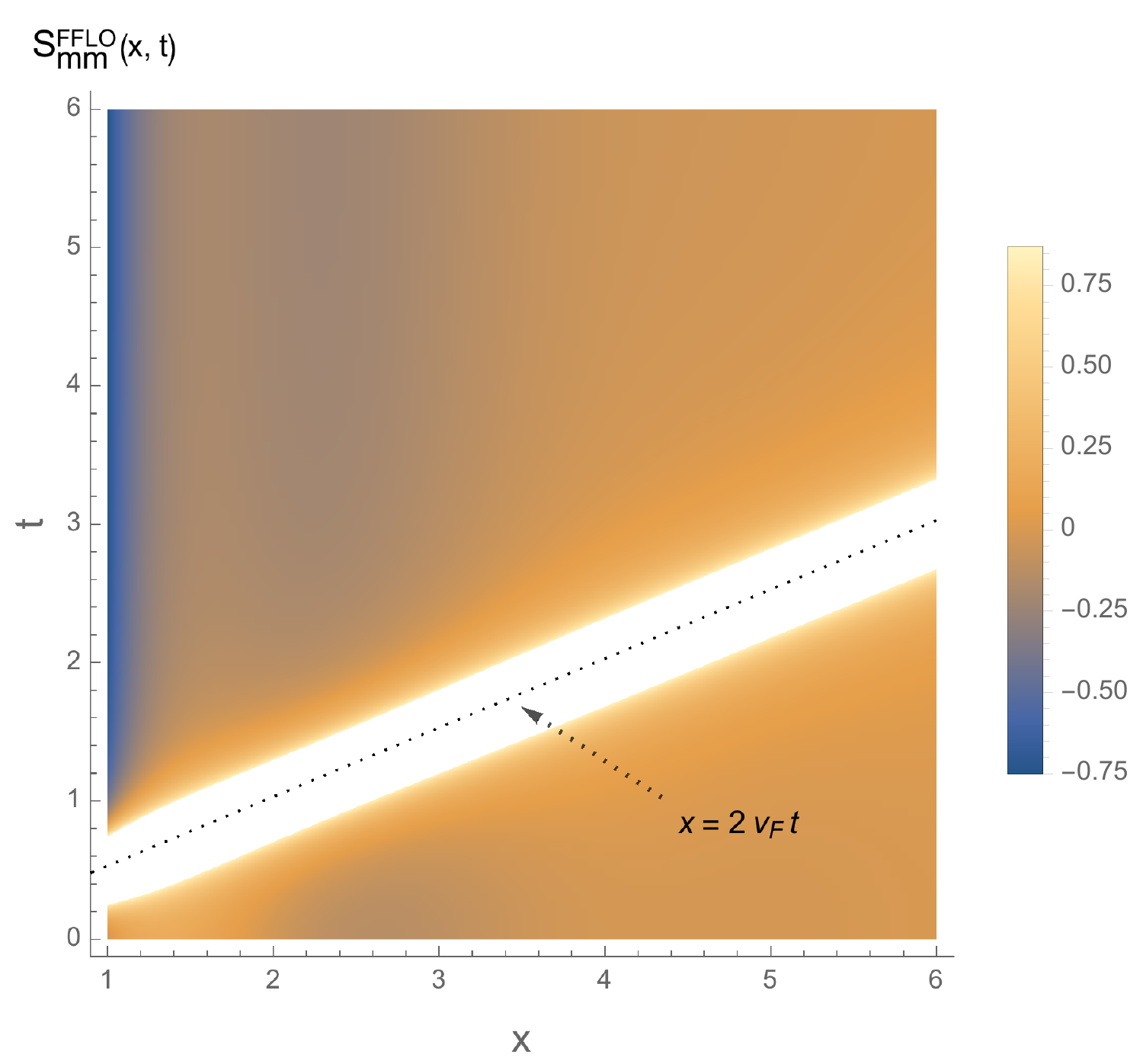}
  \caption{Space-time intensity plot of magnetization-magnetization
    correlation function $S^{\mbox{\tiny\itshape FFLO}}_{mm}(x,t)$
    following a $U\to 0$ quench at $t=0$ from the FFLO state for
    $K_\rho=1.6$ at $t=0^-$. The light-cone boundary ($x=2v_F t$, the
    dotted line is a guide to the eye) separates early-time
    ground-state correlation from those in the asymptotic large time
    stationary state.}
 \label{IntensityFFLOmm}
\end{figure}

\begin{figure}[!htb]
 \centering
  \includegraphics[width=80mm]{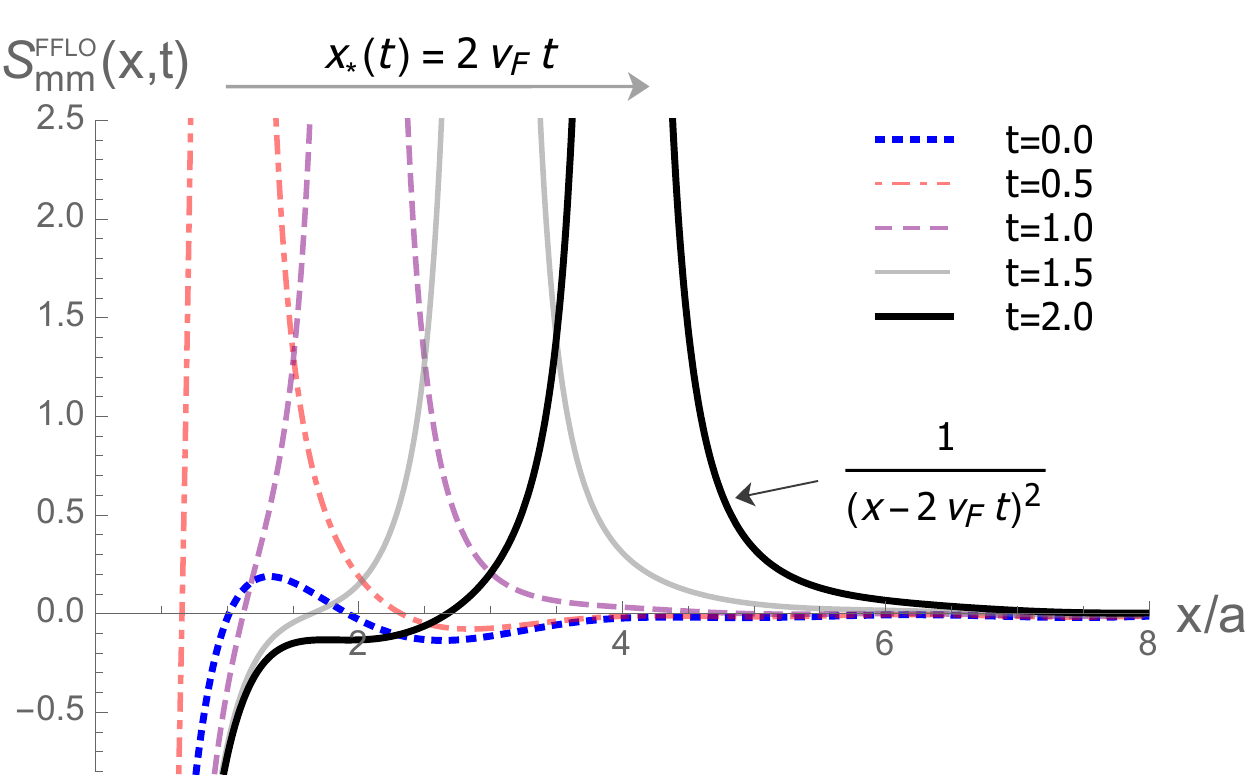}
  \caption{Magnetization-magnetization correlation function
    $S^{\mbox{\tiny\itshape FFLO}}_{mm}(x,t)$ following a $U\to 0$
    quench at $t=0$ from the FFLO state for a series of times. The
    divergence peak moves as a wave-front at the light-cone 
    $x_*(t)=2v_Ft$.}
\label{DynamicsFFLOmagnetizationmagnetizationreal}
\end{figure}

\begin{figure}[!htb]
 \centering
  \includegraphics[width=80mm]{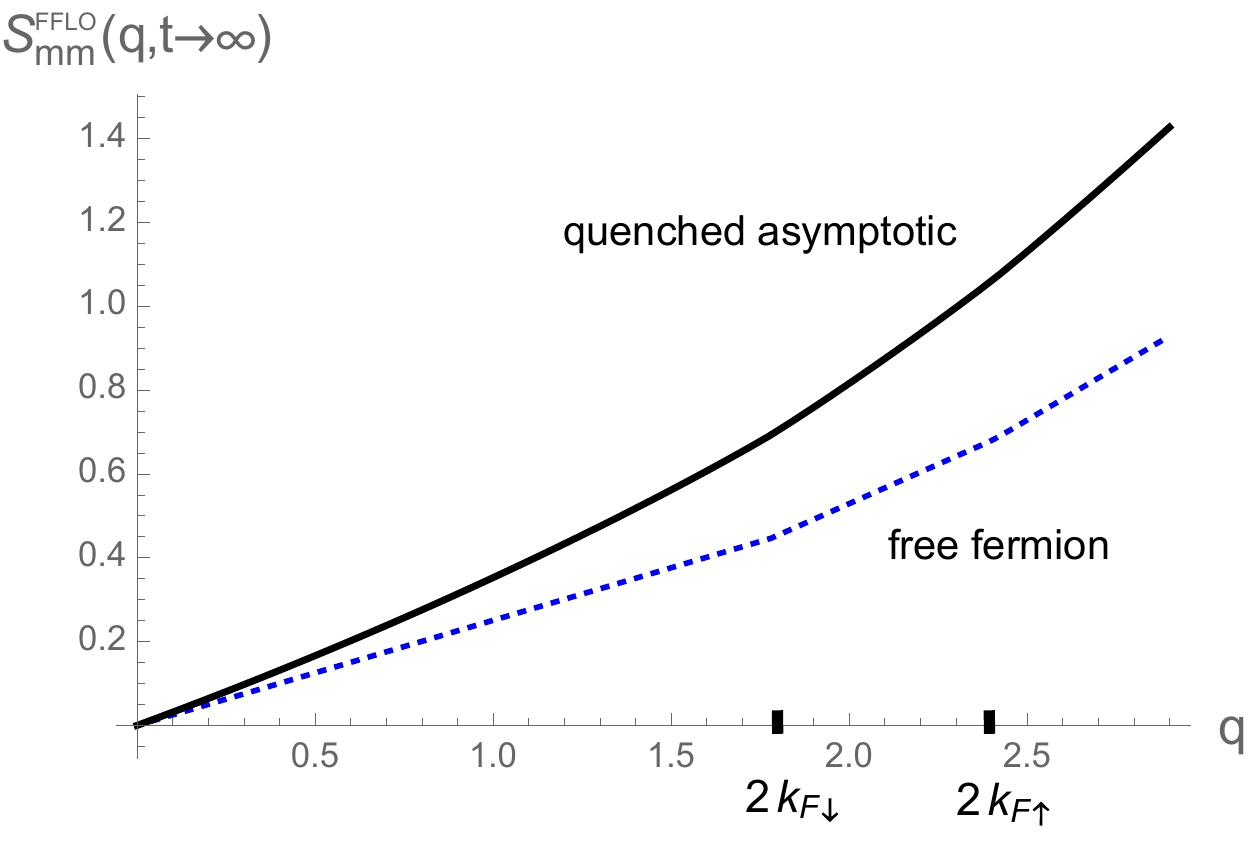}
  \caption{Asymptotic long-time limit of the magnetization structure
    function $S^{\mbox{\tiny\itshape FFLO}}_{mm}(q,t\to\infty)$
    following the quantum quench $U\to 0$ at $t=0$ from the FFLO
    ground state for $K_\rho=1.6$ at $t=0^-$, as compared to the free
    fermion result.}
 \label{DynamicsFFLOmagnetizationmagnetizationmomentum}
\end{figure}

The magnetization-magnetization correlation function is another
important observable for identification of the FFLO ground
state. Following a quench, it is given by
\begin{widetext}
\begin{align}
\label{mMcorrelationFFLOdynamics}
\begin{split}
S^{\mbox{\tiny\itshape FFLO}}_{mm}(x)&\sim-\frac{1}{4}\left[\frac{2(\kappa/2+2/\kappa)a^2}{x^{2}}+\frac{(\kappa/2-2/\kappa)a^2}{(x+2v_Ft)^{2}}+\frac{(\kappa/2-2/\kappa)a^2}{(x-2v_Ft)^{2}}\right] \\
&\quad-\frac{\cos(2k_{F\uparrow}x)+\cos(2k_{F\downarrow}x)}{2}\left(\frac{a}{x}\right)^{(K_{\rho0}+K^{-1}_{\rho0}+\frac{\kappa}{2}+\frac{2}{\kappa})/2}\left|\frac{x^2-(2v_F t)^2}{(2v_Ft)^2+a^2}\right|^{(\frac{2}{\kappa}-K_{\rho0}+K^{-1}_{\rho0}-\frac{\kappa}{2})/4},\\
&\sim
 \begin{cases}-\frac{\kappa}{2} \left(\frac{ a
     }{x}\right)^2-\left(\frac{a}{x}\right)^{K_{\rho0}+\kappa/2}\left(\frac{a}{2v_{F}
       t}\right)^{(\frac{2}{\kappa}-K_{\rho0}+K^{-1}_{\rho0}-\frac{\kappa}{2})/2}
\frac{1}{2}\left[\cos(2k_{F\uparrow}x)+\cos(2k_{F\downarrow}x)\right], &  x\gg 2v_F t ,\\ 
 \frac{1}{4}(\frac{2}{\kappa}-\frac{\kappa}{2})\frac{ a ^2}{(x-2v_F t)^2}
 , &  x\approx 2v_\sigma t,\\
-\frac{1}{2}(\frac{\kappa}{2}+\frac{2}{\kappa})\left(\frac{ a
  }{x}\right)^2-
\left(\frac{a}{x}\right)^{(K_{\rho0}+K^{-1}_{\rho0}+\frac{\kappa}{2}+\frac{2}{\kappa})/2}
\frac{1}{2}\left[\cos(2k_{F\uparrow}x)
+\cos(2k_{F\downarrow}x)\right], &  x\ll 2v_F t,\end{cases}
\end{split}
\end{align}
\end{widetext}
and is illustrated in Figs.~\ref{IntensityFFLOmm} and
\ref{DynamicsFFLOmagnetizationmagnetizationreal}. It displays a form
quite similar to the density-density correlator
$S^{\mbox{\tiny\itshape FFLO}}_{nn}(x,t)$.

Its long time limit in momentum space, $S^{\mbox{\tiny\itshape
    FFLO}}_{mm}(q,t\to\infty)$, is also illustrated in
Fig.~\ref{DynamicsFFLOmagnetizationmagnetizationmomentum}, showing
that the $k_{F\uparrow,\downarrow}$ peaks are smoothed over in the
long time limit.

\section{Summary and Conclusion}
\label{SummaryConclusion}

We used bosonization and the exact Luther-Emery mapping methods to
study the dynamics of the 1D spin-imbalanced attractive Fermi-Hubbard
model following a quench of its on-site interaction, $U\to0$,
particularly focusing on the long-time asymptotic behavior.  We
characterized the dynamics by evaluating a number of physically
accessible post-quench correlation functions, such as the spin-singlet
and triplet as well as the number and magnetization density
correlators. On the scale of light-cone time $x/(2v_F)$, these show a
dephasing-driven evolution to a stationary state that exhibits strong
qualitative dependence on the state of the initial pre-quench ground
state.

The quench from the spin-gapped BCS ground state leads to a stationary
state with a short-range correlated spin component (see Sec.~\ref{quenchHubbardBCS}), characterized by a correlation length $2\xi= 2\pi v_\sigma/U$ for $x<x_*(t)=2v_Ft$ and stationary power-law correlated
charge component. The spin component matches the exponentially decaying
spatial correlations of a noninteracting Fermi gas at finite
temperature with correlation length $\xi_T = v_F/(2\pi T)$. This
suggests that for quenches from the BCS state the spin-component
thermalizes to the effective temperature,
\begin{equation}
\label{thermaltemperature}
T_{eff} = v_F/(4\pi \xi)\sim U,
\end{equation}
while the gapless charge sector within the harmonic Luttinger liquid
analysis remains pre-thermalized \cite{Yin1,Yin2}, i.e., does not
thermalize. Thus we note that the extent to which the system appears
to thermalize can depend on the measured observable, the level to
which the gapped and gapless contributions contribute to its long-time
behavior. For example, since the number density-density correlator
[Eq.~\eqref{DynamicsBCSdensitydensityrealeqn}] is dominated by the
gapless charge sector, it shows no sign of thermalization (asymptotes
to a power-law correlated pre-thermalized form) for a quench from the
BCS ground state. On the other hand, the magnetization-magnetization
correlator \eqref{DynamicsBCSMMrealeqn} is dominated by the gapped
spin sector and appears to thermalize.

Furthermore, a quench from a ground state in the spin- (and charge-)
gapless FFLO phase leads to a power-law stationary state for both spin
and charge, thus neither thermalizes.

The thermalization and lack of it for the quench from gapped and gapless states, respectively, are consistent with the arguments in Refs. \cite{Cardy,CardyJPhys,Cardy2}, that a ``deep'' quench (defined
by large ratio of initial to final gap, $\Delta_0/\Delta_1\gg 1$),
e.g., a gapped to gapless Hamiltonian is necessary for thermalization,
a conclusion based on results obtained using the generic method of
slab construction, conformal field theory, and physical arguments. As
emphasized by Cardy \textit{et al}. \cite{Cardy2}, strictly speaking the state
itself is neither stationary nor thermal.  It is the {\em local}
observables, expressible in terms of two-point correlation functions
that thermalize in the thermodynamic limit by dephasing of an infinite
number of momentum modes. For a local observable the relevant system
is effectively open with the modes outside of the local region set by
$x_*(t)$ acting like an effective bath.


\acknowledgements 

We thank Victor Gurarie, Leon Balents, and Le Yan for stimulating
discussions. This research was supported in part by the National
Science Foundation under Grant No. DMR-1001240, through the KITP under
Grant No. NSF PHY-1125915 and by the Simons Investigator award from
the Simons Foundation. We also acknowledge partial support by the Center for Theory of
Quantum Matter at the University of Colorado. We thank the KITP for its hospitality during
our stay as part of the graduate fellowship (X.Y.) and sabbatical (L.R.)
programs, when part of this work was completed. 

\appendix

\begin{widetext}




\section{Green's functions for Luther-Emery approach}
\label{LEgreensfunctions}

In this section, we fill the technical gap leading to the ${\hat{\theta}}_\sigma$ correlation for Luther-Emery approach. Using Eq.~\eqref{spinlesscphitheta} and Wick's theorem, we obtain
\begin{align}
\begin{split}
\label{LEthetasigma}
\langle e^{i\sqrt{2}[{\hat{\theta}}_\sigma(x)-{\hat{\theta}}_\sigma(0)]}\rangle
&=\langle e^{i2[{\hat{\theta}}(x)-{\hat{\theta}}(0)]}\rangle,\\
&=(2\pi a)^2\langle \hat c^\dagger_R(x)\hat c^\dagger_L(x)\hat c_L(0)\hat c_R(0)\rangle,\\
&\sim(2\pi a)^2[-G_{12}(x)G_{21}(x)+G_{11}(x) G_{22}(x)].
\end{split}
\end{align}
Here the minus sign comes from the fermionic anti-commutation relation and 
the Green's functions of ($\hat c_R, \hat c_L$) are
 \begin{subequations}
 \label{greenfunctiongapped}
\begin{align}
G_{11}(x)&=\sum_{p} \frac{e^{ipx}}{L}\langle \hat c^\dagger_R(p)\hat c_R(p)\rangle,\\
G_{22}(x)&=\sum_{p} \frac{e^{ipx}}{L}\langle \hat c^\dagger_L(p)\hat c_L(p)\rangle,\\
G_{12}(x)&= \sum_{p} \frac{e^{ipx}}{L}\langle \hat c^\dagger_R(p)\hat c_L(p)\rangle,\\
G_{21}(x)&= \sum_{p} \frac{e^{ipx}}{L}\langle \hat c^\dagger_L(p)\hat c_R(p)\rangle,
\end{align}
\end{subequations}
which we will examine shortly how they behave asymptotically for the BCS state and the FFLO state, respectively. But before that, we notice the following to be true:
\begin{align}
\begin{split}
\langle e^{i\sqrt{2}[{\hat{\theta}}_\sigma(x)+{\hat{\theta}}_\sigma(0)]}\rangle
&=(2\pi a)^2\langle \hat c^\dagger_R(x)\hat c^\dagger_L(x)\hat c^\dagger_L(0)\hat c^\dagger_R(0)\rangle=0,
\end{split}
\end{align}
and therefore
\begin{align}
\begin{split}
\langle \cos(\sqrt{2}({\hat{\phi}}_\sigma(x))\cos(\sqrt{2}{\hat{\phi}}_\sigma(0))\rangle
&=\frac{1}{2}\Re\left[\langle e^{i\sqrt{2}[{\hat{\theta}}_\sigma(x)-{\hat{\theta}}_\sigma(0)]}\rangle+\langle e^{i\sqrt{2}[{\hat{\theta}}_\sigma(x)+{\hat{\theta}}_\sigma(0)]}\rangle\right],\\
&=\frac{1}{2}\Re\langle e^{i\sqrt{2}[{\hat{\theta}}_\sigma(x)+{\hat{\theta}}_\sigma(0)]}\rangle.
\end{split}
\end{align}
Here $\Re$ standards for the real-part.

The ${\hat{\phi}}_\sigma$ correlation turns out to be out of the reach of the LE approach. Unlike ${\hat{\theta}}_\sigma$, the ${\hat{\phi}}_\sigma$ correlation $\langle e^{i\sqrt{2}[{\hat{\phi}}_\sigma(x)-{\hat{\phi}}_\sigma(0)]}\rangle=\langle e^{i[{\hat{\phi}}(x)-{\hat{\phi}}(0)]}\rangle$ can not be related to the
correlation of the spinless fermions ($\hat c_R, \hat c_L$) in a simple way.
Especially, we have
\begin{align}
\label{spinlessphis}
\begin{split}
\langle \hat c^\dagger_R(x)\hat c_L(x)\hat c^\dagger_L(0)\hat c_R(0)\rangle&=(2\pi a)^2\langle e^{i2[{\hat{\phi}}(x)-{\hat{\phi}}(0)]}\rangle,
\end{split}
\end{align}
and is not connected to $\langle e^{i[{\hat{\phi}}(x)-{\hat{\phi}}(0)]}\rangle$ since relation \eqref{wardrelation} no longer applies for the non-quadratic Hamiltonian \eqref{SineGordonH}.

\subsection{Equilibrium BCS state}
For the BCS state, the distribution of ($\hat c_R, \hat c_L$) obeys Eq.~\eqref{LEequBCS}, by using which we easily obtain
 \begin{subequations}
 \label{GreensfunctionBCS}
\begin{align}
G_{12}(x)&=-\int^{\infty}_{-\infty}\frac{dp}{2\pi}\frac{1}{2}\sin2\beta_p
=\frac{1}{ \xi}\int^{\infty}_{0}\frac{dp}{2\pi}\frac{\cos(px/ \xi)}{\sqrt{1+p^2}}
=\frac{K_{0}(|\frac{x}{ \xi}|)}{2\pi \xi},\\
G_{11}(x)&=\int^{\infty}_{-\infty}\frac{dp}{2\pi}\frac{1-\cos2\beta_p}{2}e^{ipx}=-\frac{i}{ \xi}\int^{\infty}_{0}\frac{dp}{2\pi}\frac{p\sin(px/ \xi)}{\sqrt{1+p^2}}=i\frac{\partial_x K_0(x/ \xi)}{2\pi \xi}
=\frac{-iK_1(|\frac{x}{ \xi}|)}{2\pi \xi},\\
G_{22}(x)&=\int^{\infty}_{-\infty}\frac{dp}{2\pi}\frac{1+\cos2\beta_p}{2}e^{ipx}
=-G_{11}(x)=i\frac{K_1(|\frac{x}{ \xi}|)}{2\pi \xi},
\end{align}
\end{subequations}
with $\tan 2\beta_p=-U/(\pi v_\sigma p)$. Here $ \xi=\pi  v_\sigma/U$ is the correlation length and $K_{0,1}$ are the  modified Bessel functions satisfying
\begin{align}
\begin{split}
\label{K0largex}
K_{z}(x)\to\sqrt{\frac{\pi}{2x}}e^{-x}(1+\frac{4z^2-1}{8x}+O(\frac{1}{x^2})), \;\;\;  x> 1.
\end{split}
\end{align}

Plugging Eq.~\eqref{GreensfunctionBCS} inside \eqref{LEthetasigma}, we then find for the gapped BCS state,
\begin{align}
\begin{split}
\langle e^{i\sqrt{2}[{\hat{\theta}}_\sigma(x)-{\hat{\theta}}_\sigma(0)]}\rangle
&\sim \frac{a^2}{ x^2}e^{-2x/ \xi},
\end{split}
\end{align}
which is Eq.~\eqref{LEthetacorrelationBCS} in the main text.

The corresponding average magnetization and the long-wavelength part of the magnetization-magnetization correlation function can also be evaluated as
 \begin{align}
    \begin{split}
\bar{m}&\equiv -\frac{1}{L}\int dx\frac{\sqrt{2}}{\pi}\langle\partial_x {\hat{\phi}}_\sigma(x)\rangle=\frac{\sqrt{2}}{L}\int dx\langle\hat c^\dagger_R\hat c_R+\hat c^\dagger_L\hat c_L\rangle=\frac{\sqrt{2}}{2\pi}\int^\infty_{-\infty} dp\langle\hat c^\dagger_u(p) c_u(p)\rangle
\sim 0,
    \end{split}
  \end{align}
and
 \begin{align}
    \begin{split}
\langle \partial_x{\hat{\phi}}_\sigma(x) \partial_{x'}{\hat{\phi}}_\sigma(0)\rangle&=\sum_{n,n'=R,L}\langle \hat c^\dagger_{n}(x)\hat c_{n}(x) \hat c^\dagger_{n'}(0)\hat c_{n'}(0)\rangle
\sim \sum_{n,n'=1,2} G_{nn'}(x)G_{nn'}(-x),\\
&\sim -\frac{1}{2\pi\xi x}e^{-2x/\xi}.
    \end{split}
  \end{align}

\subsection{Equilibrium FFLO state}

For the FFLO state, the distribution of ($\hat c_R, \hat c_L$) instead obeys Eq.~\eqref{nklustatesmagnetic} and the Green's function $G_{11}(x)$ now becomes
\begin{align}
\begin{split}
G_{11}(x)
&=\int^{\tilde k_F}_{-\tilde k_F}\frac{dp}{2\pi}e^{ipx}+\int^{\infty}_{\tilde k_F}\frac{dp}{2\pi}\frac{1-\cos2\beta_p}{2}e^{ipx}
+\int^{-\tilde k_F}_{-\infty}\frac{dp}{2\pi}\frac{1-\cos2\beta_p}{2}e^{ipx},\\
&=\frac{1}{2}\int^{\tilde k_F}_{-\tilde k_F}\frac{dp}{2\pi}e^{ipx}-\frac{i}{ \xi}\int^{\infty}_{\tilde k_F \xi}\frac{dp}{2\pi}\frac{p\sin(px/ \xi)}{\sqrt{1+p^2}},\\
&=-\frac{K_1(x/ \xi)}{2\pi \xi}+\frac{i}{ \xi}\int^{\tilde k_F \xi}_{0}\frac{dp}{2\pi}\frac{p\sin(px/ \xi)}{\sqrt{1+p^2}}+\frac{\sin(\tilde k_F x)}{2\pi x}.
\end{split}
\end{align}

Focusing on long-distance asymptotic behavior, we can throw away the $K_1(x)$ part that decays exponentially. We can further simplify the second term in the above equation by noticing $\tilde k_F \xi<1$ generically true for small spin imbalance (which is the case we consider throughout the paper), and thereby obtain
\begin{align}
\begin{split}
G_{11}(x)&\sim \frac{i}{ \xi}\int^{\tilde k_F \xi}_{0}\frac{dp}{2\pi}p\sin(px/ \xi)+\frac{\sin(\tilde k_F x)}{2\pi x},\\
&\sim\frac{\sin(\tilde k_F x)-i\tilde k_F \xi\cos(\tilde k_Fx)}{2\pi x}+\frac{i \xi\sin(\tilde k_Fx)}{2\pi x^2}.
\end{split}
\end{align}
The other Green's functions can be obtained in a similar way as
\begin{align}
\begin{split}
G_{22}(x)&\sim \frac{\sin(\tilde k_F x)+i\tilde k_F \xi\cos(\tilde k_Fx)}{2\pi x}-\frac{i \xi\sin(\tilde k_Fx)}{2\pi x^2},
\end{split}
\end{align}
and
\begin{align}
\begin{split}
G_{12}(x)=G_{21}(x)
&\sim-\frac{\sin(\tilde k_F x)(1-(\tilde k_F \xi)^2/2)}{2\pi x}+\frac{\cos(\tilde k_F x)\tilde k_F\xi^2_0}{2\pi x^2}.
\end{split}
\end{align}

Plugging them back into Eq.~\eqref{LEthetasigma}, we then obtain the ${\hat{\theta}}_\sigma$ correlation for the spin-gapless FFLO state to lowest order as
\begin{align}
\begin{split}
\label{tripletspincorrelationwithh}
\langle e^{i\sqrt{2}[{\hat{\theta}}_\sigma(x)-{\hat{\theta}}_\sigma(0)]}\rangle
&\sim\frac{a^2}{x^2}[-\sin^2(\tilde k_F x)(1-(\tilde k_F \xi)^2/2)^2
+\sin^2(\tilde k_F x)+(\tilde k_F \xi)^2\cos^2(\tilde k_Fx)],\\
&\sim(\tilde k_F \xi)^2\frac{a^2}{x^2},
\end{split}
\end{align}
which is Eq.~\eqref{LEthetacorrelationFFLO} in the main text.

Similarly, the average magnetization and its correlation function's long-wavelength part are evaluated as
 \begin{align}
    \begin{split}
\bar{m}&=\frac{\sqrt{2}}{2\pi}\int^\infty_{-\infty} dp\langle\hat c^\dagger_u(p) c_u(p)\rangle=\frac{\sqrt{2}\tilde k_F}{\pi}\sim \frac{\sqrt{2}\sqrt{h^2-U^2/(\pi a)^2}}{\pi v_\sigma},
    \end{split}
  \end{align}
 and 
 \begin{align}
    \begin{split}
\langle \partial_x{\hat{\phi}}_\sigma(x) \partial_{x'}{\hat{\phi}}_\sigma(0)\rangle\sim \sum_{n,n'=1,2} G_{nn'}(x)G_{nn'}(-x) \sim -\frac{\sin^2(\tilde k_F x)}{\pi^2 x^2}.
    \end{split}
  \end{align}

\section{Semi-classical approach for equilibrium BCS state}
\label{spingappedexpansion}
In this section, we derive the ${\hat{\phi}}_\sigma$ correlation [Eq.~\eqref{BCSthetaphicorrelator}] for equilibrium BCS state within the semiclassical approach in details. Using relation,
\begin{subequations}
\label{phithetasemiclassical}
 \begin{align}
 {\hat{\phi}}_\sigma(x)=-\sqrt{K_\sigma}\frac{i\pi}{L}\sum_{p\neq0}\left(\frac{L|p|}{2\pi}\right)^{1/2}\frac{1}{p}e^{- a |p|/2-ipx}(\hat b^\dagger_p+\hat b_{-p}),\\
{\hat{\theta}} _\sigma(x)=\frac{1}{\sqrt{K_\sigma}}\frac{i\pi}{L}\sum_{p\neq0}\left(\frac{L|p|}{2\pi}\right)^{1/2}\frac{1}{|p|}e^{-a |p|/2-ipx}(\hat b^\dagger_p-\hat b_{-p}),
\end{align}
\end{subequations}
and the statistics of ($\hat b_p, \hat b_p^\dagger$) Eq.~\eqref{bdistributionphiexpansionequilibrium}, we have
\begin{align}
\begin{split}
 \langle({\hat{\phi}}_\sigma(x)-{\hat{\phi}}_\sigma(0))^2\rangle
&=K_\sigma\int^\infty_0\frac{dp}{p}e^{- a  p}[1-\cos(px)]
 \left[\langle \hat b^\dagger_p \hat b^\dagger_{-p}\rangle+\langle \hat b^\dagger_p \hat b_p\rangle+\langle \hat b_{-p} \hat b^\dagger_{-p}\rangle+\langle \hat b_{p} \hat b_{-p} \rangle\right],\\
&=K_\sigma\int^\infty_0\frac{dp}{\sqrt{p^2+1/\lambda^2}}e^{- a  p}[1-\cos(px)],\\
&=K_\sigma[\arcsinh(\lambda / a )-K_{0}(x/\lambda )],\\
&\sim \const,
\end{split}
\end{align}
where $\const=K_\sigma\arcsinh(\lambda/a)\sim O(1)$ and for long distance behavior we have ignored $K_{0}(x)$ that decays exponentially.

The ${\hat{\theta}}_\sigma$ correlation can be evaluated in a similar way. We have
\begin{align}
\begin{split}
 \langle({\hat{\theta}}_\sigma(x)-{\hat{\theta}}_\sigma(0))^2\rangle
&=\frac{1}{K_\sigma}\int^\infty_0\frac{dp}{p}e^{- a  p}[1-\cos(px)]
\left[\langle \hat b_{-p} \hat b^\dagger_{-p}\rangle+\langle \hat b^\dagger_p \hat b_p\rangle-\langle \hat b^\dagger_p \hat b^\dagger_{-p}\rangle-\langle \hat b_p \hat b_{-p}\rangle\right], \\
&=\frac{1}{K_\sigma}\int^\infty_0 \frac{dp\sqrt{p^2+1/\lambda^2}}{p^2}e^{- a  p}[1-\cos(px)],\\
&=\frac{1}{K_\sigma}\int^\infty_0 dp\left[\frac{1}{\sqrt{p^2+1/\lambda^2}}+\frac{1}{\lambda^2p^2\sqrt{p^2+1/\lambda^2}}\right]e^{- a  p}[1-\cos(px)],\\
&=\frac{1}{K_\sigma}\left[\arcsinh(\lambda / a )-K_{0}(x/\lambda )-1 -\frac{1}{2} G^{21}_{13}\left(\frac{x^2}{4\lambda^2}\bigg|\begin{matrix}
   & 3/2& \\
  0 & 1&1/2
 \end{matrix}\right)\right],\\
&\sim -x/\xi_0+\const,
\end{split}
\end{align}
with $\xi_0=2\lambda K_\sigma/\pi$ and $\const=[\arcsinh(\lambda / a )-1]/K_\sigma\sim O(1)$. Here we have used the property of MeijerG function $G^{21}_{13}$
\begin{align}
\label{meijerGfuntionasymptotic}
\begin{split}
\frac{1}{2\lambda^2}\int^\infty_0 dp\frac{1}{p^2\sqrt{p^2+1/\lambda^2}}e^{- a  p}[1-\cos(px)]
&=-\frac{1}{2} -\frac{1}{4}
 G^{21}_{13}\left(\frac{x^2}{4\lambda^2}\bigg|\begin{matrix}
   & 3/2& \\
  0 & 1&1/2
 \end{matrix}\right), \\
&\sim\frac{\pi}{4}\frac{x}{\lambda }-\frac{1}{2},  \;\;\;  x>1.
\end{split}
\end{align}



\section{Normal modes approach for charge sector dynamics}
In this section, we fill in some technical details leading to the charge sector quench dynamics results [Eq.~\eqref{quenchluttingerphitheta} and \eqref{quenchluttingerphitheta2}], obtained with the normal mode approach in the main text.  Using Eq.~\eqref{chargesectorbogoliubovtime}, we have
\label{chargemicrodynamics}
\begin{subequations}
\begin{align}
\langle b^\dagger_{p}(t)b_p(t)\rangle&=\langle b^\dagger_{-p}(t)b_{-p}(t)\rangle\nonumber,\\
&=\cosh^2\beta_1\sinh^2(\beta_1-\beta_0)
+\sinh^2\beta_1\cosh^2(\beta_1-\beta_0)
-\frac{1}{2}\sinh2\beta_1\sinh2(\beta_1-\beta_0)\cos(2v_{\rho1}pt),\\
\langle b^\dagger_{p}(t)b^\dagger_{-p}(t)\rangle&=\langle b_{p}(t)b_{-p}(t)\rangle\nonumber,\\
&=\frac{\cosh^2\beta_1}{2}\sinh2(\beta_1-\beta_0) e^{2iv_{\rho1}|p|t}
+\frac{\sinh^2\beta_1}{2}\sinh2(\beta_1-\beta_0) e^{-2iv_{\rho1}|p|t}
-\frac{1}{2}\sinh(2\beta_1)\cosh2(\beta_1-\beta_0),\\
\langle b^\dagger_{p}(t)b^\dagger_{p}(t)\rangle&=\langle b_{-p}(t)b_{-p}(t)\rangle=\langle b^\dagger_{p}(t)b_{-p}(t)\rangle=0.
\end{align}
\end{subequations}
Summing them up, we obtain
\begin{subequations}
\begin{align}
 \langle \hat b^\dagger_p \hat b^\dagger_{-p}\rangle+\langle \hat b^\dagger_p \hat b_p\rangle+\langle \hat b_{-p} \hat b^\dagger_{-p}\rangle+\langle b_{p} \hat b_{-p}\rangle
&=A_1+A_2\cos(2v_{\rho1}pt),\\
\langle b_{-p} b^\dagger_{-p}\rangle+\langle b^\dagger_p b_p\rangle-\langle b^\dagger_p b^\dagger_{-p}\rangle-\langle b_p b_{-p}\rangle
&=B_1+B_2\cos(2v_{\rho1} pt),
\end{align}
\end{subequations}
where
\begin{subequations}
\begin{align}
A_1&=\frac{1}{2}[(\cosh\beta_0-\sinh\beta_0)^2+(\cosh(2\beta_1-\beta_0)-\sinh(2\beta_1-\beta_0))^2]=\frac{1}{2}(K_{\rho0}+\frac{K^2_{\rho1}}{K_{\rho0}}),\\
A_2&=\frac{1}{2}[(\cosh\beta_0-\sinh\beta_0)^2-(\cosh(2\beta_1-\beta_0)-\sinh(2\beta_1-\beta_0))^2]=\frac{1}{2}(K_{\rho0}-\frac{K^2_{\rho1}}{K_{\rho0}}),\\
B_1&=\frac{1}{2}[(\cosh\beta_0-\sinh\beta_0)^2+(\cosh(2\beta_1-\beta_0)-\sinh(2\beta_1-\beta_0))^2]=\frac{1}{2}(\frac{1}{K_{\rho0}}+\frac{K_{\rho0}}{K^2_{\rho1}}),\\
B_2&=\frac{1}{2}[(\cosh\beta_0-\sinh\beta_0)^2-(\cosh(2\beta_1-\beta_0)-\sinh(2\beta_1-\beta_0))^2]=\frac{1}{2}(\frac{1}{K_{\rho0}}-\frac{K_{\rho0}}{K^2_{\rho1}}).
\end{align}
\end{subequations}


Thus the ${\hat{\theta}}_\rho$ and ${\hat{\phi}}_\rho$ correlations become
\begin{align}
\label{phitsquarecorrelation}
\begin{split}
\langle[{\hat{\phi}}_\rho(x,t)-{\hat{\phi}}_\rho(0,t)]^2\rangle&= \int^\infty_0\frac{dp}{p}e^{- a  p}[1-\cos(px)]\langle(\hat b^\dagger_p(t)+\hat b_{-p}(t))(\hat b^\dagger_{-p}(t)+\hat b_{p}(t))\rangle,\\
&= \int^\infty_0\frac{dp}{p}e^{- a  p}[1-\cos(px)] (A_1+A_2\cos(2v_F  pt)),\\
&=A_1\ln\left[\frac{x}{ a }\right]+\frac{1}{2}A_2\ln\left|\frac{x^2-(2v_F t)^2}{ a ^2+(2v_F t)^2}\right|,
\end{split}
\end{align}
and
\begin{align}
\label{thetatsquarecorrelation}
\begin{split}
\langle[{\hat{\theta}}_\rho(x,t)-{\hat{\theta}}_\rho(0,t)]^2\rangle&= \int^\infty_0\frac{dp}{p}e^{- a  p}[1-\cos(px)]\langle(\hat b^\dagger_p(t)-\hat b_{-p}(t))(\hat b^\dagger_{-p}(t)-\hat b_{p}(t))\rangle,\\
&= \int^\infty_0\frac{dp}{p}e^{- a  p}[1-\cos(px)](B_1+B_2\cos(2v_F  pt)),\\
&=B_1\ln\left[\frac{x}{ a }\right]+\frac{1}{2}B_2\ln\left|\frac{x^2-(2v_F t)^2}{ a ^2+(2v_F t)^2}\right|,
\end{split}
\end{align}
At $t=0$, above equations reduce to 
\begin{subequations}
\begin{align}
\langle[{\hat{\phi}}(x,t)-{\hat{\phi}}(0,t)]^2\rangle&=K_{\rho0}\ln\left[\frac{x}{ a }\right],\\
\langle[{\hat{\theta}}(x,t)-{\hat{\theta}}(0,t)]^2\rangle&=\frac{1}{K_{\rho0}}\ln\left[\frac{x}{ a }\right],
\end{align}
\end{subequations}
consistent with the initial pre-quench results Eqs.~\eqref{quadraticaveragethetab} and \eqref{quadraticaveragephib}, as expected.
Using relation \eqref{wardrelation}, we are easy to see they reproduce the results Eqs.~\eqref{quenchluttingerphitheta} and \eqref{quenchluttingerphitheta2} in the main text.

\section{Normal modes approach for spin sector BCS quench dynamics}
\label{spinmicroscopicdynamics}
In the main text, we use the Heinsenberg equations of motion approach to study the quench dynamics for the spin sector. In this appendix, we list the normal modes approach for the spin sector dynamics, for completeness purpose as well as possible generalization to other quench protocols.

Substituting ${\hat{\phi}}_\sigma(x),{\hat{\theta}}_\sigma(x)$ using relation \eqref{phithetasemiclassical} into the post-quench Hamiltonian, we obtain the post-quench Hamiltonian in terms of bosonic creation and annihilation operator as
\begin{align}
\label{LLHKbform}
\begin{split}
\hat H_\sigma&=\frac{1}{4}\sum_{p\neq0}v_F |p|\Big[(1/K_\sigma+K_\sigma)(\hat b^\dagger_p \hat b_p+\hat b^\dagger_{-p} \hat b_{-p})+(K_\sigma-1/K_\sigma)(\hat b^\dagger_p \hat b^\dagger_{-p}+\hat b_p \hat b_{-p})\Big],
\end{split}
\end{align}
which is not diagonal since the Luttinger parameter is also quenched as $K_\sigma\to1$ following the interaction quench $U\to0$. It can be diagonalized by the following Bogoliubov transformation:
\begin{align}
\begin{split}
\begin{pmatrix}\hat b_p \\ \hat b^\dagger_{-p}\end{pmatrix}=\begin{pmatrix}\cosh\beta&\sinh\beta\\ \sinh\beta&\cosh\beta\end{pmatrix}\begin{pmatrix}\hat \chi_p \\ \hat \chi^\dagger_{-p}\end{pmatrix}\equiv U_\sigma(0^+)\begin{pmatrix}\hat \chi_p \\ \hat \chi^\dagger_{-p}\end{pmatrix}
\end{split}
\end{align}
with $e^{-2\beta}=K_\sigma$.

Prior to the quench, we know from Eq.~\eqref{Bogoliubovmatrixphiexpansionequilibrium} that the initial Hamiltonian is diagonalized by
\begin{align}
\begin{split}
\begin{bmatrix}\hat b_{p}\\ \hat b^\dagger_{-p}\\\end{bmatrix}=\begin{bmatrix} u_{p}&-v_{p}\\-v_{p}&u_{p}\\\end{bmatrix}\begin{bmatrix}\hat  \alpha_{p}\\ \hat  \alpha ^\dagger_{-p}\\\end{bmatrix}\equiv U_\sigma(0^-)\begin{bmatrix}\hat  \alpha_{p}\\ \hat  \alpha ^\dagger_{-p}\\\end{bmatrix}.
\end{split}
\end{align}
The dynamics is easily encoded in the form of ($\hat\chi_p(t),\hat\chi^\dagger_p(t)$), which satisfies
\begin{align}
\label{BCSsemiclassicaldynamicstime}
\begin{split}
\begin{pmatrix}\hat \chi_p(t) \\ \hat \chi^\dagger_{-p}(t)\end{pmatrix}
=\begin{pmatrix}e^{iv_{F}|p|t}&0\\ 0&e^{-iv_{F}|p|t}\end{pmatrix}\begin{pmatrix}\hat \chi_p \\\hat\chi^\dagger_{-p}\end{pmatrix}\equiv U_{\sigma T}(t)\begin{pmatrix}\hat \chi_p \\ \hat \chi^\dagger_{-p}\end{pmatrix}.
\end{split}
\end{align}
Thus following what is done in the charge sector dynamics section, we obtain similarly
\begin{align}
\label{betatransformation}
\begin{split}
\begin{bmatrix} \hat b_p(t)\\\hat b^\dagger_{-p}(t)\\\end{bmatrix}&=U_\sigma(0^+)U_{\sigma T}(t)U_\sigma(0^+)U_\sigma(0^-)\begin{bmatrix}\hat  \alpha_{p}\\ \hat  \alpha ^\dagger_{-p}\\\end{bmatrix}.
\end{split}
\end{align}
Combining it with Eqs.~\eqref{phithetasemiclassical} and \eqref{bogoliubovuvomega}, we then find
\begin{align}
\begin{split}
 &\quad\langle({\hat{\phi}}_\sigma(x,t)-{\hat{\phi}}_\sigma(0,t))^2\rangle\\
 &=K_\sigma\int^\infty_0\frac{dp}{p}e^{- a  p}[1-\cos(px)]\left[\langle \hat b^\dagger_p(t) \hat b^\dagger_{-p}(t)\rangle+\langle \hat b^\dagger_p(t) \hat b_p(t)\rangle+\langle \hat b_{-p}(t) \hat b^\dagger_{-p}(t)\rangle+\langle b_{p}(t) \hat b_{-p}(t)\rangle\right], \\
 &=K_\sigma\int^\infty_0\frac{dp}{p}e^{- a  p}[1-\cos(px)]
 \left[(1+2v^2_p-2u_pv_p)[1+(K^{-2}_\sigma-1)(1-\cos(2v_F pt))/2]+2K^{-2}_\sigma u_p v_p[1-\cos(2v_F pt)]\right],\\
&=K_\sigma\int^\infty_0\frac{dp}{\sqrt{p^2+1/\lambda^2}}e^{- a  p}[1-\cos(px)][1+(K^{-2}_\sigma-1)(1-\cos(2v_F pt))/2],\\
&\quad+\frac{1}{2K_\sigma\lambda^2}\int^\infty_0\frac{dp}{p^2\sqrt{p^2+1/\lambda^2}}e^{- a  p}[1-\cos(px)][1-\cos(2v_F pt)],\\
&\sim \frac{K^{-1}_\sigma+K_\sigma}{2}\arcsinh(\lambda/a )+Z(x,t),
\end{split}
\end{align}
where
\begin{align}
\begin{split}
Z(x,t)&\equiv\frac{1}{2K_\sigma\lambda^2}\int^\infty_0\frac{dp}{p^2\sqrt{p^2+1/\lambda^2}}e^{- a  p}[1-\cos(px)][1-\cos(2p v_F t)],\\
&=\frac{1}{2K_\sigma\lambda^2}\int^\infty_0\frac{dp}{p^2\sqrt{p^2+1/\lambda^2}}e^{- a  p}\left(1-\cos(px)+1-\cos(2p v_F t)-\frac{1-\cos(p(x-2v_Ft))}{2}-\frac{1-\cos(p(x+2v_Ft))}{2}\right),\\
&\sim\frac{\pi}{8K_\sigma\lambda}\left(x+2v_Ft-|x-2v_Ft|\right)-\frac{1}{2K_\sigma},\\
&\sim\begin{cases} \frac{2v_F t}{2\xi}-\frac{1}{2K_\sigma}, &  x\gg 2v_Ft,
\\ \frac{x}{2\xi }-\frac{1}{2K_\sigma} , &  x\ll 2v_Ft, \end{cases}
\end{split}
\end{align}
and we have used the Meijer G function property \eqref{meijerGfuntionasymptotic} once again. The ${\hat{\phi}_\sigma}$ correlation can then be evaluated as
\begin{align}
\begin{split}
\langle e^{i\sqrt{2}[{\hat{\phi}}_\sigma(x,t)-{\hat{\phi}}_\sigma(0,t)]}\rangle
&\sim e^{- Z(x,t) }
\sim\begin{cases} e^{-\frac{2v_F t}{2\xi}}, &  x\gg 2v_F t,
\\ e^{- \frac{x}{2\xi}} , &  x\ll 2v_F t,\end{cases}
\end{split}
\end{align}
which is Eq.~\eqref{DynamicsBCSphi} in the main text.

Similarly, for the ${\hat{\theta}}_\sigma-{\hat{\theta}}_\sigma$ correlation we have
\begin{align}
\begin{split}
 &\quad\langle({\hat{\theta}}_\sigma(x,t)-{\hat{\theta}}_\sigma(0,t))^2\rangle\\
 &=\frac{1}{K_\sigma}\int^\infty_0\frac{dp}{p}e^{- a  p}[1-\cos(px)]\left[\langle \hat b^\dagger_p(t) \hat b_p(t)\rangle+\langle \hat b_{-p}(t) \hat b^\dagger_{-p}(t)\rangle-\langle \hat b^\dagger_p(t) \hat b^\dagger_{-p}(t)\rangle-\langle b_{p}(t) \hat b_{-p}(t)\rangle\right],\\
  &=\frac{1}{K_\sigma}\int^\infty_0\frac{dp}{p}e^{- a  p}[1-\cos(px)]
 \left[(1+2v^2_p-2u_pv_p)[1+(K^{2}_\sigma-1)(1-\cos(2v_F pt))/2]+2 u_p v_p[1+\cos(2v_F pt)]\right],\\
 &=\frac{1}{K_\sigma}\int^\infty_0 dp\frac{\sqrt{p^2+1/\lambda^2}}{p^2}e^{- a  p}[1-\cos(px)][1+(K^{2}_\sigma-1)(1-\cos(2v_F pt))/2],\\
&\quad+\frac{1}{2K_\sigma\lambda^2}\int^\infty_0\frac{dp}{p^2\sqrt{p^2+1/\lambda^2}}e^{- a  p}[1-\cos(px)][1+\cos(2v_F pt)],\\
&\sim \frac{K^{-1}_\sigma+K_\sigma}{2}\arcsinh(\lambda/a )+Z'(x,t),
\end{split}
\end{align}
where
\begin{align}
\begin{split}
Z'(x,t)&\equiv\frac{1}{2K_\sigma\lambda^2}\int^\infty_0\frac{dp}{p^2\sqrt{p^2+1/\lambda^2}}e^{- a  p}[1-\cos(px)][1+\cos(2p v_F t)],\\
&=\frac{1}{2K_\sigma\lambda^2}\int^\infty_0\frac{dp}{p^2\sqrt{p^2+1/\lambda^2}}e^{- a  p}\left(1-\cos(px)-(1-\cos(2p v_F t))+\frac{(1-\cos(p(x-2v_Ft))}{2}+\frac{(1-\cos(p(x+2v_Ft))}{2}\right),\\
&\sim\frac{\pi}{8K_\sigma\lambda}(3x-2v_Ft+|x-2v_Ft|)-\frac{1}{2K_\sigma},\\
&\sim\begin{cases} \frac{x-v_F t}{\xi_0 }-\frac{1}{2K_\sigma}, &  x\gg 2v_Ft,
\\ \frac{x}{2\xi_0 }-\frac{1}{2K_\sigma} , &  x\ll 2v_Ft. \end{cases}
\end{split}
\end{align}
The ${\hat{\theta}}_\sigma$ correlation thereby is
\begin{align}
\begin{split}
\langle e^{i\sqrt{2}[{\hat{\theta}}_\sigma(x)-{\hat{\theta}}_\sigma(0)]}\rangle&\propto e^{- Z'(x,t)}\sim\begin{cases} e^{-\frac{x-2v_Ft}{\xi_0 }}, &  x\gg 2v_F t,
\\ e^{-\frac{x}{2\xi_0 } }, &  x\ll 2v_F t. \end{cases}
\end{split}
\end{align}
which we recognize as Eq.~\eqref{DynamicsBCStheta} in the main text. It is also worth noting that the following correlations
\begin{subequations}
\begin{align}
\langle e^{i\sqrt{2}[{\hat{\phi}}_\sigma(x,t)+{\hat{\phi}}_\sigma(0,t)]}\rangle&=0,\\
\langle e^{i\sqrt{2}[{\hat{\theta}}_\sigma(x,t)+{\hat{\theta}}_\sigma(0,t)]}\rangle&=0,
\end{align}
\end{subequations}
are both zero.

Generalizing the above apporoach to other quench protocols such as $U\to U'\neq 0$ should also be straightforward. In that case, the Hamiltonian \eqref{LLHKbform} will contain extra terms from the finite nonlinearity $U'\cos {\hat{\phi}}_\sigma$ term, which can be again expressed in terms of quadratic operators of ($\hat b_p, \hat b^\dagger_p$) by expanding the cosine potential around its minimum. The post-quench dynamics will be governed by a matrix similar to Eq.~\eqref{BCSsemiclassicaldynamicstime}, but with $v_F$ being replaced with the post-quench spin velocity $v_{\sigma1}$. Other than that, the calculation procedure  follows exactly the same as the one discussed above. We leave this generalization to future work.

\section{Wick's theorem for Multiple fields}
\label{Wickmultiple}
For Gaussian field $\tilde{\phi}(x)$, the relation \eqref{wardrelation} can be generalized to the multiple fields case as
\begin{align}
\label{Wickmultiplefields}
\begin{split}
\langle e^{i\sum_j n_j {\hat{\phi}}(x_j)}\rangle
&=e^{-\frac{1}{2}\langle \left(\sum_i n_i {\hat{\phi}}(x_i)\right)^2\rangle}\\
&=e^{-\frac{1}{2}\sum_i n^2_i\langle  {\hat{\phi}}^2(0)\rangle -\sum_{i>j}n_i n_j\langle  \hat{\phi}(x_i)\hat{\phi}(x_j)\rangle}\\
&=e^{-\frac{1}{2}\left(\sum_i n_i\right)^2\langle  {\hat{\phi}}^2(0)\rangle+\frac{1}{2}\sum_{i>j}n_i n_j\langle ( \hat{\phi}(x_i)-  \hat{\phi}(x_j))^2\rangle}\\
&=e^{-\frac{1}{2}\left(\sum_i n_i\right)^2 \ln (L/a)+\frac{1}{2}\sum_{i>j}n_i n_j\langle ( \hat{\phi}(x_i)-  \hat{\phi}(x_j))^2\rangle}\\
&=\delta\left(\sum_i n_i\right) e^{\frac{1}{2}\sum_{i>j}n_i n_j\langle ( \hat{\phi}(x_i)-  \hat{\phi}(x_j))^2\rangle]},
\end{split}
\end{align}
where $\delta$ is the Kronicker delta function.
\end{widetext}

\end{document}